\documentclass[reprint,superscriptaddress]{revtex4-2}

\pdfoutput=1

\usepackage{graphicx}% Include figure files
\graphicspath{ {Figures/} }
\usepackage{dcolumn}% Align table columns on decimal point
\usepackage{bm}% bold math
\usepackage{xcolor}
\usepackage{colortbl}
\usepackage{soul}
\usepackage[normalem]{ulem}
\usepackage{amsmath}
\usepackage[english]{babel}
\usepackage{braket}
\usepackage{units}
\usepackage[T1]{fontenc}
\usepackage{mathtools}
\usepackage[colorlinks]{hyperref}
\usepackage{titletoc}
\usepackage{tocloft}
\usepackage{multirow}
\usepackage{rotating}
\usepackage{tabularx}

\usepackage{ifthen}
\usepackage{amssymb}
\usepackage{xcolor}

\newboolean{showcomments}
\setboolean{showcomments}{true}

\makeatletter
\newcommand{\mynote}[3]{%
  \ifthenelse{\boolean{showcomments}}{%
   \fbox{\bfseries\sffamily\scriptsize#1}%
   {\small$\blacktriangleright$\textsf{\emph{\color{#3}{#2}}}$\blacktriangleleft$}}%
  {%
   \@bsphack
   \@esphack
  }%
}
\makeatother

\definecolor{asparagus}{rgb}{0.53, 0.66, 0.42}

\makeatletter

    \def\CT@@do@color{%
      \global\let\CT@do@color\relax
            \@tempdima\wd\z@
            \advance\@tempdima\@tempdimb
            \advance\@tempdima\@tempdimc
    \advance\@tempdimb\tabcolsep
    \advance\@tempdimc\tabcolsep
    \advance\@tempdima2\tabcolsep
            \kern-\@tempdimb
            \leaders\vrule
                    \hskip\@tempdima\@plus  1fill
            \kern-\@tempdimc
            \hskip-\wd\z@ \@plus -1fill }
\makeatother

\newcommand{\ch}[1]{{\color{black}#1}}

\definecolor{black}{rgb}{0.0,0.0,0.0}

\newcommand{\Hdisp}{\hat{\mathcal{H}}_{\rm disp}}
\newcommand{\Lchain}{\mathcal{L}_{\rm envt}}
\newcommand{\Lqmodes}{\mathcal{L}_{\rm q}}
\newcommand{\Lsys}{\mathcal{L}_{\rm sys}}
\newcommand{\Lmeas}{\mathcal{L}_{\rm meas}}
\newcommand{\Smeas}{\mathcal{S}[dW]}

\newcommand{\RC}{TPP}

\newcommand{\rhou}{\hat{\rho}}
\newcommand{\rhoc}{\hat{\rho}_c}
\newcommand{\avg}[1]{\langle #1 \rangle}
\newcommand{\avgc}[1]{\langle #1 \rangle_c}

\newcommand{\proj}[2]{ |#1\rangle \! \langle #2 |}
\renewcommand{\braket}[2]{ \langle #1 | #2 \rangle }

\newcommand{\vv}[1]{ \vec{#1} }

\newcommand{\xiI}{ \xi_I }
\newcommand{\xiQ}{ \xi_Q }
\newcommand{\xiqI}{ \xi_I^{\rm qm} }
\newcommand{\xiqQ}{ \xi_Q^{\rm qm} }
\newcommand{\xicI}{ \xi_I^{\rm cl} }
\newcommand{\xicQ}{ \xi_Q^{\rm cl} }

\newcommand{\E}[1]{ \mathbb{E}[#1] }
\newcommand{\tmeas}{\mathcal{T}_{\rm meas}}

\newcommand{\sigmapred}{\sigma^{\rm est}}

\newcommand{\fidmf}{\mathcal{F}_{\rm FGDA}}
\newcommand{\fidrc}{\mathcal{F}_{\rm \RC{}}}
\newcommand{\metric}{\mathcal{E} }

\newcommand{\WO}{ \mathbf{W} }
\newcommand{\WOopt}{ \mathbf{W}^{\rm opt} }
\newcommand{\bv}{ \mathbf{b} }
\newcommand{\yv}{ \mathbf{y} }

\newcommand{\NTrain}{N_{\rm train}}
\newcommand{\NTest}{N_{\rm test}}
\newcommand{\NTraj}{N_{\rm traj}}
\newcommand{\NT}{N_{\rm T}}
\newcommand{\NO}{N_{\rm O}}

\newcommand{\Lm}{\mathbf{L}}

\newcommand{\fgauss}[1]{ {\rm G}\!\left[ #1 \right] }
\newcommand{\frc}[1]{ {\rm F}\!\left[ #1 \right] }

\begin{document}

\preprint{APS/123-QED}

\makeatletter

\title{Practical Trainable Temporal Postprocessor for Multistate Quantum Measurement}

\author{Saeed A. Khan}
\affiliation{Department of Electrical Engineering, Princeton University, Princeton, NJ 08544, USA}
\author{Ryan Kaufman}
\affiliation{Department of Physics and Astronomy, University of Pittsburgh, Pittsburgh, PA, USA}
\author{Boris Mesits}
\affiliation{Department of Physics and Astronomy, University of Pittsburgh, Pittsburgh, PA, USA}
\author{Michael Hatridge}
\affiliation{Department of Physics and Astronomy, University of Pittsburgh, Pittsburgh, PA, USA}
\author{Hakan E. T\"ureci}
\affiliation{Department of Electrical Engineering, Princeton University, Princeton, NJ 08544, USA}

% \author{Saeed A. Khan$^2$, Ryan Kaufman$^1$, Michael Hatridge$^1$, Hakan E. T\"ureci$^2$}
% \affiliation{$^1$Department of Physics and Astronomy, University of Pittsburgh, Pittsburgh, PA, USA}
% \affiliation{$^2$Department of Electrical Engineering, Princeton University, Princeton, NJ 08544, USA}

\date{\today}% It is always \today, today,
             %  but any date may be explicitly specified

% Quantum state readout typically operates under weak powers so that relevant temporal information must be extracted from measurement data that is dominated by noise, especially that with a quantum origin. Here, w

\begin{abstract}
    We develop and demonstrate a trainable temporal post-processor (\RC{}) harnessing a simple but versatile machine learning algorithm to provide optimal processing of quantum measurement data subject to arbitrary noise processes, for the readout of an arbitrary number of quantum states. We demonstrate the \RC{} on the essential task of qubit state readout, which has historically relied on temporal processing via matched filters in spite of their applicability only for specific noise conditions. Our results show that the \RC{} can reliably outperform standard filtering approaches under complex readout conditions, such as high power readout. Using simulations of quantum measurement noise sources, we show that this advantage relies on the \RC{}'s ability to learn optimal linear filters that account for general quantum noise correlations in data, such as those due to quantum jumps, or correlated noise added by a phase-preserving quantum amplifier. \ch{Furthermore, we derive an exact analytic form for the optimal \RC{} weights: this positions the \RC{} as a linearly-scaling generalization of matched filtering, valid for an arbitrary number of states under the most general readout noise conditions, all while preserving a training complexity that is essentially negligible in comparison to that of training neural networks for processing temporal quantum measurement data.} The \RC{} can be autonomously and reliably trained on measurement data and requires only linear operations, making it ideal for FPGA implementations in cQED for real-time processing of measurement data from general quantum systems.
\end{abstract}

% We show that the transformation described by the \RC{} can be expressed via an efficient semi-analytic form, providing a linearly-scaling generalization of matched filtering to an arbitrary number of states under the most general noise conditions of the readout signal emanating from the measurement chain. The \RC{} can be efficiently, autonomously, and reliably trained on measurement data, and requires only linear operations, making it ideal for FPGA implementations in cQED for real-time processing of measurement data from general quantum systems.

\maketitle

\section{Introduction}

% High fidelity qubit state readout is essential for practically all quantum information processing protocols, and its optimization has therefore been the focus of sustained research efforts. In the circuit QED (cQED) architecture, advances in the standard dispersive readout framework have been enabled primarily via quantum hardware technologies: quantum-limited amplifiers with large bandwidth and dynamic range~\cite{Roy2016, Aumentado2020}, long-lived qubits~\cite{place_new_2021}, and optimal readout protocols~\cite{walter_rapid_2017}. Nevertheless, in the push towards higher readout fidelities in complex multi-qubit processors of a practical scale, not all hardware conditions are, or remain, ideal. For example, the optimization of individual readout resonators becomes increasingly difficult with the growing scale of multi-qubit processors. More importantly, finite qubit coherence means that simply extending the measurement duration is not a viable option to enhance fidelity: faster and hence higher power measurements are needed. However, these readout powers are associated with enhanced qubit transitions, leading to the $T_1$ versus $\bar{n}$ problem~\cite{Sank2016, khezri_measurement-induced_2023, Shillito2022} and excitation to higher states~\cite{Shillito2022} outside the computational subspace. Under these conditions, it becomes imperative to be able to extract the maximum information possible from all available quantum measurement data.

High fidelity quantum measurement is essential for any quantum information processing scheme, from quantum computation to quantum machine learning. However, while measurement optimization has focused on quantum hardware advancements~\cite{Roy2016, Aumentado2020, place_new_2021}, several modern experiments operate in regimes where optimal hardware conditions are difficult to sustain, or - for machine learning with general quantum systems~\cite{angelatos_reservoir_2021, khan_physical_2021, nokkala_gaussian_2021, martinez-pena_dynamical_2021, mujal_opportunities_2021} - may not always be known. For example, in the push towards higher qubit readout fidelities with complex multi-qubit processors in circuit QED (cQED), optimization of individual readout resonators becomes increasingly difficult. More importantly, finite qubit coherence means that simply extending the measurement duration is not a viable option to enhance fidelity: faster and hence higher power measurements are needed. However, these readout powers are associated with enhanced qubit transitions, leading to the $T_1$ versus $\bar{n}$ problem~\cite{Sank2016, malekakhlagh_lifetime_2020, petrescu_lifetime_2020, hanai_intrinsic_2021, khezri_measurement-induced_2023, Shillito2022, thorbeck_readout-induced_2024} and excitation to higher states~\cite{gusenkova_quantum_2021, Shillito2022, Dumas2024} outside the computational subspace. Machine learning with quantum devices operating in unconventional regimes allows for an even broader range of complex dynamics. Quantum measurement data obtained under these conditions cannot be expected to be optimally analyzed using schemes built for more standard readout paradigms~\cite{walter_rapid_2017}. Therefore, a practical approach to extract the maximum information possible from such data is timely.

\begin{figure}[t]
    \centering
    \includegraphics[scale=1.0]{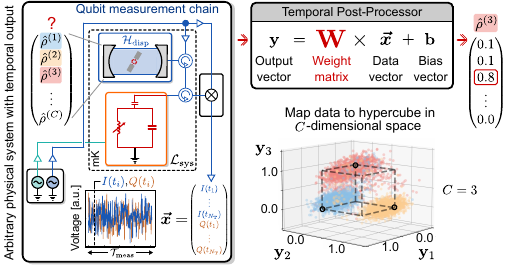}
    \caption{\textbf{Temporal post-processor (\RC{}) for multi-state classification using quantum measurement data, demonstrated for dispersive qubit readout in cQED.} The objective is to process temporal data corresponding to an unknown state (indexed $\sigma$) of an arbitrary physical system - here the state of a qubit in a quantum measurement chain - to estimate the true label $\sigma$ with maximum accuracy. The \RC{} approach uses a set of weights $\WO$ and biases $\bv$ to map the vector $\bm{\vec{x}}$ of measured data, comprising an instance of $\NO$ observables each a time series of length $\NT$, to the corners of a hypercube in $C$-dimensional space. Optimal values of $\WO$ and $\bv$ are learned by training to realize this mapping with minimal error, in a least-squares sense. Scatter plots shown in $C=3$ dimensional space are data from real qubit $p\in\{e,g,f\}$ readout after the \RC{}. }
    \label{fig:schematic}
\end{figure}

In this paper, we demonstrate a machine learning scheme to optimally process quantum measurement data for completely general quantum state classification tasks. For the most common such task of single-shot qubit state readout, standard post-processing of measurement records has remained relatively unchanged (with some exceptions~\cite{tsang_volterra_2015, lienhard_deep-neural-network_2022}): data is filtered using a ``matched filter'' (MF) constructed from the sample mean of measurement records for two states to be distinguished (for example, states $\ket{e}$ or $\ket{g}$ of a qubit). Crucially, the MF thus defined applies only to binary classification, and much more restrictively, is optimal only \ch{for idealized} conditions under which \ch{the} readout signal is subject to Gaussian white (i.e. uncorrelated) noise processes~\cite{gambetta_protocols_2007}. In many deployments where complex conditions prevail (such as multi-qubit readout) an even simpler and less optimal boxcar filter is employed, due to the ease of its construction. Our approach harnesses machine learning to provide a model-free trainable temporal post-processor (\RC{}) of quantum measurement data under \ch{the} most general noise conditions, and for an arbitrary number of states of a generic measured quantum system~(\footnote{source code available at \url{https://zenodo.org/doi/10.5281/zenodo.10020462}} for source code). We test our approach by applying it to the experimental readout of distinct qubits across a range of measurement powers. Our results demonstrate that the \RC{} reliably outperforms the standard MF whenever measured data exhibits nontrivial temporal correlations, including those of a quantum origin. We find an important such regime that has garnered significant recent attention to be that of high-power readout~\cite{gusenkova_quantum_2021, takmakov_minimizing_2021, sunada_fast_2022, bengtsson_model-based_2024, sank_system_2024}; here, we experimentally show that the \RC{} can provide a reduction in errors by up to 30\% in certain cases. Furthermore, the \RC{} achieves this improvement while requiring only linear weights applied to quantum measurement data (see Fig.~\ref{fig:schematic}): this makes it compatible with FPGA implementations for real-time hardware processing, and exacts a lower training cost~\cite{tanaka_recent_2019, gauthier_next_2021} than neural network-based machine learning schemes~\cite{lienhard_deep-neural-network_2022, luchi_enhancing_2023, maurya_scaling_2023}.

% However, for completely general the cases of qubit readout at strong drive powers and using amplifiers operated at strong pumps~[], it is no longer \textit{a priori} guaranteed that such optimal conditions hold.

% Our approach therefore focuses on an important aspect of quantum state readout that has remained relatively unchanged: the post-processing of measurement records. The standard approach is to use matched filters for classifying the $\ket{e}$ versus $\ket{g}$ state of a qubit. In many cases, an even simpler boxcar filter is employed, due to the ease of its training-free implementation. However, these matched filters do not easily extend to higher excited states of a given qubit. 

% \mjh{I think here we need to say a bit more about fidelity: unlike say QPSK where fidelity can be achieved with a bit more integration time qubit decay means readout must be fast; 99 percent fidelity means 1 percent of readout tone so efficiently capturing this information is truly vital for high fidelity readout.  Can cite old paper by Gambetta; newer Silveri, etc.}

 % However, in the cases of qubit readout at strong drive powers and using amplifiers operated at strong pumps~[], it is no longer \textit{a priori} guaranteed that such optimal conditions hold. Our scheme based on reservoir computing provably reduces to the matched filter in the case of Gaussian white noise. However, it is able to account for correlations between the noise process in more general cases, learning distinct filters that can outperform the matched filter.

Machine learning has already been established as a powerful approach to \textit{classical} temporal data processing, providing state-of-the-art fidelity in tasks such as time series prediction~\cite{canaday_rapid_2018}, and chaotic systems' forecasting~\cite{pathak_using_2017, pathak_model-free_2018, griffith_forecasting_2019} and control~\cite{canaday_model-free_2021}. Adapting this approach to quantum state classification as we do here requires its application to time-evolving \textit{quantum} signals. Signals extracted from the readout of quantum systems are often dominated by noise, making their processing distinct from that required of typical data from classical systems. More importantly, the noise in such signals can arise from truly quantum-mechanical sources, such as stochastic transitions between states of a multi-level atom (quantum `jumps'), or vacuum fluctuations in quantum modes. A key finding of our work is that the \RC{} is able to learn from precisely these quantum noise correlations in data extracted from quantum systems to improve classification fidelity. To uncover this essential principle of \RC{} learning, we first develop an interpretation of the \RC{} as the application of optimal filters to quantum measurement data. This provides a framework to quantify and visualize what is `learned' by the \RC{} from a given dataset. Secondly, the \RC{} is tested on simulated quantum measurement datasets using stochastic master equations, where quantum noise sources and hence their correlation signatures in measured data can be precisely controlled. 

% Using theoretical simulations of full quantum measurement chains for different readout conditions, we are then able to demonstrate how the filters learned by the \RC{} depend on the quantum noise characteristics of readout signals.

% Our results therefore provide a key demonstration of the practical utility of \RC{} for quantum state readout, we further provide a deeper understanding of the learning principles behind that enable this utility.

Using simulated datasets where all noise sources contribute additive Gaussian white noise - a reasonable assumption for measurement chains under asymptotically ideal conditions - we show that the \RC{} provides filters that reduce \textit{exactly} to the matched filter for binary classification. More importantly, as the \RC{} is valid for the classification of any number of states, it provides the generalization of matched filters for arbitrary state classification. We then provide a systematic analysis of \RC{} applied to quantum measurement with more complex quantum noise sources, such as quantum amplifiers adding correlated quantum noise, or noise due to state transitions. In such scenarios the \RC{} provides filters adapted to the noise characteristics: we also provide an efficient semi-analytic form for these general \RC{} filters, which can deviate substantially from filters learned under the white noise assumption, and crucially outperform the latter in qubit classification. By learning from quantum noise correlations, the \RC{} therefore utilizes a characteristic of quantum measurement data inaccessible to post-processing schemes relying on noise-agnostic matched filtering methods.

% learn from correlations not just due to classical signals, or in principle due to quantum noise in theory, but from practical systems where the majority of the noise is quantum in origin

% Autonomous and reliable use of TP

% These results demonstrate the importance of the \RC{}'s ability to learn from quantum noise correlations as a means to improve classification accuracy.

The established learning principles provide a structure and interpretability to the general applicability of the \RC{} which enhances its practical utility. First, the exact mapping to matched filters under appropriate noise conditions places the \RC{} on firm footing, guaranteed to perform at least as well as these baseline methods. Secondly, and much more importantly, the \RC{}'s ability to learn from noise (crucially, quantum noise) renders it able to then beat the MF when noise conditions change. This theoretical adaptability becomes practical due to the \RC{}'s straightforward training procedure, which is also ideal for autonomous repeated calibrations, necessary on even industrial-grade quantum processors~\cite{IBM_tuneup_1, Kelly2018, Dai2021}. Ultimately, the trainable \RC{} could provide an ideal component to optimally process quantum measurement data from general quantum devices used for machine learning, which could exhibit exotic quantum noise characteristics.

% More, the \RC{} method we demonstrate in this paper has several practical advantages which add to its utility.  First, it performs at least as well as current-best matched filter methods, second it is autonomous and so ideal for repeated calibration which is Finally, as previously noted, it makes very few assumptions about the nature of the data it processing, making it robust against non-idealities in measurement chains/qubit readout.  

The rest of this paper is organized as follows. In Sec.~\ref{sec:rc} we introduce the \RC{} framework to multi-state classification: a model-free supervised machine learning approach that can be applied to the classification of arbitrary time series. We also introduce the task used to demonstrate the \RC{} - dispersive qubit readout in the cQED architecture - and standard approaches currently used for this task. In Sec.~\ref{sec:rcfilter} we draw connections between the \RC{} approach and these standard filtering-based approaches to qubit state measurement, and provide the \RC{}'s generalization of matched filtering to arbitrary states. In Sec.~\ref{sec:rcexpt}, we apply the developed \RC{} framework to experimental data for qubit readout, showing that it can outperform standard matched-filtering at strong measurement powers relevant for high-fidelity readout. Sec.~\ref{sec:corr} explores the learning principles that enable the \RC{} to be more effective than standard matched filters using controlled simulations. We conclude with a discussion on the general applicability of \RC{} for quantum state classification and temporal processing of quantum measurement data.

% \mjh{usually at end of intro theres a bit of a `and we found good things to be excited about, are there any numerical results or statemetns of efficacy we can make here or tie to my statemetns above?}

%%%%%%%%%%%%%%%%%%%%%%%%%%%%%%%%%%%%%%%%%%%%%%%%%%%%%%%%%%%%%%%%%%%%%%%%%%%%%%%%%%%%%%%%%%%%%%%%%%%

%%%%%%%%%%%%%%%%%%%%%%%%%%%%%%%%%%%%%%%%%%%%%%%%%%%%%%%%%%%%%%%%%%%%%%%%%%%%%%%%%%%%%%%%%%%%%%%%%%%

\section{Trainable temporal post-processor for multi-state classification}
\label{sec:rc}

To overview its key features we first introduce the mathematical framework underpinning our trainable temporal post-processor (\RC{}), which is defined as follows. We consider $\NO$ continuously measured observables, each measurement yielding a time series of length $\NT$. All measured data corresponding to an unknown state with index $\sigma$ can be compiled into the vector $\bm{\vv{x}}^{(\sigma)}$ which thus exists in the space $\bm{\vv{x}}^{(\sigma)} \in \mathbb{R}^{\NO\cdot\NT}$, where $\vec{(\cdot)}$ specifies vectors containing vectorized \textit{temporal} data; examples will be provided shortly (see also Fig.~\ref{fig:schematic}). 

% \het{The vector $\bm{\vv{x}}^{(\sigma)}$ being indicated by both an arrow and a bold symbol is confusing. If there is an additional motivation for using such overloaded notation, it should stated here.}

% the measured data for the $m$th observable is a time series of length $\NT$, and is thus given by the vector $\vv{x}_m^{(p)} \in \mathbb{R}^{\NT}$, for $m \in [\NO]$. For compactness, we can introduce the vector $\bm{\vv{x}}^{(\sigma)}$ which compiles \textit{all} measured data corresponding to an unknown state with index $\sigma$, $\bm{\vv{x}}= \begin{psmallmatrix} \vv{x}_{1} \\ \vdots \\ \vv{x}_{\NO} \end{psmallmatrix}$, and hence exists in the space $\bm{\vv{x}} \in \mathbb{R}^{\NO\cdot\NT}$ (suppressing state labels for clarity). As an example, in the case of heterodyne measurement, $\NO = 2$ and $\bm{\vv{x}} = \begin{psmallmatrix} \vv{I} \\ \vv{Q} \end{psmallmatrix}$ (see Fig.~\ref{fig:schematic}).

Formally, operation of the \RC{} is then described as an input-output transformation, mapping a vector $\bm{\vv{x}}^{(\sigma)}$ from the space of measured data, $\mathbb{R}^{\NO\cdot\NT}$, to a vector $\mathbf{y} \in \mathbb{R}^{C}$ in the space of class labels; the scalar predicted class label $\sigmapred$ is given by an operation $\frc{\cdot}$ on this vector $\mathbf{y}$, so that the complete transformation is:
\begin{align}
    \sigmapred = \frc{\mathbf{y}} = \frc{ \WO\bm{\vv{x}}^{(\sigma)} + \bv }
    \label{eq:rcmap}
\end{align}

Crucially, the \RC{} transformation - defined by a trainable matrix of weights $\WO \in \mathbb{R}^{C \times \NO\cdot\NT}$ and a trainable vector of biases $\bv \in \mathbb{R}^C$ - is \textit{linear}. Machine learning using only linear trainable weights has shown remarkable success in time-dependent supervised machine learning tasks to map time series to a dynamically-evolving target function, although with a focus on classical data with weak noise~\cite{tanaka_recent_2019, gauthier_next_2021}. Here, we adapt this framework to processing of temporal measurement data from a quantum system and with a time-\textit{independent} target, as is relevant for initial state classification~\cite{gambetta_protocols_2007}. 

More precisely, $\WO$ and $\bv$ are both learned from sampled data $\bm{\vv{x}}^{(p)}$ with \textit{known} labels $p$ ($C$ in total) in a supervised learning framework. The target $\mathbf{y} \in \mathbb{R}^C$ for any instance of $\bm{\vv{x}}^{(p)}$ is taken to be a vector with only one nonzero element - a single $1$ at index $p$, defining a corner of a $C$-dimensional hypercube (referred to as one-hot encoding, see Fig.~\ref{fig:schematic}). Then, the optimal $\WOopt, \bv^{\rm opt}$ minimize a least-squares cost function to achieve this target with minimal error:
\begin{align}
    \{\WO^{\rm opt},\mathbf{b}^{\rm opt}\} = \underset{\WO,\mathbf{b}}{\rm argmin}||\mathbf{Y}- \left(\WO\mathbf{X}+\bv\right) ||^2 
    \label{eq:wopt}
\end{align}
Here $\mathbf{X}$ is the matrix containing the complete training dataset, comprising $\NTrain$ instances of $\bm{\vv{x}}^{(p)}$ for each class~$p$, while $\mathbf{Y}$ is the corresponding set of targets~(see Appendix~\ref{app:training} for full training details).  

% \het{Table I should contain the dimensions of $X$ and $Y$ as well. To be consistent this paragraph should discuss how $X$ is constructed from $\bm{\vv{x}}^{(p)}$. Should $W$ here not multiply $X$ from the right?}

A distinguishing feature of the \RC{} framework amongst other ML paradigms is that its optimization is convex and hence guaranteed to converge. We note that the function $\frc{\cdot}$ used to map the \RC{} output to an estimated class label is \textit{untrained}, and hence does not effect the training complexity; it is often taken to be the ${\rm argmax}\{\cdot\}$ function that extracts the position of the largest element in $\mathbf{y}$. However, it can also be a more general classifier, such as a Gaussian discriminator (clarified shortly). The dimensions of the various components making up the \RC{} framework are summarized in Table~\ref{tab:matShapes}.

\newcolumntype{Y}{>{\centering\arraybackslash}X}
\begin{table}[]
    \centering
    \caption{Summary of components of the \RC{} learning framework and their dimensions.}
    \begin{tabularx}{246pt}{lYl}
    \hline
    \hline
         Component & Symbol & Dimensions \\
        \hline
         \RC{} output & $\mathbf{y}$ & $\mathbb{R}^C$ \\
        Weights & $\mathbf{W}$ &  $\mathbb{R}^{C\times (\NO\cdot\NT)}$ \\
        Data & $\bm{\vec{x}}$ &  $\mathbb{R}^{\NO\cdot\NT}$ \\ 
        Bias & $\bv$ &  $\mathbb{R}^C$ \\ 
        \hline
        Data means, state $p$ & $\bm{\vec{s}}^{(p)}$ &  $\mathbb{R}^{\NO\cdot\NT}$  \\ 
        Noise process, state $p$ & $\bm{\vec{\zeta}}^{(p)}$ & $\mathbb{R}^{\NO\cdot\NT}$ \\ 
        \hline 
        ``Gram'' matrix & $\mathbf{G}$ & $\mathbb{R}^{(\NO\cdot\NT)\times(\NO\cdot\NT)}$ \\ 
        Correlation matrix & $\mathbf{V}$ & $\mathbb{R}^{(\NO\cdot\NT)\times(\NO\cdot\NT)}$
    \end{tabularx}
    \label{tab:matShapes}
\end{table}

% The second defining feature is the scope of applicability of the \RC{} framework. It natively generalizes to the classification of an arbitrary number of states $C$. Furthermore, no restriction is placed on the type of data that constitutes the vector $\bm{\vv{x}}$. In particular, no underlying physical model of the system generating the measurement data is \textit{a priori} required: any relevant information must be learned by the \RC{} from data during the training phase. This also implies that the results in this paper apply to the classification of time series that have nothing to do with qubit state measurement. Its generality and ease of training enable the \RC{} to serve as a versatile trainable classifier, suited to a variety of classification tasks. 

\subsection{Learning from noise correlations}

While Eq.~(\ref{eq:rcmap}) presents a formal mathematical formulation of the \RC{} framework in the machine learning context, we can develop further understanding of how the \RC{} learns from data to enable classification. To this end, we first note that this stochastic measurement data can be written in the very general form: 
\begin{align}
    \bm{\vec{x}}^{(\sigma)} = \bm{\vec{s}}^{(\sigma)} + \bm{\vec{\zeta}}^{(\sigma)}
    \label{eq:stochdata}
\end{align}
Here $\bm{\vec{\zeta}}^{(\sigma)}$ describes the stochasticity of the measured data: most importantly, we are interested in data where $\bm{\vec{\zeta}}^{(\sigma)}$ will be dominated by contributions from quantum noise sources. We take the noise process to have zero mean, $\mathbb{E}[\bm{\vec{\zeta}}^{(\sigma)}_j] = 0$, where $\E{\cdot}$ describes ensemble averages over distinct noise realizations (obtained for distinct measurements). Then, $\bm{\vec{s}}^{(\sigma)} = \mathbb{E}[\bm{\vec{x}}^{(\sigma)}]$ are simply the sample mean of the measured data traces for state $\sigma$. Crucially, the noise is characterized by nontrivial second-order temporal correlations, which we define as $\bm{\Sigma}^{(\sigma)}_{jk} = \mathbb{E}[\bm{\vec{\zeta}}^{(\sigma)}_j\bm{\vec{\zeta}}^{(\sigma)}_k]$. Higher-order correlations of the noise can also be generally non-zero, but are not explicitly analyzed here due to the \RC{}'s use of a quadratic loss function. 

% for heterodyne measurement, for example, this includes the noise sources from Eq.~(\ref{eq:hetI}),~(\ref{eq:hetQ}), including quantum noise.

% \begin{align}
%     \mathbf{M} = 
%         \begin{pmatrix}
%         (\bm{\vec{s}}^{(1)})^T & 1 \\
%         \vdots & \vdots \\
%         (\bm{\vec{s}}^{(C)})^T & 1
%     \end{pmatrix}\!
    % ,~
    % \mathbf{D} =     
    % \begin{pmatrix}
    %     \mathbf{G}+\mathbf{V} & \sum_c \bm{\vec{s}}^{(c)} \\
    %     \sum_c (\bm{\vec{s}}^{(c)})^T & C
    % \end{pmatrix}
% \end{align}

The use of a least-squares cost function in Eq.~(\ref{eq:wopt}) is now crucial: it means that a closed form of the optimal weights $\WOopt$ and biases $\bv^{\rm opt}$ learned by the \RC{} can be obtained (see Appendix~\ref{app:mf}). Furthermore, the form of Eq.~(\ref{eq:stochdata}) allows us to write these learned weights and biases as
\begin{align}
      \begin{pmatrix}
        \WOopt & \bv^{\rm opt} 
    \end{pmatrix}
    = 
    \mathbf{M}\mathbf{D}^{-1}.
    \label{eq:learncorr}
\end{align}
Here $\mathbf{M}$ is a matrix that depends only on the mean traces (full form in Appendix~\ref{app:mf}). In contrast, $\mathbf{D}$ is the matrix of second-order moments:
\begin{align}
    \mathbf{D} =     
    \begin{pmatrix}
        \mathbf{G}+\mathbf{V} & \sum_c \bm{\vec{s}}^{(c)} \\
        \sum_c (\bm{\vec{s}}^{(c)})^T & C
    \end{pmatrix}
    \label{eq:ddef}
\end{align}
which depends on the the ``Gram'' matrix of mean traces, $\mathbf{G} = \sum_c \bm{\vec{s}}^{(c)}(\bm{\vec{s}}^{(c)})^T$, but also on the temporal correlations via the matrix $\mathbf{V} \equiv \sum_c \bm{\Sigma}^{(c)}$. Both these quantities emerge naturally in the analysis of the resolvable expressive capacity of physical systems that are subject to noise~\cite{hu_tackling_2023}. Here, Eq.~(\ref{eq:learncorr}) implies that weights learned by the \RC{} are not determined only by data \textit{means} via $\mathbf{G}$, but are also sensitive to temporal \textit{correlations} through $\mathbf{V}$. This simple feature will distinguish the \RC{} from standard classification approaches, a result we demonstrate in the rest of our analysis.

% as closely as possible to a specified unit vector along the edge of a hypercube.

% For the purposes of this paper, we restrict to $C \leq 4$, and typically $C=2$ for qubit $\ket{g}$ versus $\ket{e}$ classification, and use only raw heterodyne measurement records for the scenario most pertinent to standard qubit readout.

% The qubit-state-dependent response of the cavity ensures that the readout signal reflected off the cavity carries qubit state information. The reflected signals are amplified by a quantum amplifier at the cryogenic stage, whose output is downconverted by mixing with a local oscillator and then digitized to finally obtain raw heterodyne measurement records. 

\subsection{Quantum noise in dispersive qubit readout}

We will demonstrate the utility of the \RC{} framework for contemporary cQED applications by focusing on readout of dispersive qubit-cavity systems. However, we emphasize that \RC{} is model-free: it can process data $\bm{\vec{x}}$ generated by an arbitrary physical system, without any knowledge of its underlying physical model. Nevertheless, we introduce a simplified theoretical model of dispersive qubit-cavity systems below, to erect a foundation for the interpretability of TPP~\cite{roscher_explainable_2020}. First, this enables us to identify the sources of quantum noise at play in dispersive qubit readout. More importantly, we use this model to generate benchmarking datasets with controlled, practically-relevant quantum noise characteristics: the \RC{}'s application to these datasets with known temporal correlations in Secs.~\ref{sec:rcfilter},~\ref{sec:corr} allows us to interpret its learning principles. The ultimate test for the \RC{} is still in its application to real qubit readout data, in Sec.~\ref{sec:rcexpt}.

% Our starting model for the readout dynamics consists of a system comprising a multi-level artificial atom (here, a transmon) dispersively coupled to a readout cavity that is driven using a coherent tone at frequency $\omega_d$. 

The standard quantum measurement chain for heterodyne readout of a multi-level artificial atom (here, a transmon) dispersively coupled to a readout cavity is depicted schematically in Fig.~\ref{fig:schematic}, and can be modeled via the stochastic master equation (SME):
\begin{align}
    d\rhoc =  \Lsys\rhoc~dt + \Lchain\rhoc~dt + \Lmeas[dW]\rhoc.
    \label{eq:sme}
\end{align}
Here the Liouvillian superoperator $\mathcal{L}_{\rm sys}$ defines the quantum system whose states are to be read out. For dispersive qubit readout, $\mathcal{L}_{\rm sys}\hat{\rho} = -i[\hat{\mathcal{H}}_{\rm disp},\hat{\rho}]$, where the dispersive Hamiltonian $\Hdisp$ for a multi-level transmon takes the form (for cavity operators in the interaction frame with respect to an incident readout tone at frequency $\omega_d$, and setting $\hbar=1$)
\begin{align}
    \Hdisp \simeq \sum_p \omega_p \proj{p}{p} -\Delta_{da}\hat{a}^{\dagger}\hat{a} + \sum_p \chi_p \hat{a}^{\dagger}\hat{a}  \proj{p}{p}. 
    \label{eq:hdisp}
\end{align}
Here $\Delta_{da} = \omega_d - \omega_a$ is the detuning between the cavity and the readout tone, while $\chi_p$ is the dispersive shift per photon when the artificial atom is in state $\ket{p}$~\cite{zhu_circuit_2013, blais_circuit_2021}. Unfortunately, the artificial atom can undergo transitions from its initial state to unmonitored loss channels, which can reduce readout fidelity; all losses through such channels are described by the general Liouvillian $\Lchain$.

% The general Liouvillian $\Lchain$ is then used to describe all losses through channels that are not directly monitored, such as transmon transitions. 

The final superoperator $\Lmeas$ defines measurement chain components that are actively monitored to read out the state of the quantum system of interest. Here, we consider continuous heterodyne monitoring of a single quantum mode of the measurement chain, generally labelled $\hat{d}$. In the simplest case, $\Lmeas$ defines readout of the cavity itself (then, $\hat{d} \to \hat{a}$); however, it can also describe the dynamics (coherent or otherwise) of any other monitored quantum devices in the measurement chain. The most pertinent example is readout of the signal mode of an (ideally linear) quantum-limited amplifier that follows the dispersive qubit-cavity system via an intermediate circulator, as shown schematically in Fig.~\ref{fig:schematic}. Most generally, $\Lmeas$ can describe the monitoring of several modes of a general quantum nonlinear processor that is embedded in the measurement chain~\cite{khan_physical_2021}. Crucially, $\Lmeas$ must include a stochastic component (indicated by the Wiener increment $dW$), describing measurement-conditioned dynamics of the dispersive qubit-cavity system under such continuous monitoring (see Appendix~\ref{app:sim}).

% , as opposed to multiple shots obtained for the same initial state which requires repeated measurements.

For a qubit in the (\textit{a priori} unknown) initial state $\ket{\sigma}$ before measurement, continuous monitoring of the measurement chain then yields a single `shot' of heterodyne records $\{I^{(\sigma)}(t), Q^{(\sigma)}(t)\}$ contingent on this state $\sigma$. The complexity of this readout task can be appreciated given the form of raw heterodyne records even under a simplified theoretical model:
\begin{subequations}
\begin{align}
    I^{(\sigma)}(t_i) &= \sqrt{\kappa}\left[ \avg{\hat{X}^{(\sigma)}(t_i)} + \xiqI(t_i) \right] + \xiI(t_i)  + \xicI(t_i), \label{eq:hetI} \\
    Q^{(\sigma)}(t_i) &= \sqrt{\kappa}\left[ \avg{\hat{P}^{(\sigma)}(t_i)} + \xiqQ(t_i) \right] +  \xiQ(t_i)  + \xicQ(t_i). \label{eq:hetQ}
\end{align}
\end{subequations}
We consider discretized temporal indices $t_i$, for $i \in [\NT]$ and $\NT = \tmeas/\Delta t$, where $\tmeas$ is the total measurement time and $\Delta t$ is the sampling time set by the digitizer. Heterodyne measurement is intended to probe the expectation values $\avg{\hat{X}^{(\sigma)}(t_i)},\avg{\hat{P}^{(\sigma)}(t_i)}$ of canonical quadratures $\hat{X} = \frac{1}{\sqrt{2}}(\hat{d}+\hat{d}^{\dagger})$, $\hat{P} = \frac{-i}{\sqrt{2}}(\hat{d}-\hat{d}^{\dagger})$ of the monitored mode $\hat{d}$; however, any individual measurement record is obscured by noise $\xi$ from various sources. 

Vacuum noise $\xi_{I/Q}(t_i)$ is associated with heterodyne measurement of even an empty cavity, and is modelled as zero-mean Gaussian white noise,
\begin{align}
    \E{\xi_{I,Q}(t_i)} = 0,~~\E{\xi_{I,Q}(t_i)\xi_{I,Q}(t_j)} = \frac{1}{\Delta t}\delta_{ij}\delta_{I,Q}  
\end{align}
More importantly, $\xi_{I}^{\rm qm}(t_i),\xi_{Q}^{\rm qm}(t_i)$ describes quantum noise contributions to measurement records, whose origin is intrinsically tied to the nature of quantum measurement. The measurement of a quantum system imposes an evolution of its state, so that a given measurement affects the outcome of subsequent measurements. This effect is described via a measurement-conditioned stochastic quantum state $\rhoc$ (referred to as a \textit{quantum trajectory}), which is distinct from the unconditional quantum state $\rhou$ formally obtained when ensemble-averaging over repeated measurements. Consequently, for any given measurement instance, observables such as the conditional quadrature expectation $\avgc{\hat{X}^{(\sigma)}(t_i)} = {\rm Tr}\{\hat{X}\rhoc^{(\sigma)}(t_i)\}$ under heterodyne monitoring can deviate from the unconditional ensemble average $\avg{\hat{X}^{(\sigma)}(t_i)}$; this difference, given by $\xi_{I}^{\rm qm}(t_i) = \avgc{\hat{X}^{(\sigma)}(t_i)}-\avg{\hat{X}^{(\sigma)}(t_i)}$, manifests as quantum noise. These terms include amplified quantum fluctuations when measuring the output field from a quantum amplifier~(see Sec.~\ref{subsec:amp}), or the influence of quantum jumps in the measured cavity field due to transitions of the dispersively coupled qubit~(see Sec.~\ref{subsec:transition}). Finally, $\xi_{I/Q}^{\rm cl}(t_i)$ describe classical noise contributions to measurement records, for example noise added by classical HEMT amplifiers. While the statistics of this noise may take different forms, it is formally distinct from heterodyne measurement noise, as it has no associated stochastic measurement superoperator in Eq.~(\ref{eq:sme}). 

The objective of the qubit readout task is then to use noisy \textit{single-shot}~\cite{mallet_single-shot_2009} temporal measurement data to obtain an estimated class label $\sigmapred$ that is ideally equal to the true class label $\sigma$. Within the \RC{} framework, then, $\NO = 2$ and $\bm{\vv{x}}^{(\sigma)} = \begin{psmallmatrix} \vv{I}^{(\sigma)} \\ \vv{Q}^{(\sigma)} \end{psmallmatrix}$, where $\vec{I}_i = I(t_i)$. The noise $\bm{\vec{\zeta}}$ in Eq.~(\ref{eq:stochdata}) then contains the terms $\xi,\xi^{\rm qm},\xi^{\rm cl}$. However, before describing \RC{} results, we first briefly review standard approaches to qubit state classification.

% Furthermore, we require , where the estimation must be performed using only a single measurement shot: such rapid readout is essential for quantum feedback and control applications~\cite{riste_initialization_2012, johnson_heralded_2012, campagne-ibarcq_persistent_2013}. 

\subsection{Standard post-processing for binary qubit state readout: matched filters}
\label{sec:dispersive}

% For a qubit state indexed by $\sigma$, we label raw heterodyne records obtained from the measurement chain described by Eq.~(\ref{eq:sme}) as $I^{(\sigma)}(t_i),Q^{(\sigma)}(t_i)$. Note that the state label $\sigma$ is unknown to the observer: post-processing of these raw heterodyne records is designed to yield a predicted label $\sigmapred$ that reveals $\sigma$.

The standard classification paradigm in cQED to obtain $\sigmapred$ from raw heterodyne records would formally be described as a filtered Gaussian discriminant analysis (FGDA) in contemporary learning theory~\cite{hastie_elements_2016}. This comprises two stages: (i) temporal filtering of each measured quadrature, and (ii) assigning a class label to filtered quadratures that maximizes the likelihood of their observation amongst all $C$ classes as determined by a Gaussian probability density function. Formally, this procedure can be written as:
\begin{align}
    \sigmapred = \fgauss{ 
    \sum_i 
    \begin{pmatrix}
        h_I(t_i)I^{(\sigma)}(t_i) \\
        h_Q(t_i)Q^{(\sigma)}(t_i)
    \end{pmatrix}
    }
    =
    \fgauss{ 
    \begin{pmatrix}
        \vec{h}_I^T\vec{I}^{(\sigma)} \! \\
        \vec{h}_Q^T\vec{Q}^{(\sigma)}
        \!
    \end{pmatrix}
    }
    \label{eq:standardmf}
\end{align}
The function $\fgauss{\cdot}$ then assigns class labels according to the aforementioned Gaussian discriminator.

A fact seldom mentioned explicitly is that both the temporal filters and the Gaussian discriminator must be constructed using a calibration dataset, analogous to the training phase of the \RC{}: a set of $\NTrain$ heterodyne records obtained when the initial qubit states are known under controlled initialization protocols. For example, for the most commonly considered case of binary qubit state classification to distinguish states $\ket{e}$ and $\ket{g}$, and under the assumption that the noise in heterodyne records is additive Gaussian white noise, an optimal filter is known: the matched filter~\cite{turin_introduction_1960, gambetta_protocols_2007,Silveri2016}. The empirical matched filter is constructed from the calibration dataset, where $(n)$ indexes distinct records, via
\begin{align}
    \vec{h}_I = \frac{1}{\NTrain}\sum_{n=1}^{\NTrain} \left( \vec{I}_{(n)}^{(e)}-\vec{I}_{(n)}^{(g)} \right)
    \label{eq:mfI}
\end{align}
with $\vec{h}_Q$ defined analogously for $I \to Q$. The function $\fgauss{\cdot}$ requires fitting Gaussian profiles to measured probability distributions of known classes, and hence uses means and variances estimated from calibration data.

While a Gaussian discriminant analysis can be applied to classification of an arbitrary number of states $C$ and beyond white noise constraints, the choice of an optimal temporal filter in these more general situations is not straightforward~\cite{Kurpiers2018}. Due to its ease of construction, often a matched filter akin to Eq.~(\ref{eq:mfI}), or an even more rudimentary boxcar filter (a uniform filter that is nonzero only when the measurement signal is on) is deployed, regardless of the complexity of the noise conditions (for example, when qubit decay is significant and more optimal filters can be found~\cite{gambetta_protocols_2007}). We will show how the \RC{} approach provides a natural generalization of matched filtering to multi-state classification, and furnishes a trainable classifier that can generalize to more complex noise environments.

\section{\RC{} learning as optimal filtering: Generalized matched filters}
\label{sec:rcfilter}

% Note that at first sight Eq.~(\ref{eq:rcmap}), which defines the \RC{} scheme for classification, appears to be analogous to Eq.~(\ref{eq:standardmf}) in the FGDA scheme. There are, in fact, close connections between the two. 

To understand how the \RC{} generalizes standard matched filtering approaches, we first show an important connection between the two schemes. Note that the learned matrix of weights $\WOopt \in \mathbb{R}^{C \times\NO\cdot\NT}$ can be equivalently expressed as:
\begin{align}
    \WOopt = 
    \begin{pmatrix}
        \bm{\vec{f}}_1^{~T} \\
        \vdots  \\
        \bm{\vec{f}}_C^{~T} 
    \end{pmatrix}
\end{align}
where $\bm{\vec{f}}_k \in \mathbb{R}^{\NO\cdot\NT}$ for $k\in [C]$. With this parameterization, Eq.~(\ref{eq:rcmap}) for the $k$th component of the vector $\mathbf{y}$ can be rewritten as:
\begin{align}
    \yv_k = \bm{\vec{f}}_k^{~T} \bm{\vec{x}} + \bv_k,~~~k \in [C]
    \label{eq:rcmapfilter}
\end{align}
When Eq.~(\ref{eq:rcmapfilter}) is compared against Eq.~(\ref{eq:standardmf}), the interpretation of $\bm{\vec{f}}_{k}$ becomes clear: this set of weights can be viewed as a temporal filter applied to the data $\bm{\vec{x}}$. \RC{} based classification can therefore be interpreted as the application of $C$ filters (one for each $k$) to obtain the estimated label $\sigmapred$. The optimal $\WOopt$ therefore defines the optimal filters that enable this estimation with minimal error. The use of $C$ optimal filters for a $C$-state classification task indicates the linear scaling of the \RC{} approach with the complexity of the task.

Remarkably, the optimal $\WOopt$ given by Eq.~(\ref{eq:learncorr}), and hence the $C$ optimal filters, can be expressed in the simple semi-analytic form:
\begin{align}
    \bm{\vec{f}}_k = \sum_{p}  C_{kp} \mathbf{V}^{-1} \bm{\vec{s}}^{(p)},~~~ k \in [C]  
    \label{eq:analyticFilters}
\end{align}
where the mean traces $\bm{\vec{s}}^{(p)}$ and correlation matrix $\mathbf{V} = \sum_p \mathbf{\Sigma}^{(p)}$ can both be empirically estimated from data under the known initial state $p$:
\begin{align}
    \bm{\vec{s}}^{(p)} &\simeq \frac{1}{\NTrain}\sum_{n=1}^{\NTrain} \bm{\vec{x}}_{(n)}^{(p)} \nonumber \\
    \mathbf{\Sigma}^{(p)} &\simeq  \frac{1}{\NTrain}\sum_{n=1}^{\NTrain} \bm{\vec{x}}_{(n)}^{(p)}\bm{\vec{x}}_{(n)}^{(p)T} - \bm{\vec{s}}^{(p)}\bm{\vec{s}}^{(p)T}
    \label{eq:empMean}
\end{align}
while the coefficients $C_{kp}$ can also be shown to depend only on $\bm{\vec{s}}^{(p)}$ and $\mathbf{V}$~(see Appendix~\ref{app:mf} for full details). Furthermore, we note that the $C$ filters are not all independent; they can be shown to satisfy the constraint (see Appendix~\ref{app:mf})
\begin{align}
    \sum_{k=1}^C \bm{\vec{f}}_k = \bm{\vec{0}},
    \label{eq:constraint}
\end{align}
where $\bm{\vec{0}} \in \mathbb{R}^{\NO\cdot\NT}$ is the null vector. This powerful constraint, which holds regardless of the statistics of the noise $\bm{\vec{\zeta}}$, implies that only $C-1$ of the $C$ filters need to be learned from training data.

\subsection{\RC{} performance under Gaussian white noise in comparison to standard FGDA}
\label{subsec:whitenoisesim}

We can now analyze the case most often assumed in cQED: that the dominant noise source in heterodyne records $I,Q$ is stationary Gaussian white noise (independent of the undetermined state), an assumption under which matched filters are optimal for binary classification. Engineering of cQED measurement chains is geared towards approaching this limit, by (i) developing large bandwidth, high dynamic range amplifiers that operate with fast response times and minimal nonlinear effects even at high gain and large input signal powers~\cite{kochetov_higher-order_2015, boutin_effect_2017},\cite{Parker2022, Remm2023, kaufman_josephson_2023, Kaufman2024}, (ii) improving qubit $T_1$ and tolerance to strong cavity drives to reduce transitions during $\tmeas$~\cite{place_new_2021}, and (iii) controlling technical noise sources such as electronic white noise from classical cryo-HEMT amplifiers and room temperature electronics. 

In this relevant limit, the correlation matrix $\mathbf{V}$ of Eq.~(\ref{eq:ddef}) becomes proportional to the identity matrix, and the resulting \RC{}-learned filters depend chiefly only on the mean traces $\bm{\vec{s}}^{(p)}$. For any $C=2$ state classification task, for example $p \in \{e,g\}$ qubit readout, we can show that $C_{ke} = -C_{kg}$, which reduces $\bm{\vec{f}_k}$ exactly to a standard binary matched filter. Remarkably, the \RC{}-learned optimal filters in the Gaussian white noise approximation then provide a semi-analytically calculable generalization of matched filters to $C$ states.

\begin{figure}[t]
    \centering
    \includegraphics[scale=1.0]{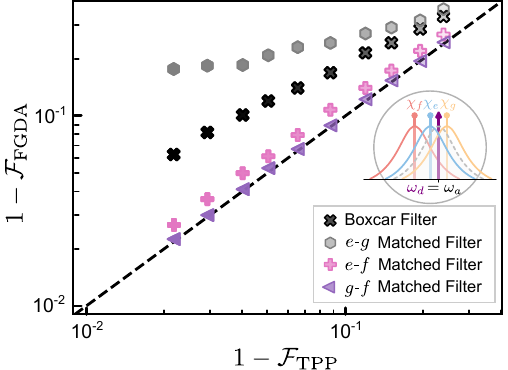}
    \caption{\textbf{Multi-state ($C=3$) classification performance of \RC{} versus FGDA under Gaussian white noise conditions.} We consider dispersive qubit readout to distinguish states $p \in \{e,g,f\}$ as a function of measurement power. For a transmon $\chi_p/\kappa \in \{-\chi, \chi, -3\chi\}$, $\chi/\kappa = 0.195$, and $\kappa/2\pi = 1.54~{\rm MHz}$. More opaque markers indicate higher measurement tone amplitudes. Inset shows induced dispersive shifts for each state (not to scale). Standard FGDA is performed using one of three MFs corresponding to each distinct state pair, as well as a boxcar filter. \RC{} filters are also followed by a Gaussian discriminator for an equivalent comparison. Only one of the binary MFs allows the FGDA to approach the \RC{}, while all other chosen filters yield a worse performance. }
    \label{fig:analyticFiltersPerf}
\end{figure}

We can now analyze the multi-state classification performance enabled by these \RC{}-learned optimal filters in comparison to the standard FGDA approach. To guarantee dispersive qubit readout data that is subject only to white noise, we use a theoretical simulation of Eq.~(\ref{eq:sme}) to generate measured heterodyne records for $C$ qubit states, under the following assumptions: (i) all qubit state transitions are neglected, (ii) any additional classical noise sources in the measurement chain are ignored, and (iii) therefore direct readout of the cavity can be considered instead of the use of a quantum amplifier and the potential quantum noise added by it. We take the cavity measurement tone to be applied for a subset of the total $\mathcal{T}_{\rm meas}$, namely for $[\mathcal{T}_{\rm on},\mathcal{T}_{\rm off}]$, and to be coincident with the cavity center frequency so that $\Delta_{da} = 0$, usual for transmon readout (for full details, see Appendix~\ref{app:dispInf}). Other system parameters can be found in the caption of Fig.~\ref{fig:analyticFiltersPerf}. 

The \RC{} can be used to generate optimal filters, and hence perform classification, for arbitrary $C$; for examples of calculated filters, see Fig.~\ref{fig:analyticFilters}, Appendix~\ref{app:mf}. For concreteness, here we analyze the classification performance enabled by the \RC{} to distinguish $C=3$ states $p\in\{e,g,f\}$. Our choice of resonantly driving the readout cavity means the sign of cavity dispersive shifts for transmon states $e$ and $f$ is the same, and is opposite to that for $g$, making them harder to distinguish (see also Fig.~\ref{fig:analyticFiltersPerf} inset). We note that the specific details of the readout scheme do not change the \RC{} learning procedure.

%   These simulations yield single-shot measurement records for any number of transmon states. Examples of these records are then shown in Fig.~\ref{fig:analyticFilters} for four distinct transmon states $p \in \{e,g,f,h\}$; for ease of visualization we only consider the $I$ quadrature.

% We use this simulated dataset as a training set to determine the \RC{}-learned filters under the white noise assumption, as defined by Eq.~(\ref{eq:analyticFilters}). While the individual measurement records are obscured by white noise, the empirically-calculated mean traces in the top right of Fig.~\ref{fig:analyticFilters} illustrate the physics at play. The mean traces grow once the measurement tone is turned on past $\mathcal{T}_{\rm on}$, and settle to a steady state depending on the induced dispersive shift $\chi_p$ and the measurement amplitude. The traces begin to fall beyond $\mathcal{T}_{\rm off}$ and eventually settle to background levels. These means, together with an estimate of the variances, determine the coefficients $C_{kp}$ that define the contribution of the mean trace $\vec{s}^{(p)}$ to the $k$th filter, and are hence sufficient to calculate optimal filters for the classification of any subset of states. 

For this three-state classification task, a unique filter choice for the FGDA is not known. While certain approaches at constructing filters have been attempted~\cite{Chen2023ThreeStateNoAmp}, boxcar filtering is still commonly employed. Another approach might be to use a matched filter that optimizes distinction of just one pair of states. There are 3 such filters in total: for discrimination of $e$-$g$ states as defined in Eq.~(\ref{eq:mfI}), as well as analogously-defined filters for $e$-$f$ and $g$-$f$ states. 

In Fig.~\ref{fig:analyticFiltersPerf}, we show classification infidelities $1-\mathcal{F}$, calculated for datasets with increasing measurement tone amplitude (more opaque markers), using both the optimal \RC{} filter and the FGDA with the four aforementioned filter choices. We emphasize again that these datasets are generated via simplified theoretical simulations guaranteeing white noise conditions, in particular ignoring any non-idealities associated with strong readout drives; under these conditions, classification performance improves steadily with increasing measurement tone amplitude, as shown. Even in this regime, we clearly observe that the FGDA infidelities for most filter choices are worse than the \RC{}. Interestingly, the poorest performer is not the boxcar filter; instead, it is the $e$-$g$ filter, which would be optimal if we were only distinguishing $\{e,g\}$ states, that yields the worst performance. This is because the $e$-$g$ filter is completely unaware of the $f$ state: it attempts to best discriminate $e$ and $g$, but in doing so substantially confuses $e$ and $f$ states that are already the hardest to distinguish. The $e$-$f$ filter corrects this major problem and hence performs better, but does not discriminate $e$ and $g$ as well as the $e$-$g$ filter would. Due to the specific driving conditions and phases, the $g$-$f$ filter unwittingly does a good job at addressing both these problems, yielding the best performance. Nevertheless, it can only match the \RC{}.

%%%%%%%%%%%%%%%%%%%%%%%%%%%%%%%%%%%%%%%%%%%%%%%%%%%%%%%%%%%%%%%%%%%%%%%%%%%%%%%%%%%%%%%%%%%%%%%%%%%

%%%%%%%%%%%%%%%%%%%%%%%%%%%%%%%%%%%%%%%%%%%%%%%%%%%%%%%%%%%%%%%%%%%%%%%%%%%%%%%%%%%%%%%%%%%%%%%%%%%

This trial-and-error approach relies on knowledge of optimal matched filtering from binary classification, but clearly cannot be optimal for $C>2$: none of the filter choices are informed by the statistical properties of measured data for \textit{all} $C$ classes to be distinguished. Alternative approaches, such as using multiple classifiers with upto $C-1$ independent filters (for an equivalent resource cost to the \RC{}) can account for all classes, but as we show in Appendix~\ref{app:multifgda} do not outperform the \RC{}, and also exhibit a dependence on readout conditions. In either case, the brute-force determination of pairwise matched filters scales at least with the number of distinct state pairs, which grows quadratically with $C$; this is before one even accounts for fine-tuning of filter coefficients (analogous to learning $C_{kp}$ in the \RC{} approach). In contrast, the \RC{} approach provides a simple automated scheme to learn optimal filters, takes data for readout of all classes into account, is model-free and thus applicable to arbitrary readout conditions, and scales only linearly with the task dimension set by $C$. 

% \subsection{\RC{} performance under general noise conditions}

However, the true strength of \RC{} learning arises when noise in measured heterodyne records no longer satisfies the additive Gaussian white noise assumption, which may arise if any of the conditions (i)-(iii) for qubit measurement chains listed earlier are not met. Departures from this ideal scenario are widely prevalent in cQED, and will be apparent in experimental results in the following section. Through the rest of this paper, we show how the trainability of the \RC{} approach enables it to learn filters tailored to these more general noise conditions, and consequently outperform the standard FGDA based on binary matched filters.

\section{\RC{}-Learning for real qubits}
\label{sec:rcexpt}
\subsection{Experimental Results}\label{subsec:exptData}
\begin{figure*}[t!]
    \centering
    \includegraphics[scale=1.0]{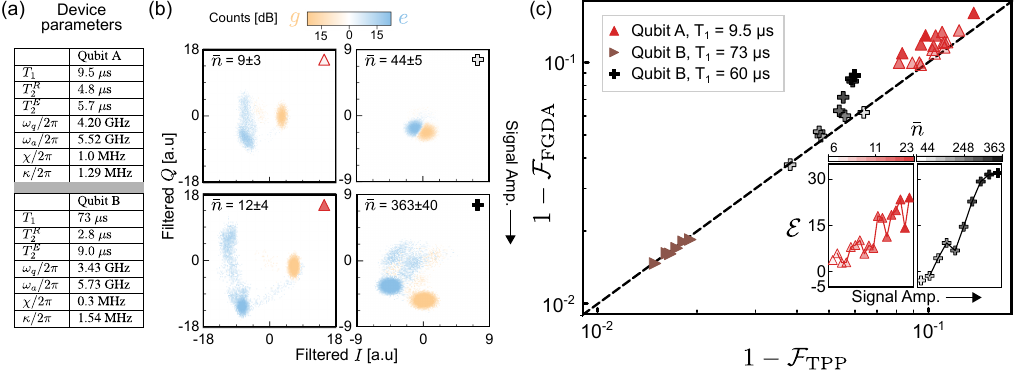}
    \caption{\textbf{Classification performance of \RC{} versus FGDA for readout of real qubits.} (a) Parameters of various dispersive qubit-cavity systems used for gathering readout data. Coherence measurements are subject to $10\%$ variation over time. (b) Representative qubit readout histograms under boxcar filtering as a function of measurement signal amplitude. (c) Readout data for three dispersive qubit-cavity systems is analyzed and the resulting classification infidelities for binary ($C=2$) state classification are plotted against each other. The dashed line marks $1-\fidmf = 1-\fidrc$. For datasets with variable shading of markers (red and black), more opaque markers indicate higher measurement tone amplitudes, with corresponding resonator photon number $\bar{n}$ indicated via colorbars. Inset: Percentage fewer errors $\metric$ computed for indicated datasets with increasing input signal amplitude.  }
    \label{fig:exptResults}
\end{figure*}

To demonstrate how the general learning capabilities of the \RC{} approach can aid qubit state classification in a practical setting, we now apply it to the readout of finite-lifetime qubits in an experimental cQED measurement chain. The essential components of the measurement chain are as depicted schematically in Fig. ~\ref{fig:schematic} and described by Eq.~(\ref{eq:sme}). The actual circuit diagram is shown in Fig.~\ref{appfig:hardwareSchematic} in Appendix~\ref{app:expmt}, and important parameters characterizing the measurement chain components are summarized in Fig.~\ref{fig:exptResults}(a).
%Table~\ref{tab:qubitInfo}. 

We consider two distinct cavity systems, for the dispersive readout of distinct single qubits A and B to discriminate states $p \in \{e,g\}$. For \textit{lossless} qubits that are read out dispersively for a fixed measurement time $\tmeas$, the ratio $\chi/\kappa$ determines the \textit{theoretical} maximum readout fidelity; in particular, an optimal value for this ratio is known under these ideal conditions~\cite{blais_circuit_2021}. However, experimental considerations mean that operating parameters must be designed with several other factors in mind. At high $\chi/\kappa$ ratios with modest or higher $\kappa$, for large $\kappa$ with modest $\chi/\kappa$ ratios, and especially when both are true, the experiment is sensitive to dephasing from the thermal occupation of the readout resonator at a rate proportional to $\bar{n}\kappa$ \cite{Schuster2005dephasing}. This can be quite limiting to the $T_2$ dephasing time of the qubit if the readout resonator is strongly coupled to the environment and/or the environment has appreciable average thermal photon occupation $\bar{n}$. In the opposite low $\chi/\kappa$ limit, the qubit is shielded from thermal dephasing, but readout becomes very difficult as the rate at which one learns about the qubit state from a steady state coherent drive is proportional to $\chi/\kappa$ \cite{blais_circuit_2021}. In this experiment, the lower-than-usual $\chi/\kappa \approx 0.2$ in qubit B represents a compromise between these two limits, while also enabling the high fidelity discrimination of multiple excited states of the transmon (See Fig.~\ref{appfig:RO_histograms}, Appendix~\ref{app:expmt}).

% For both qubits considered in the present setup, the realized $\chi/\kappa$ ratio is non-optimal for binary discrimination of $p\in\{e,g\}$. \mjh{kind of too rough on our poor design parameters; design of chi and kappa represents multiple compromises, readable qubits take a hit on T2 due to stray photons reading them out on accident, also Purcell limits depending on details of geometry; call these typical parameters?} 

%%%%%%%%%%%%%%%%%%%%%%%%%%%%%%%%%%%%%%%%%%%%%%%%%%%%%%%%%%%%%%%%%%%%%%%%%%%%%%%%%%%%%%%%%%%%%%%%%%

%%%%%%%%%%%%%%%%%%%%%%%%%%%%%%%%%%%%%%%%%%%%%%%%%%%%%%%%%%%%%%%%%%%%%%%%%%%%%%%%%%%%%%%%%%%%%%%%%%

Each readout cavity is driven in reflection, and its output signal is amplified also in reflection using a Josephson Parametric Amplifier (JPA). We employ the latest iteration of strongly-pumped and weakly-nonlinear JPAs \cite{Kaufman2024}, boasting a superior dynamic range. Such JPAs operate well below saturation even at signal powers that correspond to over 100 photons, enabling us to probe qubit readout at high measurement powers. By choosing a signal frequency at exactly half the pump frequency, we can operate the JPA in phase-sensitive mode. We can also operate the amplifier in phase-preserving mode if we detune the signal from half the pump frequency by greater than the spectral width of the pulse. Several filters are used to reject the strong JPA pump tone required to enable this operation. Circulators are used to route the output signals away from the input signals and to isolate the qubit from amplified noise. 

In ideal circumstances, the use of stronger measurement tones should increase the classification fidelity for qubit readout, as shown via simplified theoretical simulations in Fig.~\ref{fig:analyticFiltersPerf}. In practice, however, higher measurement powers are known to be associated with a variety of complex dynamical effects that can limit fidelity. Perhaps the most common observation is enhanced qubit $e\to g$ decay under strong driving (referred to as the $T_1$ versus $\bar{n}$ problem). The relative accessibility of higher excited states in transmon qubits means that at strong enough driving, general multi-level transitions to these higher levels can also be observed. There have also been predictions of chaotic dynamics and ionization~\cite{Cohen2023Chaos, Shillito2022} at certain readout resonator occupation levels, as well as complex dynamics due to qubit-induced resonator nonlinearities~\cite{khezri_measuring_2016}. The theoretical understanding of these effects, and their modeling via an SME analogous to Eq.~(\ref{eq:sme}) is an ongoing challenge.  

% We analyze how the \RC{}, which is a model-free approach, fares at qubit state classification   

In our experiments, we perform readout across this domain using two different qubits. For Qubit A, we simultaneously vary both pulse amplitude and pulse duration ($\mathcal{T}_{\rm off}-\mathcal{T}_{\rm on}$), the latter from 300~ns to 1150~ns, to together obtain roughly 9$\pm$3 to 18$\pm$5 photons in the cavity in the steady state. For Qubit B's phase-preserving dataset, measurement pulse durations vary independently from 500~ns to 900~ns, and measurement amplitudes are adjusted to drive roughly 44$\pm$5 to 363$\pm$40 photons in the cavity in the steady state; the significantly larger photon number is tolerated due to the low Qubit B $\chi/\kappa$. At the lowest pulse duration and amplitude, this corresponds to just enough discriminating power to separate the measured distributions for the two states by approximately their width in a boxcar-filtered IQ plane (namely, without the use of an empirical MF). An example of the individual readout histograms for qubits initialized in states $p\in \{e,g\}$ at this lowest measurement tone power is shown in Fig.~\ref{fig:exptResults}(b). Qubit B's phase-sensitive dataset was recorded with a pulse time of 800ns with a shaped pulse to shorten the effect of the cavity ring-up time similar to \cite{McClure2016}.

At the highest measurement powers, we are able to populate the readout cavity with up to 100 photons, calibrated by observing the frequency shift of the qubit drive frequency versus the occupation of the readout resonator. At these powers, extreme higher-state transitions become visible during the readout pulse~\cite{Sank2016}; an example is shown in Fig.~\ref{fig:exptResults}(b) (see also Fig.~\ref{appfig:RO_histograms} in Appendix~\ref{app:expmt}). There is also a notable elliptical distortion in the high-amplitude data, particularly for qubit A. We suspect that this is due to the short duration of the pulses and the inclusion of the cavity ring-up and ring-down in the integration, since the simple boxcar filter used to integrate the histograms in Fig. \ref{fig:exptResults}(b) does not rotate with the signal mean.

For such complex regimes where no simple model of the dynamics exists, the construction of an optimal filter is not known; this hence serves as an ideal testing ground for the \RC{} approach to qubit state classification. We compute the infidelities of binary classification using both the \RC{} scheme and an FGDA using the standard MF [Eq.~(\ref{eq:mfI})] under a variety of readout conditions, plotting the results against each other in Fig.~\ref{fig:exptResults}(c). 

The highest fidelity using both schemes is obtained for qubit B under conditions where its $T_1$ time is longest. This dataset was collected at a fixed, moderate measurement power; the different points correspond to a rolling of the relative JPA pump and measurement tone phase that determines the amplified quadrature under phase-sensitive operation. The dashed line marks equal classification infidelities, so that any datasets above this line yield a higher classification \textit{infidelity} with the FGDA than with the \RC{}. Here we see that both schemes exhibit very similar performance levels.

The other two datasets are obtained for readout under varying measurement powers. The depth of shading of the markers indicates the strength of measurement drives: the more opaque the marker, the higher the measurement power. We first note that the classification fidelity does not uniformly increase with signal amplitude in experiment; this is in contrast to the simplified theoretical simulations of Sec.~\ref{subsec:whitenoisesim}, and is expected due to the aforementioned dynamical effects exhibited in real qubit readout at higher readout powers (neglected in Fig.~\ref{fig:analyticFiltersPerf}).

For weaker measurement powers, we see that the \RC{} and the FGDA are once again comparable. However, a very clear trend emerges: for higher measurement powers - where measurement dynamics become much more complex as demonstrated in Fig.~\ref{fig:exptResults}(b) - the \RC{} generally outperforms the FGDA. To more precisely quantify the difference in performance between the \RC{} and FGDA, we introduce the metric $\metric$:
\begin{align}
    \metric =  \left(\frac{\fidrc-\fidmf}{1-\fidmf}\right)\!\times\! 100
\end{align}
which essentially asks: ``what percentage fewer errors does the \RC{} make when compared to the FGDA?'' We plot $\metric$ in the inset of Fig.~\ref{fig:exptResults}(c) for the two qubit readout experiments where the input power is varied. We see clearly that with increasing power, the \RC{} can significantly outperform the FGDA scheme, committing as many as 30\% fewer errors in the experiments considered. In fact, in certain cases where the FGDA predicts a reduction in classification fidelity with increasing readout power, the \RC{}'s learning advantage can even enable a qualitatively different trend, instead boosting classification performance with increasing readout power (for details, see Appendix~\ref{app:fidvamp}).

Our results demonstrate that the \RC{} approach can be successfully applied to real qubit readout across a broad spectrum of measurement conditions. Furthermore, the \RC{} can even outperform the standard FGDA in certain relevant regimes, such as for high-power readout. While the \RC{} can thus be applied as a model-free learning tool, we are also interested in understanding the principles that enable the \RC{} to outperform standard approaches using an MF. Uncovering these principles can help identify the types of classification tasks where \RC{} learning is essential. Our interpretation of \RC{} learning as optimal filtering proves a useful tool in this vein.  

%%%%%%%%%%%%%%%%%%%%%%%%%%%%%%%%%%%%%%%%%%%%%%%%%%%%%%%%%%%%%%%%%%%%%%%%%%%%%%%%%%%%%%%%%%%%%%%%%%

\begin{figure}[t]
    \centering
    \includegraphics[scale=1.0]{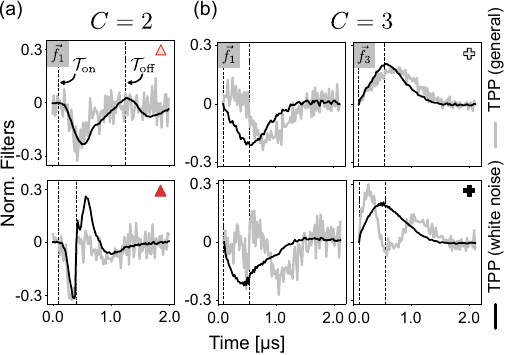}
    \caption{\textbf{Adaptation of \RC{}-learned filters with increasing measurement tone amplitude and evolving noise conditions.} Black curves are normalized \RC{} filters under the white noise assumption; for binary state classification, these are identical to standard matched filters. Gray curves are general \RC{} filters with no assumptions on noise statistics. (a) Filter $\vec{f}_1$ for binary ($C=2$) classification, and (b) filters $\vec{f}_{1,3}$ for $C=3$ state classification. In both cases, at weaker amplitudes the general \RC{} filter closely matches the \RC{} filter assuming white noise. However, for higher measurement amplitudes, a marked difference between the white noise \RC{} filter and the general \RC{} filter is observed. }
    \label{fig:filterComparisons}
\end{figure}

%%%%%%%%%%%%%%%%%%%%%%%%%%%%%%%%%%%%%%%%%%%%%%%%%%%%%%%%%%%%%%%%%%%%%%%%%%%%%%%%%%%%%%%%%%%%%%%%%%

\subsection{Adaptation of \RC{}-learned filters under strong measurement tones}

% The observed difference in performance between the \RC{} and the standard FGDA lies in the former's ability to learn from data as experimental conditions evolve. Our interpretation of \RC{} learning as the determination of optimal filters proves particularly insightful in expressing this adaptability. 

For visualization, we only analyze filters $\vec{f}_k \in \mathbb{R}^{\NT}$ for $I$ quadrature data; the complete vector $\bm{\vec{f}}_k$ includes filters for all $\NO$ observables. Recall that for a $C$ state classification task, the \RC{} learns $C$ filters; however, the sum of filters is constrained by Eq.~(\ref{eq:constraint}), so that $C-1$ filters are sufficient to describe the \RC{}'s learning capabilities. In Fig.~\ref{fig:filterComparisons}(a) we first consider filters learned by the \RC{} for a $C=2$ classification task, for select experimental datasets from Fig.~\ref{fig:exptResults} obtained under a low and a high measurement power. It therefore suffices to analyze just $\vec{f}_1$, the first filter for the $I$ quadrature, as a function of measurement power. The black curves are filters learned under the assumption of Gaussian white noise, given by Eq.~(\ref{eq:analyticFilters}); recall that for this binary case, these filters are exactly the standard MF. The gray curves, in contrast, are filters learned by the \RC{} for arbitrary noise conditions, obtained by solving Eq.~(\ref{eq:wopt}). At a low measurement tone amplitude (less opaque marker), the general \RC{} filter appears very similar to the \RC{} filter under white noise. As the measurement tone amplitude is increased, however, the \RC{}-learned filter under arbitrary noise can deviate substantially from the \RC{} filter under white noise. This is accompanied by a marked difference in performance, observed in Fig.~\ref{fig:exptResults}(c).

Crucially, the generalization of matched filters provided by \RC{}-learning as discussed in Sec.~\ref{subsec:whitenoisesim} enables a similar comparison for classification tasks of an arbitrary number of states. We show learned filters for $C=3$ state classification of $p \in \{e,g,f\}$ in Fig.~\ref{fig:filterComparisons}(b), again for a low and high measurement power. It is now sufficient to consider any two of three distinct $I$-quadrature filters; here we choose $\vec{f}_1$ and $\vec{f}_3$. Once more, the general \RC{} filters begin to deviate significantly from \RC{} filters under the white noise assumption at high powers. Most importantly, these filters provide an improvement in 3-state classification fidelity relative to the FGDA scheme (for brevity, full results are provided in~Appendix~\ref{app:3state}).

Clearly, the precise form of filters learned by the \RC{} to outperform white noise filters must be influenced by some physical phenomena that arise at strong measurement powers. However, the \RC{} is not provided with any physical description for such phenomena, which is in fact part of its model-free appeal. What then, is the mechanism through which the \RC{} can learn about such phenomena to compute optimal filters? The answer lies \ch{explicitly in Eq.~(\ref{eq:analyticFilters})}: \RC{}-learned filters are sensitive to noise correlations in data via $\mathbf{V}$. Using simulations of measurement chains where the noise structure of quantum measurement data can be precisely controlled, we show that the noise structure can strongly deviate from white noise conditions under practical settings. Crucially, the \RC{} can adapt to these changes whereas the MF cannot.

\section{\RC{} learning: Simulation results}
\label{sec:corr}

% In the present section, we use a simple, but practical example to demonstrate the features of \RC{}-based learning that enable the aforementioned advantage.

% \subsection{Learning correlations}
% \label{subsec:learnCorr}

% \begin{align}
%     \WOopt = \mathbf{Y}\mathbf{X}^T(\mathbf{X}\mathbf{X}^T - \lambda \mathbf{I})^{-1},
% \end{align}
% where the matrix $\mathbf{X}$ consists of individual vectors $\vec{x}^{(p)}$ for all $p$ classes and training records $\NTrain$, and $\lambda$ is a regularization parameter. 

% We note the dependence on the quadratic term $\mathbf{X}\mathbf{X}^T$, which is simply the (uncentered) correlation matrix calculated for every pair of temporal measured data points, and averaged over the entire multi-class training set. 

As discussed in Sec.~\ref{sec:rc}, the \RC{} weights and hence optimal filters depend on mean traces, but are also cognizant of - and can learn from - the noise structure of measured data via the temporal correlation matrix $\mathbf{V}$. This is in stark contrast to the use of a matched filter.

Crucially, data obtained from \textit{quantum} systems can exhibit temporal correlations that have a quantum-mechanical origin. In what follows, we demonstrate the ability of the \RC{} to learn these quantum correlations, using simulations of two experimental setups where such quantum noise sources arise naturally: (i) readout using phase-preserving quantum amplifiers with a finite bandwidth, so that the amplifier added noise (demanded by quantum mechanics) has a nonzero correlation time, and (ii) readout of finite lifetime qubits with multi-level transitions (quantum jumps).

\subsection{Correlated quantum noise added by finite-bandwidth phase-preserving quantum amplifiers}
\label{subsec:amp}

Quantum-limited amplifiers are a mainstay of measurement chains in cQED, needed to overcome the added classical noise of following HEMTs. Phase-preserving quantum amplifiers are necessitated by quantum mechanics to add a minimum amount of noise to the incoming cavity signal being processed. The correlation time of this added quantum noise is determined by the dynamics of the amplifier itself, namely its active linewidth reduced by anti-damping necessary for gain. For finite bandwidth amplifiers operating at large enough gains, this can lead to the addition of quantum noise with non-zero correlation time in measured heterodyne data.

% The finite amplifier bandwidth determines the correlation structure of this added noise: for large bandwidth amplifiers operating at moderate gains, the correlation time of the added quantum noise is set  

To simulate qubit readout in these circumstances, we consider a quantum measurement chain described by Eq.~(\ref{eq:sme}) now consisting of a qubit-cavity-amplifier setup. $\Lmeas$ then describes the readout of a non-degenerate (i.e. two-mode) parametric amplifier and its non-reciprocal coupling to the cavity used to monitor the qubit. We ignore qubit state transitions, so that $\Lchain$ only describes losses via unmonitored ports of the cavity and amplifier. Full details of the simulated SME are included in Appendix~\ref{app:simAmp}. 

We must consider added classical noise in the measurement chain, as this is what demands the use of a quantum amplifier in the first place. We take the added classical noise to be purely white, $\xi^{\rm cl}(t_i) = \sqrt{\bar{n}_{\rm cl}} \frac{dW}{dt}(t_i)$, with a noise power $\bar{n}_{\rm cl} = 30$, parameterized as usual in ``photon number'' units; these assumptions on the noise structure and power are taken from standard cQED experiments, including our own. Now, the obtained heterodyne measurement records, Eqs.~(\ref{eq:hetI}),~(\ref{eq:hetQ}) contain two dominant noise sources: (i) excess classical white noise, and (ii) quantum noise added by the amplifier, contained once again in quantum trajectories $\avgc{\hat{X}^{(\sigma)}(t)}$ and $\avgc{\hat{P}^{(\sigma)}(t)}$.

%%%%%%%%%%%%%%%%%%%%%%%%%%%%%%%%%%%%%%%%%%%%%%%%%%%%%%%%%%%%%%%%%%%%%%%%%%%%%%%%%%%%%%%%%%%%%%%%%%

\begin{figure}[t]
    \centering
    \includegraphics[scale=1.0]{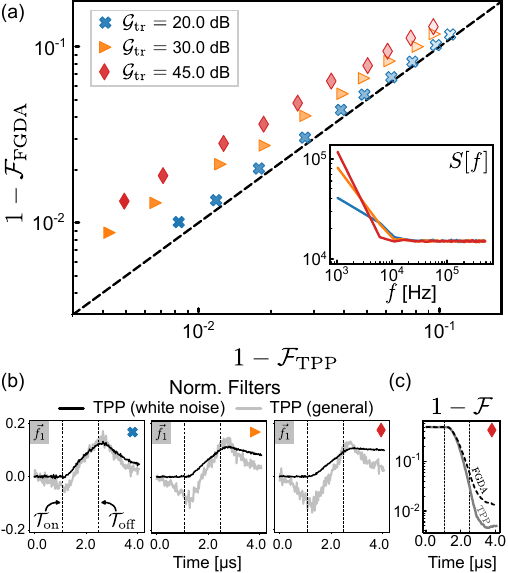}
    \caption{\textbf{Classification performance of \RC{} versus FGDA on simulated dataset of readout via a phase-preserving quantum amplifier.} (a) Classification infidelities for varying amplifier transmission gains $\mathcal{G}_{\rm tr}$ as a function of measurement signal amplitude (more opaque markers are higher amplitudes). The ratio of the bare amplifier linewidth to the cavity mode linewidth is $\gamma/\kappa = 5$. Noise PSD is shown in the inset for the different operating gains (for a linear amplifier, this is independent of the measurement signal amplitude). (b) Learned filters under white noise assumption (black) and general noise conditions (gray) for representative datasets of each value of $\mathcal{G}_{\rm tr}$. (c) Classification infidelities as a function of total time $t$. The measurement tone is only on between the two vertical dashed lines. }
    \label{fig:simResultsAmp}
\end{figure}

%%%%%%%%%%%%%%%%%%%%%%%%%%%%%%%%%%%%%%%%%%%%%%%%%%%%%%%%%%%%%%%%%%%%%%%%%%%%%%%%%%%%%%%%%%%%%%%%%%

% To quantify the correlations in measured data, we introduce the power spectral density, which is simply the Fourier transform of the autocorrelation function (by the Wiener-Khinchin theorem), $S_{m}[f] = \int d\tau~e^{-i2\pi f \tau}\E{x^m(0)x^m(\tau)}$.  

We restrict ourselves for the moment to binary classification of states $\ket{e}$ and $\ket{g}$; here, the matched filtering (MF) scheme is unambiguously defined, and serves as a concrete benchmark for comparison to the \RC{} approach. In Fig.~\ref{fig:simResultsAmp}, we compare calculated infidelities using the FGDA and \RC{} approaches for three different values of amplifier transmission gain $\mathcal{G}_{\rm tr}$, and as a function of the coherent input tone power: darker markers correspond to readout with stronger input tones. 

To understand how correlations in the measured data depend on the varying amplifier gain, we introduce the noise power spectral density (PSD) of the data (here, the $I$-quadrature) for state $\ket{p}$,
% \begin{align}
%     S^{(p)}[f] = \frac{1}{\NTrain}\sum_n^{\NTrain}\sum_{j>k}^{\NT} e^{-i2\pi f \tau_j} I_n^{(p)}(t_j)I_n^{(p)}(t_k) 
% \end{align}
% \begin{align}
%     S^{(p)}[f] = \frac{1}{\NTrain}\sum_{n=1}^{\NTrain}\sum_{j>k}^{\NT} e^{-i2\pi f \tau_j}[\bm{\vec{\zeta}}^{(p)}_{(n)}]_j[\bm{\vec{\zeta}}^{(p)}_{(n)}]_k \approx \sum_{j>k}^{\NT} e^{-i2\pi f \tau_j} \bm{\Sigma}_{jk}^{(p)}
% \end{align}
\begin{align}
    S^{(p)}[f] \approx \sum_{j>k}^{\NT} e^{-i2\pi f \tau_{jk} } \bm{\Sigma}_{jk}^{(p)}
    \label{eq:noisepsd}
\end{align}
where $\tau_{jk} = \Delta t(j-k)$. The PSD is simply the Fourier transform of the noise autocorrelation function (by the Wiener-Khinchin theorem). Through $\mathbf{V}$, the \RC{} learns from these correlations when optimizing filters. The noise PSD is plotted in the inset of Fig.~\ref{fig:simResultsAmp}; for the current readout task, this is independent of $p$. With increasing gain, the PSD deviates from the flat spectrum representative of white noise to a spectrum peaked at low frequencies, indicative of an extended correlation time. The observations also emphasize that added noise by the quantum amplifier dominates over heterodyne measurement noise $\xi$, as well as excess classical noise $\xi^{\rm cl}$.

For the lowest considered amplifier gain, we see that the FGDA and \RC{} classification performance is quite close. However, with increasing gain, the FGDA infidelity is substantially higher, up to an order of magnitude worse for the largest gain considered here. This \RC{} performance advantage is enabled by optimized filters, shown in Fig.~\ref{fig:simResultsAmp}(b). The measurement tone is only on between the two dashed vertical lines. The curves in black show white noise filters, exactly equal to the MF in this binary case. Note that these filters also change with gain: the amplifier response time increases at higher gains, so the mean traces and hence the MF derived from these traces exhibit much slower rise and fall times. The general \RC{} filter is similar to the MF at low gains, but becomes markedly distinct at higher gains.

Interestingly, one such change is that at high gains the general \RC{} filter becomes non-zero even prior to the measurement signal turning on (the first vertical dashed line). This appears odd at first sight, since there must not be any information that could enable state classification before a measurement tone probes the cavity used for dispersive qubit measurement. To validate this, in Fig.~\ref{fig:simResultsAmp}(d) we plot $1-\mathcal{F}$ calculated for an increasing length of measured data, $t \in [0,\mathcal{T}_{\rm meas}]$. We clearly see that for $t < \mathcal{T}_{\rm on}$, both the \RC{} and FGDA cannot distinguish the states, as must be the case. The non-zero segment of the general \RC{} filter before $\mathcal{T}_{\rm on}$ instead accounts for noise correlations. In particular, due to the long correlation time of noise added by the quantum amplifier, noise in data beyond $\mathcal{T}_{\rm on}$ is correlated with noise from $t < \mathcal{T}_{\rm on}$. The general \RC{} filter is aware of these correlations that the standard MF is completely oblivious to, and by accounting for them improves classification performance.

% In contrast, the MF operates under simplifying assumptions on the noise, so the resulting filter is predominantly determined by the mean traces of measured data, and is oblivious to noise correlations. 

\subsection{Correlated quantum noise due to multi-level transitions}
\label{subsec:transition}

A transmon is a multi-level artificial atom, as described by Eq.~(\ref{eq:hdisp}); as a result, it is possible to excite levels beyond the typical two-level computational subspace of $e$ and $g$ states. Such transitions manifest as stochastic quantum jumps in quantum measurement data, and are an important source of error in readout.

%%%%%%%%%%%%%%%%%%%%%%%%%%%%%%%%%%%%%%%%%%%%%%%%%%%%%%%%%%%%%%%%%%%%%%%%%%%%%%%%%%%%%%%%%%%%%%%%%%

\begin{figure}[t]
    \centering
    \includegraphics[scale=1.0]{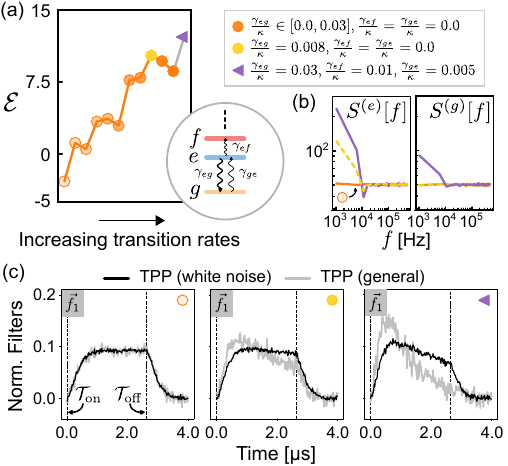}
    \caption{\textbf{Classification performance of \RC{} versus FGDA on simulated dataset of readout of a qubit experiencing multi-level transitions.} (a) $\metric$ as a function of increasing transition rate values (more opaque markers), as shown. Schematic in the inset shows transmon levels and non-zero transition rates considered. (b) Noise PSD $S^{(p)}[f]$ for three representative datasets in the inset, indicating deviation from flat (white noise) as measurement data includes more transitions. (c) \RC{} learned filters (gray) compared to matched filters (black) for representative datasets, showing adaptation with transition rates. }
    \label{fig:simTransitions}
\end{figure}

%%%%%%%%%%%%%%%%%%%%%%%%%%%%%%%%%%%%%%%%%%%%%%%%%%%%%%%%%%%%%%%%%%%%%%%%%%%%%%%%%%%%%%%%%%%%%%%%%%

To model measurement under such conditions, we now consider the dispersive heterodyne readout of a finite lifetime transmon with possible occupied levels $\{e,g,f\}$. We further allow only a subset of all possible allowed transitions between these levels, and with static rates: $\ket{e} \to \ket{g}$ at rate $\gamma_{eg}$, the reverse $\ket{g} \to \ket{e}$ at rate $\gamma_{ge}$, and $\ket{e} \to \ket{f}$ at rate $\gamma_{ef}$ (see Fig.~\ref{fig:simTransitions} inset). The transitions are described by superoperator $\Lchain$, while $\Lmeas$ describes the measurement tone incident on the cavity, and the heterodyne measurement superoperator for the same; for full details see Appendix~\ref{app:simTransitions}.

For simplicity, we now further neglect excess classical noise added by the measurement chain, dropping terms $\xi_{I/Q}^{\rm cl}(t)$. As a result, the obtained measurement records, Eqs.~(\ref{eq:hetI}),~(\ref{eq:hetQ}), contain only two noise sources: white heterodyne measurement noise, and quantum noise due to qubit state transitions imprinted on the emanated cavity field, contained in quantum trajectories of cavity quadratures $\avgc{\hat{X}^{(\sigma)}(t)}$ and $\avgc{\hat{P}^{(\sigma)}(t)}$. We then generate simulated datasets by integrating the resulting full SME, Eq.~(\ref{eq:sme}) for different values of transition rates, and consider the task of binary classification of states $p \in \{e,g\}$.

% $S^{(p)}[f] = \int d\tau~e^{-i2\pi f \tau}\E{I^{(p)}(0)I^{(p)}(\tau)} - $

We compare the performance of a trained \RC{} against that of an FGDA with an empirical MF using the metric $\metric$ in Fig.~\ref{fig:simTransitions}(a) with varying transition rates. The noise PSD is plotted in Fig.~\ref{fig:simTransitions}(b) for representative datasets. In the absence of any transitions (lightest orange), $S^{(p)}[f]$ is flat at all frequencies, regardless of the initially prepared state $p$. This is because the measured data only has heterodyne white noise. With an increase in $\gamma_{eg}$, we note that $S^{(e)}[f]$ deviates from the white noise spectrum, attaining a peak at low frequencies. In contrast, $S^{(g)}[f]$ remains unchanged as trajectories for initial states $\ket{g}$ undergo no transitions. In the most complex case where we allow for all considered transitions, $S^{(g)}[f]$ also starts to demonstrate deviation from the white noise spectrum.

From readout datasets with no transitions to readout data with increasing transition rates, we note a small but clear improvement in classification performance using the trained \RC{} in comparison to the FGDA. That the \RC{} is able to learn information in the presence of transitions that evades the MF is clear when we compare the two sets of filters in Fig.~\ref{fig:simTransitions}(c). As the transition rates increase, the MF undergoes modifications due to the changes to the means of heterodyne records. However, the \RC{} is sensitive to \ch{changes} beyond means - in the correlations of measured data - and increasingly learns a distinct filter with sharply decaying features. We note that the utility of similar exponential linear filters for finite-lifetime qubits has been the subject of earlier analytic work~\cite{gambetta_protocols_2007}. The \RC{} approach generalizes the ability to learn such filters in the presence of arbitrary transition rates and measurement tones, and for multi-state classification. 

One may note that in the absence any multi-level transitions (Fig.~\ref{fig:simTransitions}(a), first datapoint) the FGDA appears to outperform the \RC{} ($\metric < 0$); given the results of Sec.~\ref{subsec:whitenoisesim}, this may seem odd, as here the measurement noise is exactly Gaussian white noise, so the \RC{} filter reduces exactly to the MF used in the standard FGDA. The important distinction is that, unlike Sec.~\ref{subsec:whitenoisesim}, here we are deploying the \textit{general} \RC{}, which makes no \textit{a priori} assumptions about noise characteristics. In the special case where the noise is in fact Gaussian white noise, the MF is already cognizant of the correct noise statistics, while the \RC{} must learn them via training, leading to a slight under-performance that is alleviated as the size of the training dataset is increased (see also Appendix~\ref{app:training}). Of course, this freedom is precisely what enables the \RC{} to learn more efficiently when noise characteristics are not simply Gaussian and white, for example under increasing multi-level transitions. There, the \RC{} shows an improvement relative to the standard FGDA \textit{in spite} of having to learn the new noise statistics from training data. For these more complex noise conditions, the standard MF is now sub-optimal, and the FGDA performance suffers as a result. 

Finally, we emphasize that the simplified transition model considered here is chosen to highlight the ability of the \RC{} to learn quantum noise associated with quantum jumps under controlled noise conditions, where no other nontrivial noise sources (classical or quantum) exist. The \RC{} approach to learning is model-free, and its ability to learn in more general noise settings is demonstrated by its adaptation to real qubit readout in Sec.~\ref{sec:rcexpt}.

\section{Discussion and Outlook}

In this paper we have demonstrated a machine learning approach to classification of an arbitrary number of states using temporal data obtained from quantum measurement chains. While we have focused on the task of dispersive readout of multi-level transmons, the \RC{} approach applies broadly to quantum systems, and more generally physical systems, monitored over time. Our results show that the \RC{} framework for processing quantum measurement data reduces to standard approaches based on matched filtering in the precise regimes of validity of the latter. However, the \RC{} can adapt to more general readout scenarios to significantly outperform matched filtering schemes. We show this improvement for the \RC{} trained on real qubit readout data to confirm the practical utility of our scheme. 

% The \RC{} mapping is defined via a set of linear weights, making it ideal for 

Rather than treating the \RC{} as a black box, in our work we clarify the learning mechanism that enables the \RC{} to outperform matched filtering schemes. First, we develop a heuristic interpretation of the \RC{} mapping as one of applying temporal filters to measured data. \RC{} learning then amounts to learning optimal filters. Deconstructing the learning scheme, we find the \RC{} performance advantage is enabled by its ability to learn optimal filters by accounting for noise \textit{correlations} in temporal data. When this noise is purely white, the \RC{} approach provides a generalization of matched filtering to an arbitrary number of states. 

Crucially, we find that the \RC{} can efficiently learn from correlations not just due to classical signals, or in principle due to quantum noise in theory, but from practical systems where the majority of the noise is quantum in origin. In addition to real qubit readout, using theoretical simulations where the strength of quantum noise sources can be tuned precisely, such as noise due to multi-level transitions or the added noise of phase-preserving quantum amplifiers, we clearly demonstrate that the \RC{} can learn from quantum noise correlations to outperform standard matched filtering. Furthermore, our precise identification of quantum correlations as a harnessable resource can help guide future machine learning approaches to quantum signal processing.
 
The \RC{} approach, anchored by its connection to standard matched filtering, with demonstrated advantages for real qubit readout under complex readout conditions, and feasibility for FPGA implementations (to be demonstrated in future work), is ideal for integration with cQED measurement chains for the next step in readout optimization. Furthermore, the \RC{}'s generality and ability to \ch{efficiently} learn from data could pave the way for an even broader class of applications. An important potential use is as a post-processor of quantum measurement data for quantum machine learning. With the use of general quantum machines for information processing, the optimal means to extract data from their measurements may not always be known. \ch{We believe} the \RC{} is ideally suited to uncover the optimal linear post-processing step, through training that could be incorporated as part of the optimization of the quantum machine. \ch{This is because the existence of an exact analytic form for the optimal trained \RC{} weights eliminates the need for multiple training epochs, batch-wise evaluations, or gradient computations, so that training the \RC{} adds minimal complexity to the optimization of an already complex quantum measurement chain, in stark contrast to the substantial overhead of training a neural network used as a post-processor.} Finally, optimal state estimation is essential for control applications. The trainable \RC{} can form part of a framework for control applications, such as Kalman filtering for quantum systems.

\begin{acknowledgements}
    We would like to thank Leon Bello, Dan Gauthier, and Shyam Shankar for useful discussions. This work was supported by the AFOSR under Grant No. FA9550-20-1-0177, by the Army Research Office under Grant No. W911NF18-1-0144, and by the J. Insley Blair Pyne Fund. The views and conclusions contained in this document are those of the authors and should not be interpreted as representing the official policies, either expressed or implied, of the AFOSR, Army Research Office, or the U.S. Government. The U.S. Government is authorized to reproduce and distribute reprints for Government purposes notwithstanding any copyright notation herein.
\end{acknowledgements}

% \begin{enumerate}
%     \item Basics of learning to classify states with the \RC{} and distinguishing wrt. standard approaches 
%     \item Explore limits in which matched filter is learned to show that the \RC{} reduces to standard approaches suitably
%     \item briefly explain majority vote matched filter
%     \begin{enumerate}
%         \item address that it does NOT scale well
%         \item many ways to receive no answer in 4-state case
%         \item can we visualize this in the IQ plane?
%     \end{enumerate}
%     \item Real data (fake and qubit):
%     \begin{enumerate}
%         \item High fidelity $T_1 = 10$ and $T_1=80~\mu{\rm s}$ qubits
%         \item EGF data to showcase 3-state classification
%         \item Autonomous - Detects artefacts
%         \item Spontaneous downconversion regime?
%         \item Principal component analysis/Eigentasks and relation to overfitting
%         \item Define \% of errors metric
%     \end{enumerate}
%     \item Fake data:
%     \begin{enumerate}
%         \item Spontaneous downconversion regime simulation
%         \item Different variances of input signals
%     \end{enumerate}
%     \item Outro: connection to future projects
% \end{enumerate}

\bibliography{main}

%apsrev4-2.bst 2019-01-14 (MD) hand-edited version of apsrev4-1.bst
%Control: key (0)
%Control: author (8) initials jnrlst
%Control: editor formatted (1) identically to author
%Control: production of article title (0) allowed
%Control: page (0) single
%Control: year (1) truncated
%Control: production of eprint (0) enabled
\begin{thebibliography}{61}%
\makeatletter
\providecommand \@ifxundefined [1]{%
 \@ifx{#1\undefined}
}%
\providecommand \@ifnum [1]{%
 \ifnum #1\expandafter \@firstoftwo
 \else \expandafter \@secondoftwo
 \fi
}%
\providecommand \@ifx [1]{%
 \ifx #1\expandafter \@firstoftwo
 \else \expandafter \@secondoftwo
 \fi
}%
\providecommand \natexlab [1]{#1}%
\providecommand \enquote  [1]{``#1''}%
\providecommand \bibnamefont  [1]{#1}%
\providecommand \bibfnamefont [1]{#1}%
\providecommand \citenamefont [1]{#1}%
\providecommand \href@noop [0]{\@secondoftwo}%
\providecommand \href [0]{\begingroup \@sanitize@url \@href}%
\providecommand \@href[1]{\@@startlink{#1}\@@href}%
\providecommand \@@href[1]{\endgroup#1\@@endlink}%
\providecommand \@sanitize@url [0]{\catcode `\\12\catcode `\$12\catcode `\&12\catcode `\#12\catcode `\^12\catcode `\_12\catcode `\%12\relax}%
\providecommand \@@startlink[1]{}%
\providecommand \@@endlink[0]{}%
\providecommand \url  [0]{\begingroup\@sanitize@url \@url }%
\providecommand \@url [1]{\endgroup\@href {#1}{\urlprefix }}%
\providecommand \urlprefix  [0]{URL }%
\providecommand \Eprint [0]{\href }%
\providecommand \doibase [0]{https://doi.org/}%
\providecommand \selectlanguage [0]{\@gobble}%
\providecommand \bibinfo  [0]{\@secondoftwo}%
\providecommand \bibfield  [0]{\@secondoftwo}%
\providecommand \translation [1]{[#1]}%
\providecommand \BibitemOpen [0]{}%
\providecommand \bibitemStop [0]{}%
\providecommand \bibitemNoStop [0]{.\EOS\space}%
\providecommand \EOS [0]{\spacefactor3000\relax}%
\providecommand \BibitemShut  [1]{\csname bibitem#1\endcsname}%
\let\auto@bib@innerbib\@empty
%</preamble>
\bibitem [{\citenamefont {Roy}\ and\ \citenamefont {Devoret}(2016)}]{Roy2016}%
  \BibitemOpen
  \bibfield  {author} {\bibinfo {author} {\bibfnamefont {A.}~\bibnamefont {Roy}}\ and\ \bibinfo {author} {\bibfnamefont {M.}~\bibnamefont {Devoret}},\ }\bibfield  {title} {\bibinfo {title} {Introduction to parametric amplification of quantum signals with josephson circuits},\ }\href {https://doi.org/10.1016/J.CRHY.2016.07.012} {\bibfield  {journal} {\bibinfo  {journal} {Comptes Rendus Physique}\ }\textbf {\bibinfo {volume} {17}},\ \bibinfo {pages} {740} (\bibinfo {year} {2016})}\BibitemShut {NoStop}%
\bibitem [{\citenamefont {Aumentado}(2020)}]{Aumentado2020}%
  \BibitemOpen
  \bibfield  {author} {\bibinfo {author} {\bibfnamefont {J.}~\bibnamefont {Aumentado}},\ }\bibfield  {title} {\bibinfo {title} {Superconducting parametric amplifiers: The state of the art in josephson parametric amplifiers},\ }\href {https://doi.org/10.1109/MMM.2020.2993476} {\bibfield  {journal} {\bibinfo  {journal} {IEEE Microwave Magazine}\ }\textbf {\bibinfo {volume} {21}},\ \bibinfo {pages} {45} (\bibinfo {year} {2020})}\BibitemShut {NoStop}%
\bibitem [{\citenamefont {Place}\ \emph {et~al.}(2021)\citenamefont {Place}, \citenamefont {Rodgers}, \citenamefont {Mundada}, \citenamefont {Smitham}, \citenamefont {Fitzpatrick}, \citenamefont {Leng}, \citenamefont {Premkumar}, \citenamefont {Bryon}, \citenamefont {Vrajitoarea}, \citenamefont {Sussman}, \citenamefont {Cheng}, \citenamefont {Madhavan}, \citenamefont {Babla}, \citenamefont {Le}, \citenamefont {Gang}, \citenamefont {Jäck}, \citenamefont {Gyenis}, \citenamefont {Yao}, \citenamefont {Cava}, \citenamefont {de~Leon},\ and\ \citenamefont {Houck}}]{place_new_2021}%
  \BibitemOpen
  \bibfield  {author} {\bibinfo {author} {\bibfnamefont {A.~P.~M.}\ \bibnamefont {Place}}, \bibinfo {author} {\bibfnamefont {L.~V.~H.}\ \bibnamefont {Rodgers}}, \bibinfo {author} {\bibfnamefont {P.}~\bibnamefont {Mundada}}, \bibinfo {author} {\bibfnamefont {B.~M.}\ \bibnamefont {Smitham}}, \bibinfo {author} {\bibfnamefont {M.}~\bibnamefont {Fitzpatrick}}, \bibinfo {author} {\bibfnamefont {Z.}~\bibnamefont {Leng}}, \bibinfo {author} {\bibfnamefont {A.}~\bibnamefont {Premkumar}}, \bibinfo {author} {\bibfnamefont {J.}~\bibnamefont {Bryon}}, \bibinfo {author} {\bibfnamefont {A.}~\bibnamefont {Vrajitoarea}}, \bibinfo {author} {\bibfnamefont {S.}~\bibnamefont {Sussman}}, \bibinfo {author} {\bibfnamefont {G.}~\bibnamefont {Cheng}}, \bibinfo {author} {\bibfnamefont {T.}~\bibnamefont {Madhavan}}, \bibinfo {author} {\bibfnamefont {H.~K.}\ \bibnamefont {Babla}}, \bibinfo {author} {\bibfnamefont {X.~H.}\ \bibnamefont {Le}}, \bibinfo {author} {\bibfnamefont {Y.}~\bibnamefont {Gang}}, \bibinfo {author} {\bibfnamefont
  {B.}~\bibnamefont {Jäck}}, \bibinfo {author} {\bibfnamefont {A.}~\bibnamefont {Gyenis}}, \bibinfo {author} {\bibfnamefont {N.}~\bibnamefont {Yao}}, \bibinfo {author} {\bibfnamefont {R.~J.}\ \bibnamefont {Cava}}, \bibinfo {author} {\bibfnamefont {N.~P.}\ \bibnamefont {de~Leon}},\ and\ \bibinfo {author} {\bibfnamefont {A.~A.}\ \bibnamefont {Houck}},\ }\bibfield  {title} {{\selectlanguage {English}\bibinfo {title} {New material platform for superconducting transmon qubits with coherence times exceeding 0.3 milliseconds}},\ }\href {https://doi.org/10.1038/s41467-021-22030-5} {\bibfield  {journal} {\bibinfo  {journal} {Nature Communications}\ }\textbf {\bibinfo {volume} {12}},\ \bibinfo {pages} {1779} (\bibinfo {year} {2021})},\ \bibinfo {note} {number: 1 Publisher: Nature Publishing Group}\BibitemShut {NoStop}%
\bibitem [{\citenamefont {Angelatos}\ \emph {et~al.}(2021)\citenamefont {Angelatos}, \citenamefont {Khan},\ and\ \citenamefont {Türeci}}]{angelatos_reservoir_2021}%
  \BibitemOpen
  \bibfield  {author} {\bibinfo {author} {\bibfnamefont {G.}~\bibnamefont {Angelatos}}, \bibinfo {author} {\bibfnamefont {S.~A.}\ \bibnamefont {Khan}},\ and\ \bibinfo {author} {\bibfnamefont {H.~E.}\ \bibnamefont {Türeci}},\ }\bibfield  {title} {\bibinfo {title} {Reservoir {Computing} {Approach} to {Quantum} {State} {Measurement}},\ }\href {https://doi.org/10.1103/PhysRevX.11.041062} {\bibfield  {journal} {\bibinfo  {journal} {Physical Review X}\ }\textbf {\bibinfo {volume} {11}},\ \bibinfo {pages} {041062} (\bibinfo {year} {2021})}\BibitemShut {NoStop}%
\bibitem [{\citenamefont {Khan}\ \emph {et~al.}(2021)\citenamefont {Khan}, \citenamefont {Hu}, \citenamefont {Angelatos},\ and\ \citenamefont {Türeci}}]{khan_physical_2021}%
  \BibitemOpen
  \bibfield  {author} {\bibinfo {author} {\bibfnamefont {S.~A.}\ \bibnamefont {Khan}}, \bibinfo {author} {\bibfnamefont {F.}~\bibnamefont {Hu}}, \bibinfo {author} {\bibfnamefont {G.}~\bibnamefont {Angelatos}},\ and\ \bibinfo {author} {\bibfnamefont {H.~E.}\ \bibnamefont {Türeci}},\ }\bibfield  {title} {\bibinfo {title} {Physical reservoir computing using finitely-sampled quantum systems},\ }\bibfield  {journal} {\bibinfo  {journal} {arXiv:2110.13849 [quant-ph]}\ }\href {https://doi.org/10.48550/arXiv.2110.13849} {10.48550/arXiv.2110.13849} (\bibinfo {year} {2021})\BibitemShut {NoStop}%
\bibitem [{\citenamefont {Nokkala}\ \emph {et~al.}(2021)\citenamefont {Nokkala}, \citenamefont {Martínez-Peña}, \citenamefont {Giorgi}, \citenamefont {Parigi}, \citenamefont {Soriano},\ and\ \citenamefont {Zambrini}}]{nokkala_gaussian_2021}%
  \BibitemOpen
  \bibfield  {author} {\bibinfo {author} {\bibfnamefont {J.}~\bibnamefont {Nokkala}}, \bibinfo {author} {\bibfnamefont {R.}~\bibnamefont {Martínez-Peña}}, \bibinfo {author} {\bibfnamefont {G.~L.}\ \bibnamefont {Giorgi}}, \bibinfo {author} {\bibfnamefont {V.}~\bibnamefont {Parigi}}, \bibinfo {author} {\bibfnamefont {M.~C.}\ \bibnamefont {Soriano}},\ and\ \bibinfo {author} {\bibfnamefont {R.}~\bibnamefont {Zambrini}},\ }\bibfield  {title} {{\selectlanguage {English}\bibinfo {title} {Gaussian states of continuous-variable quantum systems provide universal and versatile reservoir computing}},\ }\href {https://doi.org/10.1038/s42005-021-00556-w} {\bibfield  {journal} {\bibinfo  {journal} {Communications Physics}\ }\textbf {\bibinfo {volume} {4}},\ \bibinfo {pages} {1} (\bibinfo {year} {2021})}\BibitemShut {NoStop}%
\bibitem [{\citenamefont {Martínez-Peña}\ \emph {et~al.}(2021)\citenamefont {Martínez-Peña}, \citenamefont {Giorgi}, \citenamefont {Nokkala}, \citenamefont {Soriano},\ and\ \citenamefont {Zambrini}}]{martinez-pena_dynamical_2021}%
  \BibitemOpen
  \bibfield  {author} {\bibinfo {author} {\bibfnamefont {R.}~\bibnamefont {Martínez-Peña}}, \bibinfo {author} {\bibfnamefont {G.~L.}\ \bibnamefont {Giorgi}}, \bibinfo {author} {\bibfnamefont {J.}~\bibnamefont {Nokkala}}, \bibinfo {author} {\bibfnamefont {M.~C.}\ \bibnamefont {Soriano}},\ and\ \bibinfo {author} {\bibfnamefont {R.}~\bibnamefont {Zambrini}},\ }\bibfield  {title} {\bibinfo {title} {Dynamical {Phase} {Transitions} in {Quantum} {Reservoir} {Computing}},\ }\href {https://doi.org/10.1103/PhysRevLett.127.100502} {\bibfield  {journal} {\bibinfo  {journal} {Physical Review Letters}\ }\textbf {\bibinfo {volume} {127}},\ \bibinfo {pages} {100502} (\bibinfo {year} {2021})}\BibitemShut {NoStop}%
\bibitem [{\citenamefont {Mujal}\ \emph {et~al.}(2021)\citenamefont {Mujal}, \citenamefont {Martínez-Peña}, \citenamefont {Nokkala}, \citenamefont {García-Beni}, \citenamefont {Giorgi}, \citenamefont {Soriano},\ and\ \citenamefont {Zambrini}}]{mujal_opportunities_2021}%
  \BibitemOpen
  \bibfield  {author} {\bibinfo {author} {\bibfnamefont {P.}~\bibnamefont {Mujal}}, \bibinfo {author} {\bibfnamefont {R.}~\bibnamefont {Martínez-Peña}}, \bibinfo {author} {\bibfnamefont {J.}~\bibnamefont {Nokkala}}, \bibinfo {author} {\bibfnamefont {J.}~\bibnamefont {García-Beni}}, \bibinfo {author} {\bibfnamefont {G.~L.}\ \bibnamefont {Giorgi}}, \bibinfo {author} {\bibfnamefont {M.~C.}\ \bibnamefont {Soriano}},\ and\ \bibinfo {author} {\bibfnamefont {R.}~\bibnamefont {Zambrini}},\ }\bibfield  {title} {{\selectlanguage {English}\bibinfo {title} {Opportunities in {Quantum} {Reservoir} {Computing} and {Extreme} {Learning} {Machines}}},\ }\href {https://doi.org/10.1002/qute.202100027} {\bibfield  {journal} {\bibinfo  {journal} {Advanced Quantum Technologies}\ }\textbf {\bibinfo {volume} {4}},\ \bibinfo {pages} {2100027} (\bibinfo {year} {2021})}\BibitemShut {NoStop}%
\bibitem [{\citenamefont {Sank}\ \emph {et~al.}(2016)\citenamefont {Sank}, \citenamefont {Chen}, \citenamefont {Khezri}, \citenamefont {Kelly}, \citenamefont {Barends}, \citenamefont {Campbell}, \citenamefont {Chen}, \citenamefont {Chiaro}, \citenamefont {Dunsworth}, \citenamefont {Fowler}, \citenamefont {Jeffrey}, \citenamefont {Lucero}, \citenamefont {Megrant}, \citenamefont {Mutus}, \citenamefont {Neeley}, \citenamefont {Neill}, \citenamefont {O'Malley}, \citenamefont {Quintana}, \citenamefont {Roushan}, \citenamefont {Vainsencher}, \citenamefont {White}, \citenamefont {Wenner}, \citenamefont {Korotkov},\ and\ \citenamefont {Martinis}}]{Sank2016}%
  \BibitemOpen
  \bibfield  {author} {\bibinfo {author} {\bibfnamefont {D.}~\bibnamefont {Sank}}, \bibinfo {author} {\bibfnamefont {Z.}~\bibnamefont {Chen}}, \bibinfo {author} {\bibfnamefont {M.}~\bibnamefont {Khezri}}, \bibinfo {author} {\bibfnamefont {J.}~\bibnamefont {Kelly}}, \bibinfo {author} {\bibfnamefont {R.}~\bibnamefont {Barends}}, \bibinfo {author} {\bibfnamefont {B.}~\bibnamefont {Campbell}}, \bibinfo {author} {\bibfnamefont {Y.}~\bibnamefont {Chen}}, \bibinfo {author} {\bibfnamefont {B.}~\bibnamefont {Chiaro}}, \bibinfo {author} {\bibfnamefont {A.}~\bibnamefont {Dunsworth}}, \bibinfo {author} {\bibfnamefont {A.}~\bibnamefont {Fowler}}, \bibinfo {author} {\bibfnamefont {E.}~\bibnamefont {Jeffrey}}, \bibinfo {author} {\bibfnamefont {E.}~\bibnamefont {Lucero}}, \bibinfo {author} {\bibfnamefont {A.}~\bibnamefont {Megrant}}, \bibinfo {author} {\bibfnamefont {J.}~\bibnamefont {Mutus}}, \bibinfo {author} {\bibfnamefont {M.}~\bibnamefont {Neeley}}, \bibinfo {author} {\bibfnamefont {C.}~\bibnamefont {Neill}}, \bibinfo
  {author} {\bibfnamefont {P.~J.}\ \bibnamefont {O'Malley}}, \bibinfo {author} {\bibfnamefont {C.}~\bibnamefont {Quintana}}, \bibinfo {author} {\bibfnamefont {P.}~\bibnamefont {Roushan}}, \bibinfo {author} {\bibfnamefont {A.}~\bibnamefont {Vainsencher}}, \bibinfo {author} {\bibfnamefont {T.}~\bibnamefont {White}}, \bibinfo {author} {\bibfnamefont {J.}~\bibnamefont {Wenner}}, \bibinfo {author} {\bibfnamefont {A.~N.}\ \bibnamefont {Korotkov}},\ and\ \bibinfo {author} {\bibfnamefont {J.~M.}\ \bibnamefont {Martinis}},\ }\bibfield  {title} {\bibinfo {title} {Measurement-induced state transitions in a superconducting qubit: Beyond the rotating wave approximation},\ }\href {https://doi.org/10.1103/PHYSREVLETT.117.190503/FIGURES/4/MEDIUM} {\bibfield  {journal} {\bibinfo  {journal} {Physical Review Letters}\ }\textbf {\bibinfo {volume} {117}},\ \bibinfo {pages} {190503} (\bibinfo {year} {2016})}\BibitemShut {NoStop}%
\bibitem [{\citenamefont {Malekakhlagh}\ \emph {et~al.}(2020)\citenamefont {Malekakhlagh}, \citenamefont {Petrescu},\ and\ \citenamefont {Türeci}}]{malekakhlagh_lifetime_2020}%
  \BibitemOpen
  \bibfield  {author} {\bibinfo {author} {\bibfnamefont {M.}~\bibnamefont {Malekakhlagh}}, \bibinfo {author} {\bibfnamefont {A.}~\bibnamefont {Petrescu}},\ and\ \bibinfo {author} {\bibfnamefont {H.~E.}\ \bibnamefont {Türeci}},\ }\bibfield  {title} {\bibinfo {title} {Lifetime renormalization of weakly anharmonic superconducting qubits. {I}. {Role} of number nonconserving terms},\ }\href {https://doi.org/10.1103/PhysRevB.101.134509} {\bibfield  {journal} {\bibinfo  {journal} {Physical Review B}\ }\textbf {\bibinfo {volume} {101}},\ \bibinfo {pages} {134509} (\bibinfo {year} {2020})},\ \bibinfo {note} {publisher: American Physical Society}\BibitemShut {NoStop}%
\bibitem [{\citenamefont {Petrescu}\ \emph {et~al.}(2020)\citenamefont {Petrescu}, \citenamefont {Malekakhlagh},\ and\ \citenamefont {Türeci}}]{petrescu_lifetime_2020}%
  \BibitemOpen
  \bibfield  {author} {\bibinfo {author} {\bibfnamefont {A.}~\bibnamefont {Petrescu}}, \bibinfo {author} {\bibfnamefont {M.}~\bibnamefont {Malekakhlagh}},\ and\ \bibinfo {author} {\bibfnamefont {H.~E.}\ \bibnamefont {Türeci}},\ }\bibfield  {title} {\bibinfo {title} {Lifetime renormalization of driven weakly anharmonic superconducting qubits. {II}. {The} readout problem},\ }\href {https://doi.org/10.1103/PhysRevB.101.134510} {\bibfield  {journal} {\bibinfo  {journal} {Physical Review B}\ }\textbf {\bibinfo {volume} {101}},\ \bibinfo {pages} {134510} (\bibinfo {year} {2020})}\BibitemShut {NoStop}%
\bibitem [{\citenamefont {Hanai}\ \emph {et~al.}(2021)\citenamefont {Hanai}, \citenamefont {McDonald},\ and\ \citenamefont {Clerk}}]{hanai_intrinsic_2021}%
  \BibitemOpen
  \bibfield  {author} {\bibinfo {author} {\bibfnamefont {R.}~\bibnamefont {Hanai}}, \bibinfo {author} {\bibfnamefont {A.}~\bibnamefont {McDonald}},\ and\ \bibinfo {author} {\bibfnamefont {A.}~\bibnamefont {Clerk}},\ }\bibfield  {title} {\bibinfo {title} {Intrinsic mechanisms for drive-dependent {Purcell} decay in superconducting quantum circuits},\ }\href {https://doi.org/10.1103/PhysRevResearch.3.043228} {\bibfield  {journal} {\bibinfo  {journal} {Physical Review Research}\ }\textbf {\bibinfo {volume} {3}},\ \bibinfo {pages} {043228} (\bibinfo {year} {2021})}\BibitemShut {NoStop}%
\bibitem [{\citenamefont {Khezri}\ \emph {et~al.}(2023)\citenamefont {Khezri}, \citenamefont {Opremcak}, \citenamefont {Chen}, \citenamefont {Miao}, \citenamefont {McEwen}, \citenamefont {Bengtsson}, \citenamefont {White}, \citenamefont {Naaman}, \citenamefont {Sank}, \citenamefont {Korotkov}, \citenamefont {Chen},\ and\ \citenamefont {Smelyanskiy}}]{khezri_measurement-induced_2023}%
  \BibitemOpen
  \bibfield  {author} {\bibinfo {author} {\bibfnamefont {M.}~\bibnamefont {Khezri}}, \bibinfo {author} {\bibfnamefont {A.}~\bibnamefont {Opremcak}}, \bibinfo {author} {\bibfnamefont {Z.}~\bibnamefont {Chen}}, \bibinfo {author} {\bibfnamefont {K.~C.}\ \bibnamefont {Miao}}, \bibinfo {author} {\bibfnamefont {M.}~\bibnamefont {McEwen}}, \bibinfo {author} {\bibfnamefont {A.}~\bibnamefont {Bengtsson}}, \bibinfo {author} {\bibfnamefont {T.}~\bibnamefont {White}}, \bibinfo {author} {\bibfnamefont {O.}~\bibnamefont {Naaman}}, \bibinfo {author} {\bibfnamefont {D.}~\bibnamefont {Sank}}, \bibinfo {author} {\bibfnamefont {A.~N.}\ \bibnamefont {Korotkov}}, \bibinfo {author} {\bibfnamefont {Y.}~\bibnamefont {Chen}},\ and\ \bibinfo {author} {\bibfnamefont {V.}~\bibnamefont {Smelyanskiy}},\ }\bibfield  {title} {\bibinfo {title} {Measurement-induced state transitions in a superconducting qubit: {Within} the rotating-wave approximation},\ }\href {https://doi.org/10.1103/PhysRevApplied.20.054008} {\bibfield  {journal} {\bibinfo
  {journal} {Physical Review Applied}\ }\textbf {\bibinfo {volume} {20}},\ \bibinfo {pages} {054008} (\bibinfo {year} {2023})}\BibitemShut {NoStop}%
\bibitem [{\citenamefont {Shillito}\ \emph {et~al.}(2022)\citenamefont {Shillito}, \citenamefont {Petrescu}, \citenamefont {Cohen}, \citenamefont {Beall}, \citenamefont {Hauru}, \citenamefont {Ganahl}, \citenamefont {Lewis}, \citenamefont {Vidal},\ and\ \citenamefont {Blais}}]{Shillito2022}%
  \BibitemOpen
  \bibfield  {author} {\bibinfo {author} {\bibfnamefont {R.}~\bibnamefont {Shillito}}, \bibinfo {author} {\bibfnamefont {A.}~\bibnamefont {Petrescu}}, \bibinfo {author} {\bibfnamefont {J.}~\bibnamefont {Cohen}}, \bibinfo {author} {\bibfnamefont {J.}~\bibnamefont {Beall}}, \bibinfo {author} {\bibfnamefont {M.}~\bibnamefont {Hauru}}, \bibinfo {author} {\bibfnamefont {M.}~\bibnamefont {Ganahl}}, \bibinfo {author} {\bibfnamefont {A.~G.}\ \bibnamefont {Lewis}}, \bibinfo {author} {\bibfnamefont {G.}~\bibnamefont {Vidal}},\ and\ \bibinfo {author} {\bibfnamefont {A.}~\bibnamefont {Blais}},\ }\bibfield  {title} {\bibinfo {title} {Dynamics of transmon ionization},\ }\href {https://doi.org/10.1103/PHYSREVAPPLIED.18.034031/FIGURES/8/MEDIUM} {\bibfield  {journal} {\bibinfo  {journal} {Physical Review Applied}\ }\textbf {\bibinfo {volume} {18}},\ \bibinfo {pages} {034031} (\bibinfo {year} {2022})}\BibitemShut {NoStop}%
\bibitem [{\citenamefont {Thorbeck}\ \emph {et~al.}(2024)\citenamefont {Thorbeck}, \citenamefont {Xiao}, \citenamefont {Kamal},\ and\ \citenamefont {Govia}}]{thorbeck_readout-induced_2024}%
  \BibitemOpen
  \bibfield  {author} {\bibinfo {author} {\bibfnamefont {T.}~\bibnamefont {Thorbeck}}, \bibinfo {author} {\bibfnamefont {Z.}~\bibnamefont {Xiao}}, \bibinfo {author} {\bibfnamefont {A.}~\bibnamefont {Kamal}},\ and\ \bibinfo {author} {\bibfnamefont {L.~C.}\ \bibnamefont {Govia}},\ }\bibfield  {title} {\bibinfo {title} {Readout-{Induced} {Suppression} and {Enhancement} of {Superconducting} {Qubit} {Lifetimes}},\ }\href {https://doi.org/10.1103/PhysRevLett.132.090602} {\bibfield  {journal} {\bibinfo  {journal} {Physical Review Letters}\ }\textbf {\bibinfo {volume} {132}},\ \bibinfo {pages} {090602} (\bibinfo {year} {2024})}\BibitemShut {NoStop}%
\bibitem [{\citenamefont {Gusenkova}\ \emph {et~al.}(2021)\citenamefont {Gusenkova}, \citenamefont {Spiecker}, \citenamefont {Gebauer}, \citenamefont {Willsch}, \citenamefont {Willsch}, \citenamefont {Valenti}, \citenamefont {Karcher}, \citenamefont {Grünhaupt}, \citenamefont {Takmakov}, \citenamefont {Winkel}, \citenamefont {Rieger}, \citenamefont {Ustinov}, \citenamefont {Roch}, \citenamefont {Wernsdorfer}, \citenamefont {Michielsen}, \citenamefont {Sander},\ and\ \citenamefont {Pop}}]{gusenkova_quantum_2021}%
  \BibitemOpen
  \bibfield  {author} {\bibinfo {author} {\bibfnamefont {D.}~\bibnamefont {Gusenkova}}, \bibinfo {author} {\bibfnamefont {M.}~\bibnamefont {Spiecker}}, \bibinfo {author} {\bibfnamefont {R.}~\bibnamefont {Gebauer}}, \bibinfo {author} {\bibfnamefont {M.}~\bibnamefont {Willsch}}, \bibinfo {author} {\bibfnamefont {D.}~\bibnamefont {Willsch}}, \bibinfo {author} {\bibfnamefont {F.}~\bibnamefont {Valenti}}, \bibinfo {author} {\bibfnamefont {N.}~\bibnamefont {Karcher}}, \bibinfo {author} {\bibfnamefont {L.}~\bibnamefont {Grünhaupt}}, \bibinfo {author} {\bibfnamefont {I.}~\bibnamefont {Takmakov}}, \bibinfo {author} {\bibfnamefont {P.}~\bibnamefont {Winkel}}, \bibinfo {author} {\bibfnamefont {D.}~\bibnamefont {Rieger}}, \bibinfo {author} {\bibfnamefont {A.~V.}\ \bibnamefont {Ustinov}}, \bibinfo {author} {\bibfnamefont {N.}~\bibnamefont {Roch}}, \bibinfo {author} {\bibfnamefont {W.}~\bibnamefont {Wernsdorfer}}, \bibinfo {author} {\bibfnamefont {K.}~\bibnamefont {Michielsen}}, \bibinfo {author} {\bibfnamefont
  {O.}~\bibnamefont {Sander}},\ and\ \bibinfo {author} {\bibfnamefont {I.~M.}\ \bibnamefont {Pop}},\ }\bibfield  {title} {\bibinfo {title} {Quantum {Nondemolition} {Dispersive} {Readout} of a {Superconducting} {Artificial} {Atom} {Using} {Large} {Photon} {Numbers}},\ }\href {https://doi.org/10.1103/PhysRevApplied.15.064030} {\bibfield  {journal} {\bibinfo  {journal} {Physical Review Applied}\ }\textbf {\bibinfo {volume} {15}},\ \bibinfo {pages} {064030} (\bibinfo {year} {2021})},\ \bibinfo {note} {publisher: American Physical Society}\BibitemShut {NoStop}%
\bibitem [{\citenamefont {Dumas}\ \emph {et~al.}(2024)\citenamefont {Dumas}, \citenamefont {Groleau-Paré}, \citenamefont {McDonald}, \citenamefont {Muñoz-Arias}, \citenamefont {Lledó}, \citenamefont {D'Anjou},\ and\ \citenamefont {Blais}}]{Dumas2024}%
  \BibitemOpen
  \bibfield  {author} {\bibinfo {author} {\bibfnamefont {M.~F.}\ \bibnamefont {Dumas}}, \bibinfo {author} {\bibfnamefont {B.}~\bibnamefont {Groleau-Paré}}, \bibinfo {author} {\bibfnamefont {A.}~\bibnamefont {McDonald}}, \bibinfo {author} {\bibfnamefont {M.~H.}\ \bibnamefont {Muñoz-Arias}}, \bibinfo {author} {\bibfnamefont {C.}~\bibnamefont {Lledó}}, \bibinfo {author} {\bibfnamefont {B.}~\bibnamefont {D'Anjou}},\ and\ \bibinfo {author} {\bibfnamefont {A.}~\bibnamefont {Blais}},\ }\bibfield  {title} {\bibinfo {title} {Unified picture of measurement-induced ionization in the transmon},\ }\href {https://arxiv.org/abs/2402.06615v1} {\  (\bibinfo {year} {2024})}\BibitemShut {NoStop}%
\bibitem [{\citenamefont {Walter}\ \emph {et~al.}(2017)\citenamefont {Walter}, \citenamefont {Kurpiers}, \citenamefont {Gasparinetti}, \citenamefont {Magnard}, \citenamefont {Potočnik}, \citenamefont {Salathé}, \citenamefont {Pechal}, \citenamefont {Mondal}, \citenamefont {Oppliger}, \citenamefont {Eichler},\ and\ \citenamefont {Wallraff}}]{walter_rapid_2017}%
  \BibitemOpen
  \bibfield  {author} {\bibinfo {author} {\bibfnamefont {T.}~\bibnamefont {Walter}}, \bibinfo {author} {\bibfnamefont {P.}~\bibnamefont {Kurpiers}}, \bibinfo {author} {\bibfnamefont {S.}~\bibnamefont {Gasparinetti}}, \bibinfo {author} {\bibfnamefont {P.}~\bibnamefont {Magnard}}, \bibinfo {author} {\bibfnamefont {A.}~\bibnamefont {Potočnik}}, \bibinfo {author} {\bibfnamefont {Y.}~\bibnamefont {Salathé}}, \bibinfo {author} {\bibfnamefont {M.}~\bibnamefont {Pechal}}, \bibinfo {author} {\bibfnamefont {M.}~\bibnamefont {Mondal}}, \bibinfo {author} {\bibfnamefont {M.}~\bibnamefont {Oppliger}}, \bibinfo {author} {\bibfnamefont {C.}~\bibnamefont {Eichler}},\ and\ \bibinfo {author} {\bibfnamefont {A.}~\bibnamefont {Wallraff}},\ }\bibfield  {title} {\bibinfo {title} {Rapid {High}-{Fidelity} {Single}-{Shot} {Dispersive} {Readout} of {Superconducting} {Qubits}},\ }\href {https://doi.org/10.1103/PhysRevApplied.7.054020} {\bibfield  {journal} {\bibinfo  {journal} {Physical Review Applied}\ }\textbf {\bibinfo {volume}
  {7}},\ \bibinfo {pages} {054020} (\bibinfo {year} {2017})},\ \bibinfo {note} {publisher: American Physical Society}\BibitemShut {NoStop}%
\bibitem [{\citenamefont {Tsang}(2015)}]{tsang_volterra_2015}%
  \BibitemOpen
  \bibfield  {author} {\bibinfo {author} {\bibfnamefont {M.}~\bibnamefont {Tsang}},\ }\bibfield  {title} {\bibinfo {title} {Volterra filters for quantum estimation and detection},\ }\href {https://doi.org/10.1103/PhysRevA.92.062119} {\bibfield  {journal} {\bibinfo  {journal} {Physical Review A}\ }\textbf {\bibinfo {volume} {92}},\ \bibinfo {pages} {062119} (\bibinfo {year} {2015})}\BibitemShut {NoStop}%
\bibitem [{\citenamefont {Lienhard}\ \emph {et~al.}(2022)\citenamefont {Lienhard}, \citenamefont {Vepsäläinen}, \citenamefont {Govia}, \citenamefont {Hoffer}, \citenamefont {Qiu}, \citenamefont {Ristè}, \citenamefont {Ware}, \citenamefont {Kim}, \citenamefont {Winik}, \citenamefont {Melville}, \citenamefont {Niedzielski}, \citenamefont {Yoder}, \citenamefont {Ribeill}, \citenamefont {Ohki}, \citenamefont {Krovi}, \citenamefont {Orlando}, \citenamefont {Gustavsson},\ and\ \citenamefont {Oliver}}]{lienhard_deep-neural-network_2022}%
  \BibitemOpen
  \bibfield  {author} {\bibinfo {author} {\bibfnamefont {B.}~\bibnamefont {Lienhard}}, \bibinfo {author} {\bibfnamefont {A.}~\bibnamefont {Vepsäläinen}}, \bibinfo {author} {\bibfnamefont {L.~C.}\ \bibnamefont {Govia}}, \bibinfo {author} {\bibfnamefont {C.~R.}\ \bibnamefont {Hoffer}}, \bibinfo {author} {\bibfnamefont {J.~Y.}\ \bibnamefont {Qiu}}, \bibinfo {author} {\bibfnamefont {D.}~\bibnamefont {Ristè}}, \bibinfo {author} {\bibfnamefont {M.}~\bibnamefont {Ware}}, \bibinfo {author} {\bibfnamefont {D.}~\bibnamefont {Kim}}, \bibinfo {author} {\bibfnamefont {R.}~\bibnamefont {Winik}}, \bibinfo {author} {\bibfnamefont {A.}~\bibnamefont {Melville}}, \bibinfo {author} {\bibfnamefont {B.}~\bibnamefont {Niedzielski}}, \bibinfo {author} {\bibfnamefont {J.}~\bibnamefont {Yoder}}, \bibinfo {author} {\bibfnamefont {G.~J.}\ \bibnamefont {Ribeill}}, \bibinfo {author} {\bibfnamefont {T.~A.}\ \bibnamefont {Ohki}}, \bibinfo {author} {\bibfnamefont {H.~K.}\ \bibnamefont {Krovi}}, \bibinfo {author} {\bibfnamefont {T.~P.}\
  \bibnamefont {Orlando}}, \bibinfo {author} {\bibfnamefont {S.}~\bibnamefont {Gustavsson}},\ and\ \bibinfo {author} {\bibfnamefont {W.~D.}\ \bibnamefont {Oliver}},\ }\bibfield  {title} {\bibinfo {title} {Deep-{Neural}-{Network} {Discrimination} of {Multiplexed} {Superconducting}-{Qubit} {States}},\ }\href {https://doi.org/10.1103/PhysRevApplied.17.014024} {\bibfield  {journal} {\bibinfo  {journal} {Physical Review Applied}\ }\textbf {\bibinfo {volume} {17}},\ \bibinfo {pages} {014024} (\bibinfo {year} {2022})},\ \bibinfo {note} {publisher: American Physical Society}\BibitemShut {NoStop}%
\bibitem [{\citenamefont {Gambetta}\ \emph {et~al.}(2007)\citenamefont {Gambetta}, \citenamefont {Braff}, \citenamefont {Wallraff}, \citenamefont {Girvin},\ and\ \citenamefont {Schoelkopf}}]{gambetta_protocols_2007}%
  \BibitemOpen
  \bibfield  {author} {\bibinfo {author} {\bibfnamefont {J.}~\bibnamefont {Gambetta}}, \bibinfo {author} {\bibfnamefont {W.~A.}\ \bibnamefont {Braff}}, \bibinfo {author} {\bibfnamefont {A.}~\bibnamefont {Wallraff}}, \bibinfo {author} {\bibfnamefont {S.~M.}\ \bibnamefont {Girvin}},\ and\ \bibinfo {author} {\bibfnamefont {R.~J.}\ \bibnamefont {Schoelkopf}},\ }\bibfield  {title} {\bibinfo {title} {Protocols for optimal readout of qubits using a continuous quantum nondemolition measurement},\ }\href {https://doi.org/10.1103/PhysRevA.76.012325} {\bibfield  {journal} {\bibinfo  {journal} {Physical Review A}\ }\textbf {\bibinfo {volume} {76}},\ \bibinfo {pages} {012325} (\bibinfo {year} {2007})},\ \bibinfo {note} {publisher: American Physical Society}\BibitemShut {NoStop}%
\bibitem [{Note1()}]{Note1}%
  \BibitemOpen
  \bibinfo {note} {Source code available at \protect \url {https://zenodo.org/doi/10.5281/zenodo.10020462}}\BibitemShut {NoStop}%
\bibitem [{\citenamefont {Takmakov}\ \emph {et~al.}(2021)\citenamefont {Takmakov}, \citenamefont {Winkel}, \citenamefont {Foroughi}, \citenamefont {Planat}, \citenamefont {Gusenkova}, \citenamefont {Spiecker}, \citenamefont {Rieger}, \citenamefont {Grünhaupt}, \citenamefont {Ustinov}, \citenamefont {Wernsdorfer}, \citenamefont {Pop},\ and\ \citenamefont {Roch}}]{takmakov_minimizing_2021}%
  \BibitemOpen
  \bibfield  {author} {\bibinfo {author} {\bibfnamefont {I.}~\bibnamefont {Takmakov}}, \bibinfo {author} {\bibfnamefont {P.}~\bibnamefont {Winkel}}, \bibinfo {author} {\bibfnamefont {F.}~\bibnamefont {Foroughi}}, \bibinfo {author} {\bibfnamefont {L.}~\bibnamefont {Planat}}, \bibinfo {author} {\bibfnamefont {D.}~\bibnamefont {Gusenkova}}, \bibinfo {author} {\bibfnamefont {M.}~\bibnamefont {Spiecker}}, \bibinfo {author} {\bibfnamefont {D.}~\bibnamefont {Rieger}}, \bibinfo {author} {\bibfnamefont {L.}~\bibnamefont {Grünhaupt}}, \bibinfo {author} {\bibfnamefont {A.}~\bibnamefont {Ustinov}}, \bibinfo {author} {\bibfnamefont {W.}~\bibnamefont {Wernsdorfer}}, \bibinfo {author} {\bibfnamefont {I.}~\bibnamefont {Pop}},\ and\ \bibinfo {author} {\bibfnamefont {N.}~\bibnamefont {Roch}},\ }\bibfield  {title} {\bibinfo {title} {Minimizing the {Discrimination} {Time} for {Quantum} {States} of an {Artificial} {Atom}},\ }\href {https://doi.org/10.1103/PhysRevApplied.15.064029} {\bibfield  {journal} {\bibinfo  {journal}
  {Physical Review Applied}\ }\textbf {\bibinfo {volume} {15}},\ \bibinfo {pages} {064029} (\bibinfo {year} {2021})}\BibitemShut {NoStop}%
\bibitem [{\citenamefont {Sunada}\ \emph {et~al.}(2022)\citenamefont {Sunada}, \citenamefont {Kono}, \citenamefont {Ilves}, \citenamefont {Tamate}, \citenamefont {Sugiyama}, \citenamefont {Tabuchi},\ and\ \citenamefont {Nakamura}}]{sunada_fast_2022}%
  \BibitemOpen
  \bibfield  {author} {\bibinfo {author} {\bibfnamefont {Y.}~\bibnamefont {Sunada}}, \bibinfo {author} {\bibfnamefont {S.}~\bibnamefont {Kono}}, \bibinfo {author} {\bibfnamefont {J.}~\bibnamefont {Ilves}}, \bibinfo {author} {\bibfnamefont {S.}~\bibnamefont {Tamate}}, \bibinfo {author} {\bibfnamefont {T.}~\bibnamefont {Sugiyama}}, \bibinfo {author} {\bibfnamefont {Y.}~\bibnamefont {Tabuchi}},\ and\ \bibinfo {author} {\bibfnamefont {Y.}~\bibnamefont {Nakamura}},\ }\bibfield  {title} {\bibinfo {title} {Fast {Readout} and {Reset} of a {Superconducting} {Qubit} {Coupled} to a {Resonator} with an {Intrinsic} {Purcell} {Filter}},\ }\href {https://doi.org/10.1103/PhysRevApplied.17.044016} {\bibfield  {journal} {\bibinfo  {journal} {Physical Review Applied}\ }\textbf {\bibinfo {volume} {17}},\ \bibinfo {pages} {044016} (\bibinfo {year} {2022})},\ \bibinfo {note} {publisher: American Physical Society}\BibitemShut {NoStop}%
\bibitem [{\citenamefont {Bengtsson}\ \emph {et~al.}(2024)\citenamefont {Bengtsson}, \citenamefont {Opremcak}, \citenamefont {Khezri}, \citenamefont {Sank}, \citenamefont {Bourassa}, \citenamefont {Satzinger}, \citenamefont {Hong}, \citenamefont {Erickson}, \citenamefont {Lester}, \citenamefont {Miao}, \citenamefont {Korotkov}, \citenamefont {Kelly}, \citenamefont {Chen},\ and\ \citenamefont {Klimov}}]{bengtsson_model-based_2024}%
  \BibitemOpen
  \bibfield  {author} {\bibinfo {author} {\bibfnamefont {A.}~\bibnamefont {Bengtsson}}, \bibinfo {author} {\bibfnamefont {A.}~\bibnamefont {Opremcak}}, \bibinfo {author} {\bibfnamefont {M.}~\bibnamefont {Khezri}}, \bibinfo {author} {\bibfnamefont {D.}~\bibnamefont {Sank}}, \bibinfo {author} {\bibfnamefont {A.}~\bibnamefont {Bourassa}}, \bibinfo {author} {\bibfnamefont {K.~J.}\ \bibnamefont {Satzinger}}, \bibinfo {author} {\bibfnamefont {S.}~\bibnamefont {Hong}}, \bibinfo {author} {\bibfnamefont {C.}~\bibnamefont {Erickson}}, \bibinfo {author} {\bibfnamefont {B.~J.}\ \bibnamefont {Lester}}, \bibinfo {author} {\bibfnamefont {K.~C.}\ \bibnamefont {Miao}}, \bibinfo {author} {\bibfnamefont {A.~N.}\ \bibnamefont {Korotkov}}, \bibinfo {author} {\bibfnamefont {J.}~\bibnamefont {Kelly}}, \bibinfo {author} {\bibfnamefont {Z.}~\bibnamefont {Chen}},\ and\ \bibinfo {author} {\bibfnamefont {P.~V.}\ \bibnamefont {Klimov}},\ }\bibfield  {title} {\bibinfo {title} {Model-{Based} {Optimization} of {Superconducting} {Qubit}
  {Readout}},\ }\href {https://doi.org/10.1103/PhysRevLett.132.100603} {\bibfield  {journal} {\bibinfo  {journal} {Physical Review Letters}\ }\textbf {\bibinfo {volume} {132}},\ \bibinfo {pages} {100603} (\bibinfo {year} {2024})}\BibitemShut {NoStop}%
\bibitem [{\citenamefont {Sank}\ \emph {et~al.}(2024)\citenamefont {Sank}, \citenamefont {Opremcak}, \citenamefont {Bengtsson}, \citenamefont {Khezri}, \citenamefont {Chen}, \citenamefont {Naaman},\ and\ \citenamefont {Korotkov}}]{sank_system_2024}%
  \BibitemOpen
  \bibfield  {author} {\bibinfo {author} {\bibfnamefont {D.}~\bibnamefont {Sank}}, \bibinfo {author} {\bibfnamefont {A.}~\bibnamefont {Opremcak}}, \bibinfo {author} {\bibfnamefont {A.}~\bibnamefont {Bengtsson}}, \bibinfo {author} {\bibfnamefont {M.}~\bibnamefont {Khezri}}, \bibinfo {author} {\bibfnamefont {Z.}~\bibnamefont {Chen}}, \bibinfo {author} {\bibfnamefont {O.}~\bibnamefont {Naaman}},\ and\ \bibinfo {author} {\bibfnamefont {A.}~\bibnamefont {Korotkov}},\ }\bibfield  {title} {\bibinfo {title} {System {Characterization} of {Dispersive} {Readout} in {Superconducting} {Qubits}}\ }\href {https://doi.org/10.48550/arXiv.2402.00413} {10.48550/arXiv.2402.00413} (\bibinfo {year} {2024}),\ \bibinfo {note} {arXiv:2402.00413 [quant-ph]}\BibitemShut {NoStop}%
\bibitem [{\citenamefont {Tanaka}\ \emph {et~al.}(2019)\citenamefont {Tanaka}, \citenamefont {Yamane}, \citenamefont {Héroux}, \citenamefont {Nakane}, \citenamefont {Kanazawa}, \citenamefont {Takeda}, \citenamefont {Numata}, \citenamefont {Nakano},\ and\ \citenamefont {Hirose}}]{tanaka_recent_2019}%
  \BibitemOpen
  \bibfield  {author} {\bibinfo {author} {\bibfnamefont {G.}~\bibnamefont {Tanaka}}, \bibinfo {author} {\bibfnamefont {T.}~\bibnamefont {Yamane}}, \bibinfo {author} {\bibfnamefont {J.~B.}\ \bibnamefont {Héroux}}, \bibinfo {author} {\bibfnamefont {R.}~\bibnamefont {Nakane}}, \bibinfo {author} {\bibfnamefont {N.}~\bibnamefont {Kanazawa}}, \bibinfo {author} {\bibfnamefont {S.}~\bibnamefont {Takeda}}, \bibinfo {author} {\bibfnamefont {H.}~\bibnamefont {Numata}}, \bibinfo {author} {\bibfnamefont {D.}~\bibnamefont {Nakano}},\ and\ \bibinfo {author} {\bibfnamefont {A.}~\bibnamefont {Hirose}},\ }\bibfield  {title} {\bibinfo {title} {Recent advances in physical reservoir computing: {A} review},\ }\href {https://doi.org/10.1016/j.neunet.2019.03.005} {\bibfield  {journal} {\bibinfo  {journal} {Neural Networks}\ }\textbf {\bibinfo {volume} {115}},\ \bibinfo {pages} {100} (\bibinfo {year} {2019})}\BibitemShut {NoStop}%
\bibitem [{\citenamefont {Gauthier}\ \emph {et~al.}(2021)\citenamefont {Gauthier}, \citenamefont {Bollt}, \citenamefont {Griffith},\ and\ \citenamefont {Barbosa}}]{gauthier_next_2021}%
  \BibitemOpen
  \bibfield  {author} {\bibinfo {author} {\bibfnamefont {D.~J.}\ \bibnamefont {Gauthier}}, \bibinfo {author} {\bibfnamefont {E.}~\bibnamefont {Bollt}}, \bibinfo {author} {\bibfnamefont {A.}~\bibnamefont {Griffith}},\ and\ \bibinfo {author} {\bibfnamefont {W.~A.~S.}\ \bibnamefont {Barbosa}},\ }\bibfield  {title} {{\selectlanguage {English}\bibinfo {title} {Next generation reservoir computing}},\ }\href {https://doi.org/10.1038/s41467-021-25801-2} {\bibfield  {journal} {\bibinfo  {journal} {Nature Communications}\ }\textbf {\bibinfo {volume} {12}},\ \bibinfo {pages} {5564} (\bibinfo {year} {2021})},\ \bibinfo {note} {number: 1 Publisher: Nature Publishing Group}\BibitemShut {NoStop}%
\bibitem [{\citenamefont {Luchi}\ \emph {et~al.}(2023)\citenamefont {Luchi}, \citenamefont {Trevisanutto}, \citenamefont {Roggero}, \citenamefont {DuBois}, \citenamefont {Rosen}, \citenamefont {Turro}, \citenamefont {Amitrano},\ and\ \citenamefont {Pederiva}}]{luchi_enhancing_2023}%
  \BibitemOpen
  \bibfield  {author} {\bibinfo {author} {\bibfnamefont {P.}~\bibnamefont {Luchi}}, \bibinfo {author} {\bibfnamefont {P.~E.}\ \bibnamefont {Trevisanutto}}, \bibinfo {author} {\bibfnamefont {A.}~\bibnamefont {Roggero}}, \bibinfo {author} {\bibfnamefont {J.~L.}\ \bibnamefont {DuBois}}, \bibinfo {author} {\bibfnamefont {Y.~J.}\ \bibnamefont {Rosen}}, \bibinfo {author} {\bibfnamefont {F.}~\bibnamefont {Turro}}, \bibinfo {author} {\bibfnamefont {V.}~\bibnamefont {Amitrano}},\ and\ \bibinfo {author} {\bibfnamefont {F.}~\bibnamefont {Pederiva}},\ }\bibfield  {title} {\bibinfo {title} {Enhancing {Qubit} {Readout} with {Autoencoders}},\ }\href {https://doi.org/10.1103/PhysRevApplied.20.014045} {\bibfield  {journal} {\bibinfo  {journal} {Physical Review Applied}\ }\textbf {\bibinfo {volume} {20}},\ \bibinfo {pages} {014045} (\bibinfo {year} {2023})}\BibitemShut {NoStop}%
\bibitem [{\citenamefont {Maurya}\ \emph {et~al.}(2023)\citenamefont {Maurya}, \citenamefont {Mude}, \citenamefont {Oliver}, \citenamefont {Lienhard},\ and\ \citenamefont {Tannu}}]{maurya_scaling_2023}%
  \BibitemOpen
  \bibfield  {author} {\bibinfo {author} {\bibfnamefont {S.}~\bibnamefont {Maurya}}, \bibinfo {author} {\bibfnamefont {C.~N.}\ \bibnamefont {Mude}}, \bibinfo {author} {\bibfnamefont {W.~D.}\ \bibnamefont {Oliver}}, \bibinfo {author} {\bibfnamefont {B.}~\bibnamefont {Lienhard}},\ and\ \bibinfo {author} {\bibfnamefont {S.}~\bibnamefont {Tannu}},\ }\bibfield  {title} {\bibinfo {title} {Scaling {Qubit} {Readout} with {Hardware} {Efficient} {Machine} {Learning} {Architectures}},\ }in\ \href {https://doi.org/10.1145/3579371.3589042} {\emph {\bibinfo {booktitle} {Proceedings of the 50th {Annual} {International} {Symposium} on {Computer} {Architecture}}}},\ \bibinfo {series and number} {{ISCA} '23}\ (\bibinfo  {publisher} {Association for Computing Machinery},\ \bibinfo {address} {New York, NY, USA},\ \bibinfo {year} {2023})\ pp.\ \bibinfo {pages} {1--13}\BibitemShut {NoStop}%
\bibitem [{\citenamefont {Canaday}\ \emph {et~al.}(2018)\citenamefont {Canaday}, \citenamefont {Griffith},\ and\ \citenamefont {Gauthier}}]{canaday_rapid_2018}%
  \BibitemOpen
  \bibfield  {author} {\bibinfo {author} {\bibfnamefont {D.}~\bibnamefont {Canaday}}, \bibinfo {author} {\bibfnamefont {A.}~\bibnamefont {Griffith}},\ and\ \bibinfo {author} {\bibfnamefont {D.~J.}\ \bibnamefont {Gauthier}},\ }\bibfield  {title} {\bibinfo {title} {Rapid time series prediction with a hardware-based reservoir computer},\ }\href {https://doi.org/10.1063/1.5048199} {\bibfield  {journal} {\bibinfo  {journal} {Chaos: An Interdisciplinary Journal of Nonlinear Science}\ }\textbf {\bibinfo {volume} {28}},\ \bibinfo {pages} {123119} (\bibinfo {year} {2018})},\ \bibinfo {note} {publisher: American Institute of Physics}\BibitemShut {NoStop}%
\bibitem [{\citenamefont {Pathak}\ \emph {et~al.}(2017)\citenamefont {Pathak}, \citenamefont {Lu}, \citenamefont {Hunt}, \citenamefont {Girvan},\ and\ \citenamefont {Ott}}]{pathak_using_2017}%
  \BibitemOpen
  \bibfield  {author} {\bibinfo {author} {\bibfnamefont {J.}~\bibnamefont {Pathak}}, \bibinfo {author} {\bibfnamefont {Z.}~\bibnamefont {Lu}}, \bibinfo {author} {\bibfnamefont {B.~R.}\ \bibnamefont {Hunt}}, \bibinfo {author} {\bibfnamefont {M.}~\bibnamefont {Girvan}},\ and\ \bibinfo {author} {\bibfnamefont {E.}~\bibnamefont {Ott}},\ }\bibfield  {title} {\bibinfo {title} {Using machine learning to replicate chaotic attractors and calculate {Lyapunov} exponents from data},\ }\href {https://doi.org/10.1063/1.5010300} {\bibfield  {journal} {\bibinfo  {journal} {Chaos: An Interdisciplinary Journal of Nonlinear Science}\ }\textbf {\bibinfo {volume} {27}},\ \bibinfo {pages} {121102} (\bibinfo {year} {2017})}\BibitemShut {NoStop}%
\bibitem [{\citenamefont {Pathak}\ \emph {et~al.}(2018)\citenamefont {Pathak}, \citenamefont {Hunt}, \citenamefont {Girvan}, \citenamefont {Lu},\ and\ \citenamefont {Ott}}]{pathak_model-free_2018}%
  \BibitemOpen
  \bibfield  {author} {\bibinfo {author} {\bibfnamefont {J.}~\bibnamefont {Pathak}}, \bibinfo {author} {\bibfnamefont {B.}~\bibnamefont {Hunt}}, \bibinfo {author} {\bibfnamefont {M.}~\bibnamefont {Girvan}}, \bibinfo {author} {\bibfnamefont {Z.}~\bibnamefont {Lu}},\ and\ \bibinfo {author} {\bibfnamefont {E.}~\bibnamefont {Ott}},\ }\bibfield  {title} {\bibinfo {title} {Model-{Free} {Prediction} of {Large} {Spatiotemporally} {Chaotic} {Systems} from {Data}: {A} {Reservoir} {Computing} {Approach}},\ }\href {https://doi.org/10.1103/PhysRevLett.120.024102} {\bibfield  {journal} {\bibinfo  {journal} {Physical Review Letters}\ }\textbf {\bibinfo {volume} {120}},\ \bibinfo {pages} {024102} (\bibinfo {year} {2018})}\BibitemShut {NoStop}%
\bibitem [{\citenamefont {Griffith}\ \emph {et~al.}(2019)\citenamefont {Griffith}, \citenamefont {Pomerance},\ and\ \citenamefont {Gauthier}}]{griffith_forecasting_2019}%
  \BibitemOpen
  \bibfield  {author} {\bibinfo {author} {\bibfnamefont {A.}~\bibnamefont {Griffith}}, \bibinfo {author} {\bibfnamefont {A.}~\bibnamefont {Pomerance}},\ and\ \bibinfo {author} {\bibfnamefont {D.~J.}\ \bibnamefont {Gauthier}},\ }\bibfield  {title} {\bibinfo {title} {Forecasting chaotic systems with very low connectivity reservoir computers},\ }\href {https://doi.org/10.1063/1.5120710} {\bibfield  {journal} {\bibinfo  {journal} {Chaos: An Interdisciplinary Journal of Nonlinear Science}\ }\textbf {\bibinfo {volume} {29}},\ \bibinfo {pages} {123108} (\bibinfo {year} {2019})}\BibitemShut {NoStop}%
\bibitem [{\citenamefont {Canaday}\ \emph {et~al.}(2021)\citenamefont {Canaday}, \citenamefont {Pomerance},\ and\ \citenamefont {Gauthier}}]{canaday_model-free_2021}%
  \BibitemOpen
  \bibfield  {author} {\bibinfo {author} {\bibfnamefont {D.}~\bibnamefont {Canaday}}, \bibinfo {author} {\bibfnamefont {A.}~\bibnamefont {Pomerance}},\ and\ \bibinfo {author} {\bibfnamefont {D.~J.}\ \bibnamefont {Gauthier}},\ }\bibfield  {title} {{\selectlanguage {English}\bibinfo {title} {Model-free control of dynamical systems with deep reservoir computing}},\ }\href {https://doi.org/10.1088/2632-072X/ac24f3} {\bibfield  {journal} {\bibinfo  {journal} {Journal of Physics: Complexity}\ }\textbf {\bibinfo {volume} {2}},\ \bibinfo {pages} {035025} (\bibinfo {year} {2021})},\ \bibinfo {note} {publisher: IOP Publishing}\BibitemShut {NoStop}%
\bibitem [{\citenamefont {Sheldon}\ \emph {et~al.}(2016)\citenamefont {Sheldon}, \citenamefont {Magesan}, \citenamefont {Chow},\ and\ \citenamefont {Gambetta}}]{IBM_tuneup_1}%
  \BibitemOpen
  \bibfield  {author} {\bibinfo {author} {\bibfnamefont {S.}~\bibnamefont {Sheldon}}, \bibinfo {author} {\bibfnamefont {E.}~\bibnamefont {Magesan}}, \bibinfo {author} {\bibfnamefont {J.~M.}\ \bibnamefont {Chow}},\ and\ \bibinfo {author} {\bibfnamefont {J.~M.}\ \bibnamefont {Gambetta}},\ }\bibfield  {title} {\bibinfo {title} {Procedure for systematically tuning up cross-talk in the cross-resonance gate},\ }\href {https://doi.org/10.1103/PHYSREVA.93.060302/FIGURES/6/MEDIUM} {\bibfield  {journal} {\bibinfo  {journal} {Physical Review A}\ }\textbf {\bibinfo {volume} {93}},\ \bibinfo {pages} {060302} (\bibinfo {year} {2016})}\BibitemShut {NoStop}%
\bibitem [{\citenamefont {Kelly}\ \emph {et~al.}(2018)\citenamefont {Kelly}, \citenamefont {Malley}, \citenamefont {Neeley}, \citenamefont {Neven},\ and\ \citenamefont {Google}}]{Kelly2018}%
  \BibitemOpen
  \bibfield  {author} {\bibinfo {author} {\bibfnamefont {J.}~\bibnamefont {Kelly}}, \bibinfo {author} {\bibfnamefont {P.~O.}\ \bibnamefont {Malley}}, \bibinfo {author} {\bibfnamefont {M.}~\bibnamefont {Neeley}}, \bibinfo {author} {\bibfnamefont {H.}~\bibnamefont {Neven}},\ and\ \bibinfo {author} {\bibfnamefont {J.~M.~M.}\ \bibnamefont {Google}},\ }\bibfield  {title} {\bibinfo {title} {Physical qubit calibration on a directed acyclic graph},\ }\href {https://arxiv.org/abs/1803.03226v1} {\bibfield  {journal} {\bibinfo  {journal} {arXiv:1803.03226 [quant-ph]}\ } (\bibinfo {year} {2018})}\BibitemShut {NoStop}%
\bibitem [{\citenamefont {Dai}\ \emph {et~al.}(2021)\citenamefont {Dai}, \citenamefont {Tennant}, \citenamefont {Trappen}, \citenamefont {Martinez}, \citenamefont {Melanson}, \citenamefont {Yurtalan}, \citenamefont {Tang}, \citenamefont {Novikov}, \citenamefont {Grover}, \citenamefont {Disseler}, \citenamefont {Basham}, \citenamefont {Das}, \citenamefont {Kim}, \citenamefont {Melville}, \citenamefont {Niedzielski}, \citenamefont {Weber}, \citenamefont {Yoder}, \citenamefont {Lidar},\ and\ \citenamefont {Lupascu}}]{Dai2021}%
  \BibitemOpen
  \bibfield  {author} {\bibinfo {author} {\bibfnamefont {X.}~\bibnamefont {Dai}}, \bibinfo {author} {\bibfnamefont {D.~M.}\ \bibnamefont {Tennant}}, \bibinfo {author} {\bibfnamefont {R.}~\bibnamefont {Trappen}}, \bibinfo {author} {\bibfnamefont {A.~J.}\ \bibnamefont {Martinez}}, \bibinfo {author} {\bibfnamefont {D.}~\bibnamefont {Melanson}}, \bibinfo {author} {\bibfnamefont {M.~A.}\ \bibnamefont {Yurtalan}}, \bibinfo {author} {\bibfnamefont {Y.}~\bibnamefont {Tang}}, \bibinfo {author} {\bibfnamefont {S.}~\bibnamefont {Novikov}}, \bibinfo {author} {\bibfnamefont {J.~A.}\ \bibnamefont {Grover}}, \bibinfo {author} {\bibfnamefont {S.~M.}\ \bibnamefont {Disseler}}, \bibinfo {author} {\bibfnamefont {J.~I.}\ \bibnamefont {Basham}}, \bibinfo {author} {\bibfnamefont {R.}~\bibnamefont {Das}}, \bibinfo {author} {\bibfnamefont {D.~K.}\ \bibnamefont {Kim}}, \bibinfo {author} {\bibfnamefont {A.~J.}\ \bibnamefont {Melville}}, \bibinfo {author} {\bibfnamefont {B.~M.}\ \bibnamefont {Niedzielski}}, \bibinfo {author}
  {\bibfnamefont {S.~J.}\ \bibnamefont {Weber}}, \bibinfo {author} {\bibfnamefont {J.~L.}\ \bibnamefont {Yoder}}, \bibinfo {author} {\bibfnamefont {D.~A.}\ \bibnamefont {Lidar}},\ and\ \bibinfo {author} {\bibfnamefont {A.}~\bibnamefont {Lupascu}},\ }\bibfield  {title} {\bibinfo {title} {Calibration of flux crosstalk in large-scale flux-tunable superconducting quantum circuits},\ }\href {https://doi.org/10.1103/PRXQuantum.2.040313} {\bibfield  {journal} {\bibinfo  {journal} {PRX Quantum}\ }\textbf {\bibinfo {volume} {2}},\ \bibinfo {pages} {040313} (\bibinfo {year} {2021})}\BibitemShut {NoStop}%
\bibitem [{\citenamefont {Hu}\ \emph {et~al.}(2023)\citenamefont {Hu}, \citenamefont {Angelatos}, \citenamefont {Khan}, \citenamefont {Vives}, \citenamefont {Türeci}, \citenamefont {Bello}, \citenamefont {Rowlands}, \citenamefont {Ribeill},\ and\ \citenamefont {Türeci}}]{hu_tackling_2023}%
  \BibitemOpen
  \bibfield  {author} {\bibinfo {author} {\bibfnamefont {F.}~\bibnamefont {Hu}}, \bibinfo {author} {\bibfnamefont {G.}~\bibnamefont {Angelatos}}, \bibinfo {author} {\bibfnamefont {S.~A.}\ \bibnamefont {Khan}}, \bibinfo {author} {\bibfnamefont {M.}~\bibnamefont {Vives}}, \bibinfo {author} {\bibfnamefont {E.}~\bibnamefont {Türeci}}, \bibinfo {author} {\bibfnamefont {L.}~\bibnamefont {Bello}}, \bibinfo {author} {\bibfnamefont {G.~E.}\ \bibnamefont {Rowlands}}, \bibinfo {author} {\bibfnamefont {G.~J.}\ \bibnamefont {Ribeill}},\ and\ \bibinfo {author} {\bibfnamefont {H.~E.}\ \bibnamefont {Türeci}},\ }\bibfield  {title} {\bibinfo {title} {Tackling {Sampling} {Noise} in {Physical} {Systems} for {Machine} {Learning} {Applications}: {Fundamental} {Limits} and {Eigentasks}},\ }\href {https://doi.org/10.1103/PhysRevX.13.041020} {\bibfield  {journal} {\bibinfo  {journal} {Physical Review X}\ }\textbf {\bibinfo {volume} {13}},\ \bibinfo {pages} {041020} (\bibinfo {year} {2023})}\BibitemShut {NoStop}%
\bibitem [{\citenamefont {Roscher}\ \emph {et~al.}(2020)\citenamefont {Roscher}, \citenamefont {Bohn}, \citenamefont {Duarte},\ and\ \citenamefont {Garcke}}]{roscher_explainable_2020}%
  \BibitemOpen
  \bibfield  {author} {\bibinfo {author} {\bibfnamefont {R.}~\bibnamefont {Roscher}}, \bibinfo {author} {\bibfnamefont {B.}~\bibnamefont {Bohn}}, \bibinfo {author} {\bibfnamefont {M.~F.}\ \bibnamefont {Duarte}},\ and\ \bibinfo {author} {\bibfnamefont {J.}~\bibnamefont {Garcke}},\ }\bibfield  {title} {\bibinfo {title} {Explainable {Machine} {Learning} for {Scientific} {Insights} and {Discoveries}},\ }\href {https://doi.org/10.1109/ACCESS.2020.2976199} {\bibfield  {journal} {\bibinfo  {journal} {IEEE Access}\ }\textbf {\bibinfo {volume} {8}},\ \bibinfo {pages} {42200} (\bibinfo {year} {2020})},\ \bibinfo {note} {conference Name: IEEE Access}\BibitemShut {NoStop}%
\bibitem [{\citenamefont {Zhu}\ \emph {et~al.}(2013)\citenamefont {Zhu}, \citenamefont {Ferguson}, \citenamefont {Manucharyan},\ and\ \citenamefont {Koch}}]{zhu_circuit_2013}%
  \BibitemOpen
  \bibfield  {author} {\bibinfo {author} {\bibfnamefont {G.}~\bibnamefont {Zhu}}, \bibinfo {author} {\bibfnamefont {D.~G.}\ \bibnamefont {Ferguson}}, \bibinfo {author} {\bibfnamefont {V.~E.}\ \bibnamefont {Manucharyan}},\ and\ \bibinfo {author} {\bibfnamefont {J.}~\bibnamefont {Koch}},\ }\bibfield  {title} {\bibinfo {title} {Circuit {QED} with fluxonium qubits: {Theory} of the dispersive regime},\ }\href {https://doi.org/10.1103/PhysRevB.87.024510} {\bibfield  {journal} {\bibinfo  {journal} {Physical Review B}\ }\textbf {\bibinfo {volume} {87}},\ \bibinfo {pages} {024510} (\bibinfo {year} {2013})}\BibitemShut {NoStop}%
\bibitem [{\citenamefont {Blais}\ \emph {et~al.}(2021)\citenamefont {Blais}, \citenamefont {Grimsmo}, \citenamefont {Girvin},\ and\ \citenamefont {Wallraff}}]{blais_circuit_2021}%
  \BibitemOpen
  \bibfield  {author} {\bibinfo {author} {\bibfnamefont {A.}~\bibnamefont {Blais}}, \bibinfo {author} {\bibfnamefont {A.~L.}\ \bibnamefont {Grimsmo}}, \bibinfo {author} {\bibfnamefont {S.}~\bibnamefont {Girvin}},\ and\ \bibinfo {author} {\bibfnamefont {A.}~\bibnamefont {Wallraff}},\ }\bibfield  {title} {\bibinfo {title} {Circuit quantum electrodynamics},\ }\href {https://doi.org/10.1103/RevModPhys.93.025005} {\bibfield  {journal} {\bibinfo  {journal} {Reviews of Modern Physics}\ }\textbf {\bibinfo {volume} {93}},\ \bibinfo {pages} {025005} (\bibinfo {year} {2021})}\BibitemShut {NoStop}%
\bibitem [{\citenamefont {Mallet}\ \emph {et~al.}(2009)\citenamefont {Mallet}, \citenamefont {Ong}, \citenamefont {Palacios-Laloy}, \citenamefont {Nguyen}, \citenamefont {Bertet}, \citenamefont {Vion},\ and\ \citenamefont {Esteve}}]{mallet_single-shot_2009}%
  \BibitemOpen
  \bibfield  {author} {\bibinfo {author} {\bibfnamefont {F.}~\bibnamefont {Mallet}}, \bibinfo {author} {\bibfnamefont {F.~R.}\ \bibnamefont {Ong}}, \bibinfo {author} {\bibfnamefont {A.}~\bibnamefont {Palacios-Laloy}}, \bibinfo {author} {\bibfnamefont {F.}~\bibnamefont {Nguyen}}, \bibinfo {author} {\bibfnamefont {P.}~\bibnamefont {Bertet}}, \bibinfo {author} {\bibfnamefont {D.}~\bibnamefont {Vion}},\ and\ \bibinfo {author} {\bibfnamefont {D.}~\bibnamefont {Esteve}},\ }\bibfield  {title} {{\selectlanguage {English}\bibinfo {title} {Single-shot qubit readout in circuit quantum electrodynamics}},\ }\href {https://doi.org/10.1038/nphys1400} {\bibfield  {journal} {\bibinfo  {journal} {Nature Physics}\ }\textbf {\bibinfo {volume} {5}},\ \bibinfo {pages} {791} (\bibinfo {year} {2009})},\ \bibinfo {note} {number: 11 Publisher: Nature Publishing Group}\BibitemShut {NoStop}%
\bibitem [{\citenamefont {Hastie}\ \emph {et~al.}(2016)\citenamefont {Hastie}, \citenamefont {Tibshirani},\ and\ \citenamefont {Friedman}}]{hastie_elements_2016}%
  \BibitemOpen
  \bibfield  {author} {\bibinfo {author} {\bibfnamefont {T.}~\bibnamefont {Hastie}}, \bibinfo {author} {\bibfnamefont {R.}~\bibnamefont {Tibshirani}},\ and\ \bibinfo {author} {\bibfnamefont {J.}~\bibnamefont {Friedman}},\ }\href@noop {} {{\selectlanguage {English}\emph {\bibinfo {title} {The {Elements} of {Statistical} {Learning}: {Data} {Mining}, {Inference}, and {Prediction}, {Second} {Edition}}}}},\ \bibinfo {edition} {2nd}\ ed.\ (\bibinfo  {publisher} {Springer},\ \bibinfo {address} {New York, NY},\ \bibinfo {year} {2016})\BibitemShut {NoStop}%
\bibitem [{\citenamefont {Turin}(1960)}]{turin_introduction_1960}%
  \BibitemOpen
  \bibfield  {author} {\bibinfo {author} {\bibfnamefont {G.}~\bibnamefont {Turin}},\ }\bibfield  {title} {\bibinfo {title} {An introduction to matched filters},\ }\href {https://doi.org/10.1109/TIT.1960.1057571} {\bibfield  {journal} {\bibinfo  {journal} {IRE Transactions on Information Theory}\ }\textbf {\bibinfo {volume} {6}},\ \bibinfo {pages} {311} (\bibinfo {year} {1960})},\ \bibinfo {note} {conference Name: IRE Transactions on Information Theory}\BibitemShut {NoStop}%
\bibitem [{\citenamefont {Silveri}\ \emph {et~al.}(2016)\citenamefont {Silveri}, \citenamefont {Zalys-Geller}, \citenamefont {Hatridge}, \citenamefont {Leghtas}, \citenamefont {Devoret},\ and\ \citenamefont {Girvin}}]{Silveri2016}%
  \BibitemOpen
  \bibfield  {author} {\bibinfo {author} {\bibfnamefont {M.}~\bibnamefont {Silveri}}, \bibinfo {author} {\bibfnamefont {E.}~\bibnamefont {Zalys-Geller}}, \bibinfo {author} {\bibfnamefont {M.}~\bibnamefont {Hatridge}}, \bibinfo {author} {\bibfnamefont {Z.}~\bibnamefont {Leghtas}}, \bibinfo {author} {\bibfnamefont {M.~H.}\ \bibnamefont {Devoret}},\ and\ \bibinfo {author} {\bibfnamefont {S.~M.}\ \bibnamefont {Girvin}},\ }\bibfield  {title} {\bibinfo {title} {Theory of remote entanglement via quantum-limited phase-preserving amplification},\ }\href {https://doi.org/10.1103/PHYSREVA.93.062310/FIGURES/5/MEDIUM} {\bibfield  {journal} {\bibinfo  {journal} {Physical Review A}\ }\textbf {\bibinfo {volume} {93}},\ \bibinfo {pages} {062310} (\bibinfo {year} {2016})}\BibitemShut {NoStop}%
\bibitem [{\citenamefont {Kurpiers}\ \emph {et~al.}(2018)\citenamefont {Kurpiers}, \citenamefont {Magnard}, \citenamefont {Walter}, \citenamefont {Royer}, \citenamefont {Pechal}, \citenamefont {Heinsoo}, \citenamefont {Salathé}, \citenamefont {Akin}, \citenamefont {Storz}, \citenamefont {Besse}, \citenamefont {Gasparinetti}, \citenamefont {Blais},\ and\ \citenamefont {Wallraff}}]{Kurpiers2018}%
  \BibitemOpen
  \bibfield  {author} {\bibinfo {author} {\bibfnamefont {P.}~\bibnamefont {Kurpiers}}, \bibinfo {author} {\bibfnamefont {P.}~\bibnamefont {Magnard}}, \bibinfo {author} {\bibfnamefont {T.}~\bibnamefont {Walter}}, \bibinfo {author} {\bibfnamefont {B.}~\bibnamefont {Royer}}, \bibinfo {author} {\bibfnamefont {M.}~\bibnamefont {Pechal}}, \bibinfo {author} {\bibfnamefont {J.}~\bibnamefont {Heinsoo}}, \bibinfo {author} {\bibfnamefont {Y.}~\bibnamefont {Salathé}}, \bibinfo {author} {\bibfnamefont {A.}~\bibnamefont {Akin}}, \bibinfo {author} {\bibfnamefont {S.}~\bibnamefont {Storz}}, \bibinfo {author} {\bibfnamefont {J.~C.}\ \bibnamefont {Besse}}, \bibinfo {author} {\bibfnamefont {S.}~\bibnamefont {Gasparinetti}}, \bibinfo {author} {\bibfnamefont {A.}~\bibnamefont {Blais}},\ and\ \bibinfo {author} {\bibfnamefont {A.}~\bibnamefont {Wallraff}},\ }\bibfield  {title} {\bibinfo {title} {Deterministic quantum state transfer and remote entanglement using microwave photons},\ }\href
  {https://doi.org/10.1038/s41586-018-0195-y} {\bibfield  {journal} {\bibinfo  {journal} {Nature 2018 558:7709}\ }\textbf {\bibinfo {volume} {558}},\ \bibinfo {pages} {264} (\bibinfo {year} {2018})}\BibitemShut {NoStop}%
\bibitem [{\citenamefont {Kochetov}\ and\ \citenamefont {Fedorov}(2015)}]{kochetov_higher-order_2015}%
  \BibitemOpen
  \bibfield  {author} {\bibinfo {author} {\bibfnamefont {B.~A.}\ \bibnamefont {Kochetov}}\ and\ \bibinfo {author} {\bibfnamefont {A.}~\bibnamefont {Fedorov}},\ }\bibfield  {title} {\bibinfo {title} {Higher-order nonlinear effects in a {Josephson} parametric amplifier},\ }\href {https://doi.org/10.1103/PhysRevB.92.224304} {\bibfield  {journal} {\bibinfo  {journal} {Physical Review B}\ }\textbf {\bibinfo {volume} {92}},\ \bibinfo {pages} {224304} (\bibinfo {year} {2015})}\BibitemShut {NoStop}%
\bibitem [{\citenamefont {Boutin}\ \emph {et~al.}(2017)\citenamefont {Boutin}, \citenamefont {Toyli}, \citenamefont {Venkatramani}, \citenamefont {Eddins}, \citenamefont {Siddiqi},\ and\ \citenamefont {Blais}}]{boutin_effect_2017}%
  \BibitemOpen
  \bibfield  {author} {\bibinfo {author} {\bibfnamefont {S.}~\bibnamefont {Boutin}}, \bibinfo {author} {\bibfnamefont {D.~M.}\ \bibnamefont {Toyli}}, \bibinfo {author} {\bibfnamefont {A.~V.}\ \bibnamefont {Venkatramani}}, \bibinfo {author} {\bibfnamefont {A.~W.}\ \bibnamefont {Eddins}}, \bibinfo {author} {\bibfnamefont {I.}~\bibnamefont {Siddiqi}},\ and\ \bibinfo {author} {\bibfnamefont {A.}~\bibnamefont {Blais}},\ }\bibfield  {title} {\bibinfo {title} {Effect of {Higher}-{Order} {Nonlinearities} on {Amplification} and {Squeezing} in {Josephson} {Parametric} {Amplifiers}},\ }\href {https://doi.org/10.1103/PhysRevApplied.8.054030} {\bibfield  {journal} {\bibinfo  {journal} {Physical Review Applied}\ }\textbf {\bibinfo {volume} {8}},\ \bibinfo {pages} {054030} (\bibinfo {year} {2017})},\ \bibinfo {note} {publisher: American Physical Society}\BibitemShut {NoStop}%
\bibitem [{\citenamefont {Parker}\ \emph {et~al.}(2022)\citenamefont {Parker}, \citenamefont {Savytskyi}, \citenamefont {Vine}, \citenamefont {Laucht}, \citenamefont {Duty}, \citenamefont {Morello}, \citenamefont {Grimsmo},\ and\ \citenamefont {Pla}}]{Parker2022}%
  \BibitemOpen
  \bibfield  {author} {\bibinfo {author} {\bibfnamefont {D.~J.}\ \bibnamefont {Parker}}, \bibinfo {author} {\bibfnamefont {M.}~\bibnamefont {Savytskyi}}, \bibinfo {author} {\bibfnamefont {W.}~\bibnamefont {Vine}}, \bibinfo {author} {\bibfnamefont {A.}~\bibnamefont {Laucht}}, \bibinfo {author} {\bibfnamefont {T.}~\bibnamefont {Duty}}, \bibinfo {author} {\bibfnamefont {A.}~\bibnamefont {Morello}}, \bibinfo {author} {\bibfnamefont {A.~L.}\ \bibnamefont {Grimsmo}},\ and\ \bibinfo {author} {\bibfnamefont {J.~J.}\ \bibnamefont {Pla}},\ }\bibfield  {title} {\bibinfo {title} {Degenerate parametric amplification via three-wave mixing using kinetic inductance},\ }\href {https://doi.org/10.1103/PHYSREVAPPLIED.17.034064/FIGURES/23/MEDIUM} {\bibfield  {journal} {\bibinfo  {journal} {Physical Review Applied}\ }\textbf {\bibinfo {volume} {17}},\ \bibinfo {pages} {034064} (\bibinfo {year} {2022})}\BibitemShut {NoStop}%
\bibitem [{\citenamefont {Remm}\ \emph {et~al.}(2023)\citenamefont {Remm}, \citenamefont {Krinner}, \citenamefont {Lacroix}, \citenamefont {Hellings}, \citenamefont {Swiadek}, \citenamefont {Norris}, \citenamefont {Eichler},\ and\ \citenamefont {Wallraff}}]{Remm2023}%
  \BibitemOpen
  \bibfield  {author} {\bibinfo {author} {\bibfnamefont {A.}~\bibnamefont {Remm}}, \bibinfo {author} {\bibfnamefont {S.}~\bibnamefont {Krinner}}, \bibinfo {author} {\bibfnamefont {N.}~\bibnamefont {Lacroix}}, \bibinfo {author} {\bibfnamefont {C.}~\bibnamefont {Hellings}}, \bibinfo {author} {\bibfnamefont {F.}~\bibnamefont {Swiadek}}, \bibinfo {author} {\bibfnamefont {G.~J.}\ \bibnamefont {Norris}}, \bibinfo {author} {\bibfnamefont {C.}~\bibnamefont {Eichler}},\ and\ \bibinfo {author} {\bibfnamefont {A.}~\bibnamefont {Wallraff}},\ }\bibfield  {title} {\bibinfo {title} {Intermodulation distortion in a josephson traveling-wave parametric amplifier},\ }\href {https://doi.org/10.1103/PHYSREVAPPLIED.20.034027/FIGURES/12/MEDIUM} {\bibfield  {journal} {\bibinfo  {journal} {Physical Review Applied}\ }\textbf {\bibinfo {volume} {20}},\ \bibinfo {pages} {034027} (\bibinfo {year} {2023})}\BibitemShut {NoStop}%
\bibitem [{\citenamefont {Kaufman}\ \emph {et~al.}(2023)\citenamefont {Kaufman}, \citenamefont {White}, \citenamefont {Dykman}, \citenamefont {Iorio}, \citenamefont {Sterling}, \citenamefont {Hong}, \citenamefont {Opremcak}, \citenamefont {Bengtsson}, \citenamefont {Faoro}, \citenamefont {Bardin}, \citenamefont {Burger}, \citenamefont {Gasca},\ and\ \citenamefont {Naaman}}]{kaufman_josephson_2023}%
  \BibitemOpen
  \bibfield  {author} {\bibinfo {author} {\bibfnamefont {R.}~\bibnamefont {Kaufman}}, \bibinfo {author} {\bibfnamefont {T.}~\bibnamefont {White}}, \bibinfo {author} {\bibfnamefont {M.~I.}\ \bibnamefont {Dykman}}, \bibinfo {author} {\bibfnamefont {A.}~\bibnamefont {Iorio}}, \bibinfo {author} {\bibfnamefont {G.}~\bibnamefont {Sterling}}, \bibinfo {author} {\bibfnamefont {S.}~\bibnamefont {Hong}}, \bibinfo {author} {\bibfnamefont {A.}~\bibnamefont {Opremcak}}, \bibinfo {author} {\bibfnamefont {A.}~\bibnamefont {Bengtsson}}, \bibinfo {author} {\bibfnamefont {L.}~\bibnamefont {Faoro}}, \bibinfo {author} {\bibfnamefont {J.~C.}\ \bibnamefont {Bardin}}, \bibinfo {author} {\bibfnamefont {T.}~\bibnamefont {Burger}}, \bibinfo {author} {\bibfnamefont {R.}~\bibnamefont {Gasca}},\ and\ \bibinfo {author} {\bibfnamefont {O.}~\bibnamefont {Naaman}},\ }\bibfield  {title} {\bibinfo {title} {Josephson parametric amplifier with {Chebyshev} gain profile and high saturation},\ }\href
  {https://doi.org/10.1103/PhysRevApplied.20.054058} {\bibfield  {journal} {\bibinfo  {journal} {Physical Review Applied}\ }\textbf {\bibinfo {volume} {20}},\ \bibinfo {pages} {054058} (\bibinfo {year} {2023})}\BibitemShut {NoStop}%
\bibitem [{\citenamefont {Kaufman}\ \emph {et~al.}(2024)\citenamefont {Kaufman}, \citenamefont {Liu}, \citenamefont {Cicak}, \citenamefont {Mesits}, \citenamefont {Xia}, \citenamefont {Zhou}, \citenamefont {Nowicki}, \citenamefont {Pekker}, \citenamefont {Aumentado},\ and\ \citenamefont {Hatridge}}]{Kaufman2024}%
  \BibitemOpen
  \bibfield  {author} {\bibinfo {author} {\bibfnamefont {R.}~\bibnamefont {Kaufman}}, \bibinfo {author} {\bibfnamefont {C.}~\bibnamefont {Liu}}, \bibinfo {author} {\bibfnamefont {K.}~\bibnamefont {Cicak}}, \bibinfo {author} {\bibfnamefont {B.}~\bibnamefont {Mesits}}, \bibinfo {author} {\bibfnamefont {M.}~\bibnamefont {Xia}}, \bibinfo {author} {\bibfnamefont {C.}~\bibnamefont {Zhou}}, \bibinfo {author} {\bibfnamefont {M.}~\bibnamefont {Nowicki}}, \bibinfo {author} {\bibfnamefont {D.}~\bibnamefont {Pekker}}, \bibinfo {author} {\bibfnamefont {J.}~\bibnamefont {Aumentado}},\ and\ \bibinfo {author} {\bibfnamefont {M.}~\bibnamefont {Hatridge}},\ }\href@noop {} {\bibfield  {journal} {\bibinfo  {journal} {In Preparation}\ } (\bibinfo {year} {2024})}\BibitemShut {NoStop}%
\bibitem [{\citenamefont {Chen}\ \emph {et~al.}(2023)\citenamefont {Chen}, \citenamefont {Li}, \citenamefont {Lu}, \citenamefont {Warren}, \citenamefont {Križan}, \citenamefont {Kosen}, \citenamefont {Rommel}, \citenamefont {Ahmed}, \citenamefont {Osman}, \citenamefont {Biznárová}, \citenamefont {Roudsari}, \citenamefont {Lienhard}, \citenamefont {Caputo}, \citenamefont {Grigoras}, \citenamefont {Grönberg}, \citenamefont {Govenius}, \citenamefont {Kockum}, \citenamefont {Delsing}, \citenamefont {Bylander},\ and\ \citenamefont {Tancredi}}]{Chen2023ThreeStateNoAmp}%
  \BibitemOpen
  \bibfield  {author} {\bibinfo {author} {\bibfnamefont {L.}~\bibnamefont {Chen}}, \bibinfo {author} {\bibfnamefont {H.~X.}\ \bibnamefont {Li}}, \bibinfo {author} {\bibfnamefont {Y.}~\bibnamefont {Lu}}, \bibinfo {author} {\bibfnamefont {C.~W.}\ \bibnamefont {Warren}}, \bibinfo {author} {\bibfnamefont {C.~J.}\ \bibnamefont {Križan}}, \bibinfo {author} {\bibfnamefont {S.}~\bibnamefont {Kosen}}, \bibinfo {author} {\bibfnamefont {M.}~\bibnamefont {Rommel}}, \bibinfo {author} {\bibfnamefont {S.}~\bibnamefont {Ahmed}}, \bibinfo {author} {\bibfnamefont {A.}~\bibnamefont {Osman}}, \bibinfo {author} {\bibfnamefont {J.}~\bibnamefont {Biznárová}}, \bibinfo {author} {\bibfnamefont {A.~F.}\ \bibnamefont {Roudsari}}, \bibinfo {author} {\bibfnamefont {B.}~\bibnamefont {Lienhard}}, \bibinfo {author} {\bibfnamefont {M.}~\bibnamefont {Caputo}}, \bibinfo {author} {\bibfnamefont {K.}~\bibnamefont {Grigoras}}, \bibinfo {author} {\bibfnamefont {L.}~\bibnamefont {Grönberg}}, \bibinfo {author} {\bibfnamefont {J.}~\bibnamefont
  {Govenius}}, \bibinfo {author} {\bibfnamefont {A.~F.}\ \bibnamefont {Kockum}}, \bibinfo {author} {\bibfnamefont {P.}~\bibnamefont {Delsing}}, \bibinfo {author} {\bibfnamefont {J.}~\bibnamefont {Bylander}},\ and\ \bibinfo {author} {\bibfnamefont {G.}~\bibnamefont {Tancredi}},\ }\bibfield  {title} {\bibinfo {title} {Transmon qubit readout fidelity at the threshold for quantum error correction without a quantum-limited amplifier},\ }\href {https://doi.org/10.1038/s41534-023-00689-6} {\bibfield  {journal} {\bibinfo  {journal} {npj Quantum Information 2023 9:1}\ }\textbf {\bibinfo {volume} {9}},\ \bibinfo {pages} {1} (\bibinfo {year} {2023})}\BibitemShut {NoStop}%
\bibitem [{\citenamefont {Schuster}\ \emph {et~al.}(2005)\citenamefont {Schuster}, \citenamefont {Wallraff}, \citenamefont {Blais}, \citenamefont {Frunzio}, \citenamefont {Huang}, \citenamefont {Majer}, \citenamefont {Girvin},\ and\ \citenamefont {Schoelkopf}}]{Schuster2005dephasing}%
  \BibitemOpen
  \bibfield  {author} {\bibinfo {author} {\bibfnamefont {D.~I.}\ \bibnamefont {Schuster}}, \bibinfo {author} {\bibfnamefont {A.}~\bibnamefont {Wallraff}}, \bibinfo {author} {\bibfnamefont {A.}~\bibnamefont {Blais}}, \bibinfo {author} {\bibfnamefont {L.}~\bibnamefont {Frunzio}}, \bibinfo {author} {\bibfnamefont {R.~S.}\ \bibnamefont {Huang}}, \bibinfo {author} {\bibfnamefont {J.}~\bibnamefont {Majer}}, \bibinfo {author} {\bibfnamefont {S.~M.}\ \bibnamefont {Girvin}},\ and\ \bibinfo {author} {\bibfnamefont {R.~J.}\ \bibnamefont {Schoelkopf}},\ }\bibfield  {title} {\bibinfo {title} {Ac stark shift and dephasing of a superconducting qubit strongly coupled to a cavity field},\ }\href {https://doi.org/10.1103/PHYSREVLETT.94.123602/FIGURES/5/MEDIUM} {\bibfield  {journal} {\bibinfo  {journal} {Physical Review Letters}\ }\textbf {\bibinfo {volume} {94}},\ \bibinfo {pages} {123602} (\bibinfo {year} {2005})}\BibitemShut {NoStop}%
\bibitem [{\citenamefont {Cohen}\ \emph {et~al.}(2023)\citenamefont {Cohen}, \citenamefont {Petrescu}, \citenamefont {Shillito},\ and\ \citenamefont {Blais}}]{Cohen2023Chaos}%
  \BibitemOpen
  \bibfield  {author} {\bibinfo {author} {\bibfnamefont {J.}~\bibnamefont {Cohen}}, \bibinfo {author} {\bibfnamefont {A.}~\bibnamefont {Petrescu}}, \bibinfo {author} {\bibfnamefont {R.}~\bibnamefont {Shillito}},\ and\ \bibinfo {author} {\bibfnamefont {A.}~\bibnamefont {Blais}},\ }\bibfield  {title} {\bibinfo {title} {Reminiscence of classical chaos in driven transmons},\ }\href {https://doi.org/10.1103/PRXQUANTUM.4.020312/FIGURES/20/MEDIUM} {\bibfield  {journal} {\bibinfo  {journal} {PRX Quantum}\ }\textbf {\bibinfo {volume} {4}},\ \bibinfo {pages} {020312} (\bibinfo {year} {2023})}\BibitemShut {NoStop}%
\bibitem [{\citenamefont {Khezri}\ \emph {et~al.}(2016)\citenamefont {Khezri}, \citenamefont {Mlinar}, \citenamefont {Dressel},\ and\ \citenamefont {Korotkov}}]{khezri_measuring_2016}%
  \BibitemOpen
  \bibfield  {author} {\bibinfo {author} {\bibfnamefont {M.}~\bibnamefont {Khezri}}, \bibinfo {author} {\bibfnamefont {E.}~\bibnamefont {Mlinar}}, \bibinfo {author} {\bibfnamefont {J.}~\bibnamefont {Dressel}},\ and\ \bibinfo {author} {\bibfnamefont {A.~N.}\ \bibnamefont {Korotkov}},\ }\bibfield  {title} {\bibinfo {title} {Measuring a transmon qubit in circuit {QED}: {Dressed} squeezed states},\ }\href {https://doi.org/10.1103/PhysRevA.94.012347} {\bibfield  {journal} {\bibinfo  {journal} {Physical Review A}\ }\textbf {\bibinfo {volume} {94}},\ \bibinfo {pages} {012347} (\bibinfo {year} {2016})}\BibitemShut {NoStop}%
\bibitem [{\citenamefont {McClure}\ \emph {et~al.}(2016)\citenamefont {McClure}, \citenamefont {Paik}, \citenamefont {Bishop}, \citenamefont {Steffen}, \citenamefont {Chow},\ and\ \citenamefont {Gambetta}}]{McClure2016}%
  \BibitemOpen
  \bibfield  {author} {\bibinfo {author} {\bibfnamefont {D.~T.}\ \bibnamefont {McClure}}, \bibinfo {author} {\bibfnamefont {H.}~\bibnamefont {Paik}}, \bibinfo {author} {\bibfnamefont {L.~S.}\ \bibnamefont {Bishop}}, \bibinfo {author} {\bibfnamefont {M.}~\bibnamefont {Steffen}}, \bibinfo {author} {\bibfnamefont {J.~M.}\ \bibnamefont {Chow}},\ and\ \bibinfo {author} {\bibfnamefont {J.~M.}\ \bibnamefont {Gambetta}},\ }\bibfield  {title} {\bibinfo {title} {Rapid driven reset of a qubit readout resonator},\ }\href {https://doi.org/10.1103/PHYSREVAPPLIED.5.011001/FIGURES/4/MEDIUM} {\bibfield  {journal} {\bibinfo  {journal} {Physical Review Applied}\ }\textbf {\bibinfo {volume} {5}},\ \bibinfo {pages} {011001} (\bibinfo {year} {2016})}\BibitemShut {NoStop}%
\bibitem [{\citenamefont {Metelmann}\ and\ \citenamefont {Clerk}(2015)}]{metelmann_nonreciprocal_2015}%
  \BibitemOpen
  \bibfield  {author} {\bibinfo {author} {\bibfnamefont {A.}~\bibnamefont {Metelmann}}\ and\ \bibinfo {author} {\bibfnamefont {A.}~\bibnamefont {Clerk}},\ }\bibfield  {title} {\bibinfo {title} {Nonreciprocal {Photon} {Transmission} and {Amplification} via {Reservoir} {Engineering}},\ }\href {https://doi.org/10.1103/PhysRevX.5.021025} {\bibfield  {journal} {\bibinfo  {journal} {Physical Review X}\ }\textbf {\bibinfo {volume} {5}},\ \bibinfo {pages} {021025} (\bibinfo {year} {2015})}\BibitemShut {NoStop}%
\bibitem [{\citenamefont {Larger}\ \emph {et~al.}(2017)\citenamefont {Larger}, \citenamefont {Baylón-Fuentes}, \citenamefont {Martinenghi}, \citenamefont {Udaltsov}, \citenamefont {Chembo},\ and\ \citenamefont {Jacquot}}]{larger_high-speed_2017}%
  \BibitemOpen
  \bibfield  {author} {\bibinfo {author} {\bibfnamefont {L.}~\bibnamefont {Larger}}, \bibinfo {author} {\bibfnamefont {A.}~\bibnamefont {Baylón-Fuentes}}, \bibinfo {author} {\bibfnamefont {R.}~\bibnamefont {Martinenghi}}, \bibinfo {author} {\bibfnamefont {V.~S.}\ \bibnamefont {Udaltsov}}, \bibinfo {author} {\bibfnamefont {Y.~K.}\ \bibnamefont {Chembo}},\ and\ \bibinfo {author} {\bibfnamefont {M.}~\bibnamefont {Jacquot}},\ }\bibfield  {title} {\bibinfo {title} {High-{Speed} {Photonic} {Reservoir} {Computing} {Using} a {Time}-{Delay}-{Based} {Architecture}: {Million} {Words} per {Second} {Classification}},\ }\href {https://doi.org/10.1103/PhysRevX.7.011015} {\bibfield  {journal} {\bibinfo  {journal} {Physical Review X}\ }\textbf {\bibinfo {volume} {7}},\ \bibinfo {pages} {011015} (\bibinfo {year} {2017})},\ \bibinfo {note} {publisher: American Physical Society}\BibitemShut {NoStop}%
\bibitem [{\citenamefont {Deng}(2012)}]{deng_mnist_2012}%
  \BibitemOpen
  \bibfield  {author} {\bibinfo {author} {\bibfnamefont {L.}~\bibnamefont {Deng}},\ }\bibfield  {title} {\bibinfo {title} {The {MNIST} {Database} of {Handwritten} {Digit} {Images} for {Machine} {Learning} {Research} [{Best} of the {Web}]},\ }\href {https://doi.org/10.1109/MSP.2012.2211477} {\bibfield  {journal} {\bibinfo  {journal} {IEEE Signal Processing Magazine}\ }\textbf {\bibinfo {volume} {29}},\ \bibinfo {pages} {141} (\bibinfo {year} {2012})},\ \bibinfo {note} {conference Name: IEEE Signal Processing Magazine}\BibitemShut {NoStop}%
\end{thebibliography}%

\appendix

\begin{widetext}
    \startcontents[appendices]
\printcontents[appendices]{l}{1}{\section*{Appendices}\setcounter{tocdepth}{2}}
% \printcontents[appendices]{l}{1}{\setcounter{tocdepth}{2}}

\setcounter{page}{1}

\makeatletter
\let\toc@pre\relax
\let\toc@post\relax
\makeatother

\newpage

\newpage

% \section{Supplementary plots}

%%%%%%%%%%%%%%%%%%%%%%%%%%%%%%%%%%%%%%%%%%%%%%%%%%%%%%%%%%%%%%%%%%%%%%%%%%%%%%%%%%%%%%%%%%%%%%%%%%%

% \begin{figure}[t]
%     \centering
%     \includegraphics[scale=1.0]{Figures/realQubitResults-3.pdf}
%     \caption{\textbf{ML for 2-state classification of a single dispersively-coupled qubit under strong drive tones.} (a) Classification fidelity $\fid$ using ML-based (blue) and MF-based (gray) classifiers, as a function of temporal pulse length, and inversely proportional to pulse amplitude, as indicated in the inset. The ML always performs at least as well the MF-based classifier, and increasingly outperforms the MF at higher signal amplitudes (shorter pulses), as indicated by the increasing difference in achieved $\fid$ (shaded region). (b) Characterizing the difference in $\fid$ observed in (a) by quantifying the excess errors $\varepsilon$ made by the MF-based classifier in comparison to ML-based classifier (percentage) as the temporal pulse length (and hence signal amplitude) is varied. At the highest signal amplitude considered, the MF-based classified makes over 30\% more errors than the ML-based classifer. }
%     \label{fig:simQubitResults}
% \end{figure}

%%%%%%%%%%%%%%%%%%%%%%%%%%%%%%%%%%%%%%%%%%%%%%%%%%%%%%%%%%%%%%%%%%%%%%%%%%%%%%%%%%%%%%%%%%%%%%%%%%%

\section{Experimental Setup}
\label{app:expmt}
In this appendix section, we show a few more examples of readout IQ histograms as well as a more detailed circuit diagram for the measurement chain. Shown in Fig. \ref{appfig:RO_histograms} below, we see two examples of the extremes of the measurement data for readout of Qubit B used to generate Fig.~\ref{fig:exptResults}. Part (a) shows a lower power readout pulse performed for a short 300ns time, where the cavity barely has time to reach a steady state before the drive is turned off. Consequently, information from both the ring up and ring down must be integrated to achieve the SNR shown in this figure. Despite this measure, there is still significant infidelity from the lack of separation of the gaussian signals. In the second case, the displacement voltage is larger, and the pulse is three times as long, resulting in significantly increased separation of the gaussian signals and enabling discrimination of the $\ket{g}, \ket{e},\ket{f}$ and $\ket{h}$ states. However, the large powers required induce transitions between these states, resulting in the trails between them as the measurement integrates a mixture of different cavity states at different times.

\begin{figure}[h!]
\includegraphics[width = 6in]{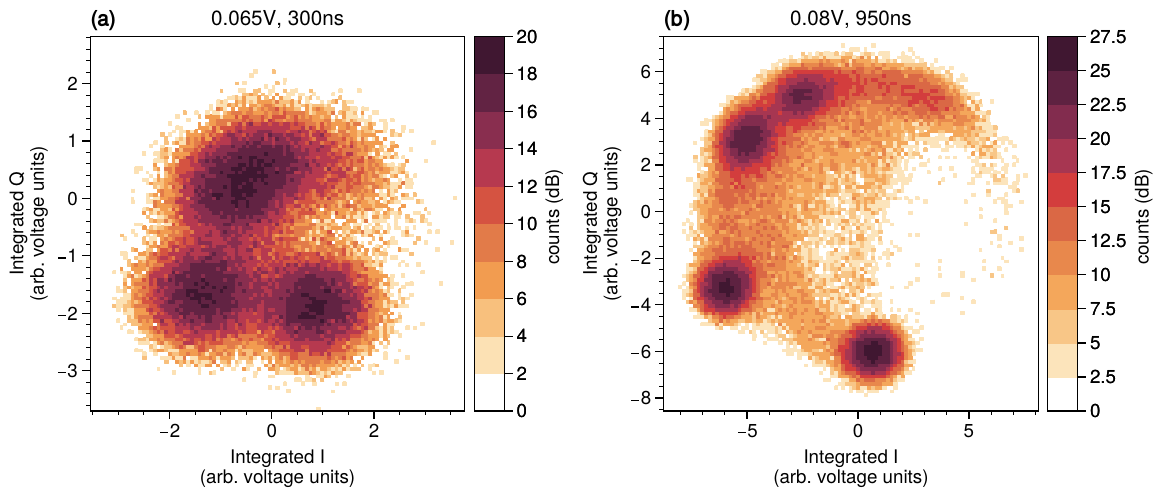}
\caption{
Comparison between boxcar-integrated IQ results of (a) a lower power pulse applied for a short time, corresponding to $\bar{n} = 116$ readout photons, and (b) a higher power pulse applied for a longer time, corresponding to a larger $\bar{n} = 176$ readout photons. State transitions are visible as ``trails'' leading between the primary symbols in (b). Counts are shown in logarithmic units to emphasize low-count trails. }
\label{appfig:RO_histograms}
\end{figure}

In Fig. A\ref{appfig:hardwareSchematic}, the hardware schematic of the measurements in section \ref{sec:rcexpt} are shown. The measurement setup is fairly standard, using single sideband upconversion to send signals into the dilution refrigerator, moving through three stages of attenuation with 20dB attenuation at 4K, 20dB attenuation at the 100mK stage, and approximately 45dB of attenuation at the base stage of the refrigerator, with 10dB of the base stage attenuation coming from a particularly well-thermalized copper body attenuator. The signal interacts with the qubit and cavity system, is routed by two circulation stages to the amplifier, amplified in reflection, and then is routed once again back through the circulators to the remaining stages of amplification at 4K and room temperature accordingly. From there it is downconverted by the same local oscillator to 50MHz, filtered, amplified once more at low frequency, digitized at 1GS/s, and finally demodulated and integrated to acquire a readout histogram such as the ones shown in Fig.~\ref{appfig:RO_histograms}.
\begin{figure}[h!]
\includegraphics[width = 6in]{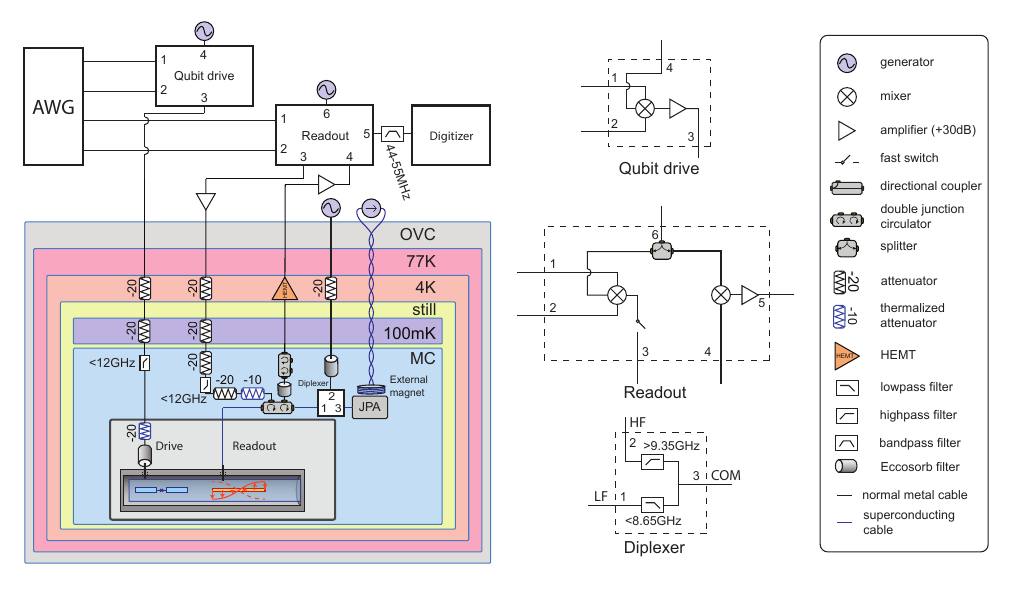}
\caption{
Upconversion and downconversion schematic for drive pulses sent first to the qubit readout resonator, driven in reflection, then routed to the amplifier and to the HEMT via two circulators.
}
\label{appfig:hardwareSchematic}
\end{figure}

\clearpage

\section{Simulating heterodyne measurement records obtained from quantum measurement chains for dispersive qubit readout}
\label{app:sim}

In this appendix section, we describe the SMEs used to model various quantum measurement chains and generated datasets analyzed in the main text. For convenience we reproduce the general SME of Eq.~(\ref{eq:sme}):
\begin{align}
    d\rhoc = \Lsys \rhoc~dt + \Lchain\rhoc + \Lmeas[dW]\rhoc
    \label{appeq:sme}
\end{align}
For all the considered models of quantum measurement chains for the fixed task of dispersive qubit readout, $\Lsys$ remains the same, as identified in the main text:
\begin{align}
    \Lsys\rhoc = -i[\hat{\mathcal{H}}_{\rm disp},\rhoc]
    \label{appeq:lsys}
\end{align}
where $\hat{\mathcal{H}}_{\rm disp}$ is the dispersive cQED Hamiltonian for a multi-level artificial atom,
\begin{align}
    \Hdisp \simeq \sum_p \omega_p \proj{p}{p} -\Delta_{da}\hat{a}^{\dagger}\hat{a} + \sum_p \chi_p \hat{a}^{\dagger}\hat{a}  \proj{p}{p} 
\end{align}
The superoperators $\Lchain$ and $\Lmeas[dW]$ will depend on the specific model considered.

% The initial state of the measurement chain for the readout of a qubit initialized in state $\ket{\sigma}$ is described as:
% \begin{equation}
%     \rhou(0) = \proj{\sigma}{\sigma}\otimes\hat{\varrho}_c(0)
%     \label{appeq:ansInit}
% \end{equation}
% where $\hat{\varrho}_c(t)$ is the conditional density matrix defining the quantum state of all quantum modes in the measurement chain \textit{other} than the qubit. 

\subsection{Dispersive readout with no qubit transitions and using a cavity}
\label{app:dispInf}

For qubit readout in the absence of any state transitions, $\Lchain \to 0$. As a result, the SME of Eq.~(\ref{appeq:sme}) takes the simpler form:
\begin{align}
    d\rhoc = \Lsys \rhoc~dt + \Lmeas[dW]\rhoc
    \label{appeq:sme1}
\end{align}
Here $\Lsys$ is given by Eq.~(\ref{appeq:lsys}). The superoperator $\Lmeas$ describes quantum modes in the measurement chain that are used to measure the quantum system of interest. This superoperator can be expressed in the general form:
\begin{align}
    \Lmeas[dW]\rhoc = \Lqmodes\rhoc + \Smeas\rhoc
    \label{appeq:sme1}
\end{align}
Here $\Lqmodes$ defines the unconditional dynamics of quantum modes used for measurement; here, it takes the explicit form:
\begin{align}
    \Lqmodes \rhou = -i[\eta(\hat{a}+\hat{a}^{\dagger}),\rhou] + \kappa \mathcal{D}[\hat{a}]\rhou 
\end{align}
which describes the measurement tone used for cavity readout, and the cavity losses due to its monitored port. Importantly, $\Lqmodes$ is independent of the qubit sector. 

Then, $\Smeas$ is the stochastic measurement superoperator that describes conditional evolution under continuous heterodyne monitoring:
\begin{align}
    \Smeas \rhoc = \sqrt\frac{\kappa}{2}\left(\hat{a}\rhoc + \rhoc \hat{a}^{\dagger} - \avg{\hat{a}+\hat{a}^{\dagger}}\rhoc \right)dW_I + \sqrt\frac{\kappa}{2}\left( -i\hat{a}\rhoc + i\rhoc \hat{a}^{\dagger} - \avg{-i\hat{a}+i\hat{a}^{\dagger}}\rhoc \right)dW_Q
    \label{appeq:smeas1}
\end{align}

These explicit forms of superoperators fully define Eq.~(\ref{appeq:sme}) in this regime without qubit transitions. However, this assumption can be used to further simplify the form of the SME. In particular, in the absence of transitions, the quantum state of the measurement chain is given by the ansatz:
\begin{equation}
    \rhou(t) = \proj{p}{p}\otimes\hat{\varrho}_c(t)
    \label{appeq:ansInf}
\end{equation}
where $\hat{\varrho}_c(t)$ is the conditional density matrix defining the quantum state of all quantum modes in the measurement chain \textit{other} than the qubit (namely, the cavity mode). The above implies that the qubit state is completely unchanged during the readout time. The only evolution is in the state of the modes used to readout the qubit, namely the cavity modes.

By now tracing out the qubit subspace in Eq.~(\ref{appeq:sme}), we can obtain an SME for $\hat{\varrho}_c(t)$ alone, under the ansatz of Eq.~(\ref{appeq:ansInf}). The Hamiltonian contribution from the dispersive qubit Hamiltonian yields:
\begin{align}
{\rm tr}_Q\{ \Hdisp \proj{p}{p}\otimes\hat{\varrho}_c \}  &= {\rm tr}_Q\Big\{ \sum_j \omega_j \ket{j}\!\braket{j}{p}\!\bra{p}\otimes\hat{\varrho}_c \Big\} -{\rm tr}_Q\Big\{ \proj{p}{p}\otimes (\Delta_{da}\hat{a}^{\dagger}\hat{a}\hat{\varrho}_c) \Big\} + {\rm tr}_Q\Big\{ \sum_j \chi_j \hat{a}^{\dagger}\hat{a}  \ket{j}\!\underbrace{\braket{j}{p}}_{\delta_{jp}}\!\bra{p}\otimes\hat{\varrho}_c \Big\} \nonumber \\
&= \omega_{p}\hat{\varrho}_c  - \Delta_{da}\hat{a}^{\dagger}\hat{a}\hat{\varrho}_c + \chi_{p}\hat{a}^{\dagger}\hat{a}\hat{\varrho}_c
\end{align}
and by conjugation,
\begin{align}
{\rm tr}_Q\{ \proj{p}{p}\otimes\hat{\varrho}_c \Hdisp  \}  &=  \hat{\varrho}_c\omega_{p}  - \hat{\varrho}_c \Delta_{da}\hat{a}^{\dagger}\hat{a} + \hat{\varrho}_c\chi_{p}\hat{a}^{\dagger}\hat{a}
\end{align}
following which we arrive at:
\begin{align}
{\rm tr}_Q\{-i[\Hdisp,\rhoc] \} = -i \left([-\Delta_{da}\hat{a}^{\dagger}\hat{a},\hat{\varrho}_c]  + [\chi_{p}\hat{a}^{\dagger}\hat{a},\hat{\varrho}_c] \right) = -i[ \left(-\Delta_{da}+\chi_{p}\right)\hat{a}^{\dagger}\hat{a},\hat{\varrho}_c ] \equiv -i[\hat{\mathcal{H}}_{\rm cav},\hat{\varrho}_c ]
\end{align}
where we have defined $\hat{\mathcal{H}}_{\rm cav}$ as the cavity Hamiltonian alone:
\begin{align}
    \hat{\mathcal{H}}_{\rm cav} = \left(-\Delta_{da}+\chi_{p}\right)\hat{a}^{\dagger}\hat{a} = (\omega_a + \chi_{p} - \omega_d)  \hat{a}^{\dagger}\hat{a}
\end{align}
We can perform a similar simplification on terms due to $\Lmeas$. For the ansatz in Eq.~(\ref{appeq:ansInf}), we find for $\Lqmodes$:
\begin{align}
    {\rm tr}_Q\{ \Lqmodes (\proj{p}{p}\otimes\hat{\varrho}_c) \} &=  {\rm tr}_Q\{  \proj{p}{p}\otimes\Lqmodes\hat{\varrho}_c \} = {\rm tr}_Q\{  \proj{p}{p} \} \otimes\Lqmodes\hat{\varrho}_c = \Lqmodes\hat{\varrho}_c
\end{align}
As $\Lqmodes$ was independent of the qubit subsector, it remains unchanged following the partial trace over this subsector. The stochastic measurement operator $\Smeas$ is again independent of the qubit subspace. Hence tracing out the qubit sector yields:
\begin{align}
    \sqrt{\kappa}~{\rm tr}_Q\{ \Smeas\proj{p}{p}\otimes\hat{\varrho}_c \} = \sqrt{\kappa}~{\rm tr}_Q\{ \proj{p}{p} \} \otimes\Smeas\hat{\varrho}_c = \sqrt{\kappa}\Smeas\hat{\varrho}_c
\end{align}

The final \textit{cavity-only} SME in the absence of any qubit transitions takes the form:
\begin{align}
    d\hat{\varrho}_c = -i[\hat{\mathcal{H}}_{\rm cav},\hat{\varrho}_c]dt + \Lmeas[dW] \hat{\varrho}_c
    \label{appeq:smecav}
\end{align}
The resulting SME preserves Gaussian states and can be solved exactly using a truncated equations of motion (TEOMs) approach.

\subsection{Dispersive readout with no qubit transitions and using a quantum-limited amplifier with added noise}
\label{app:simAmp}

In the absence of any state transitions, $\Lchain \to 0$, and the SME of Eq.~(\ref{appeq:sme}) takes the simpler form:
\begin{align}
    d\rhoc = \Lsys \rhoc~dt + \Lmeas[dW]\rhoc
    \label{appeq:sme2}
\end{align}
Again $\Lsys$ is given by Eq.~(\ref{appeq:lsys}), and $\Lmeas$ takes the form:
\begin{align}
    \Lmeas[dW]\rhoc = \Lqmodes\rhoc + \Smeas\rhoc
    \label{appeq:lmeas2}
\end{align}
Now, $\Lqmodes$ for the unconditional dynamics of quantum modes used for measurement takes the explicit form:
\begin{align}
    \Lqmodes \rhou = -i[\eta(\hat{a}+\hat{a}^{\dagger}),\rhou] + \kappa' \mathcal{D}[\hat{a}]\rhou + \mathcal{L}_{c}\rhoc + \mathcal{L}_{\rm amp}\rhou 
\end{align}
The first term again describes the measurement tone used for cavity readout, and the second describes cavity losses. However, the cavity's open port is now directed to a phase-preserving amplifier downstream. The superoperator $\mathcal{L}_{\rm amp}$ is the Liouvillian defining this quantum amplifier, which we take to be a two-mode non-degenerate parametric amplifier providing phase-preserving gain:
\begin{align}
    \mathcal{L}_{\rm amp} \rhoc = -i\left[\frac{-i g_{\rm amp}}{2}\hat{d}\hat{c} + h.c.,\rhoc \right] + \gamma_d\mathcal{D}[\hat{d}]\rhoc + \gamma\mathcal{D}[\hat{c}]\rhoc
\end{align}
The superoperator $\mathcal{L}_{c}$ then defines the non-reciprocal coupling between the cavity mode and amplifier's signal mode $\hat{d}$,
\begin{align}
    \mathcal{L}_{c}\rhoc = -i \left[ \frac{i g}{2} \hat{d}\hat{a}^{\dagger} + h.c.,\rhoc \right] + \Gamma \mathcal{D}[\hat{a}+\hat{d}]\rhoc
\end{align}
To ensure non-reciprocal coupling so that fields from the cavity that carry qubit state information are transmitted to the amplifier for readout, but transmission in the reverse direction is forbidden, we require $g = \Gamma$~\cite{metelmann_nonreciprocal_2015}.

Finally $\Smeas$ describes conditional evolution under continuous heterodyne monitoring, now of the amplifier's signal mode:
\begin{align}
    \Smeas \rhoc = \sqrt{\frac{\gamma_d}{2}}\left(\hat{d}\rhoc + \rhoc \hat{d}^{\dagger} - \avg{\hat{d}+\hat{d}^{\dagger}}\rhoc \right)dW_I + \sqrt{\frac{\gamma_d}{2}}\left( -i\hat{d}\rhoc + i\rhoc \hat{d}^{\dagger} - \avg{-i\hat{d}+i\hat{d}^{\dagger}}\rhoc \right)dW_Q
    \label{appeq:smeas1}
\end{align}

We now summarize the actual parameter choices used to generate quantum amplifier simulated datasets in the main text. We define the total cavity loss rate $\kappa = \kappa' + \Gamma$. Then, we choose cavity parameters so that $\kappa' = \Gamma = 0.5\kappa$, and the dispersive shift $\chi/\kappa = 0.5$. Recall that perfect non-reciprocal coupling in the desired direction requires $g =\Gamma = 0.5\kappa$. Lastly, amplifier parameters are chosen so that $\gamma = \gamma_d + \Gamma = 5\kappa$, yielding the ratio of cold amplifier linewidth to cavity linewidth $\gamma/\kappa = 5$ used in the main text, and also implying that $\gamma_d = 4.5\kappa$. 

In the absence of qubit transitions, Eq.~(\ref{appeq:ansInf}) holds once again, as $\Lmeas$ is completely independent of the qubit sector. Hence this sector may be traced out exactly as in the previous subsection. We thus arrive at a \textit{cavity-amplifier-only} SME in the absence of any qubit transitions:
\begin{align}
    d\hat{\varrho}_c = -i[\hat{\mathcal{H}}_{\rm cav},\hat{\varrho}_c]dt + \Lmeas[dW] \hat{\varrho}_c
\end{align}
for $\Lmeas$ now given by Eq.~(\ref{appeq:lmeas2}). The resulting SME again preserves Gaussian states and can be solved exactly using a truncated equations of motion (TEOMs) approach.

\subsection{Dispersive readout including multi-level transitions using a cavity}
\label{app:simTransitions}

For qubit readout allowing for state transitions, we must now include $\Lchain$ in the SME:
\begin{align}
    d\rhoc = \Lsys \rhoc~dt + \Lchain\rhoc + \Lmeas[dW]\rhoc
    \label{appeq:sme3}
\end{align}
Again $\Lsys$ is given by Eq.~(\ref{appeq:lsys}). Now the nontrivial superoperator $\Lchain$ takes the form:
\begin{align}
    \Lchain\rhou = \sum_{j\neq k} \gamma_{jk}\mathcal{D}[\proj{k}{j}]\rhou
\end{align}
where $\gamma_{jk}$ is the rate of transition from qubit state $\ket{j}$ to state $\ket{k}$.

As we still consider readout using a cavity, the remaining terms in Eq.~(\ref{appeq:sme3}) are as in Eq.~(\ref{appeq:sme1}); in particular, $\Lmeas$ takes the form:
\begin{align}
    \Lmeas[dW]\rhoc = \Lqmodes\rhoc + \Smeas\rhoc
    \label{appeq:sme1}
\end{align}
where $\Lqmodes$ is given by:
\begin{align}
    \Lqmodes \rhou = -i[\eta(\hat{a}+\hat{a}^{\dagger}),\rhou] + \kappa \mathcal{D}[\hat{a}]\rhou 
\end{align}
while $\Smeas$ is given by:
\begin{align}
    \Smeas \rhoc = \sqrt\frac{\kappa}{2}\left(\hat{a}\rhoc + \rhoc \hat{a}^{\dagger} - \avg{\hat{a}+\hat{a}^{\dagger}}\rhoc \right)dW_I + \sqrt\frac{\kappa}{2}\left( -i\hat{a}\rhoc + i\rhoc \hat{a}^{\dagger} - \avg{-i\hat{a}+i\hat{a}^{\dagger}}\rhoc \right)dW_Q
    \label{appeq:smeas3}
\end{align}

We emphasize that now the quantum state of the measurement chain can not generally be expressed in the form of Eq.~(\ref{appeq:ansInf}). Hence Eq.~(\ref{appeq:sme3}) is integrated in the joint qubit-cavity Hilbert space to generate simulated measurement datasets. 

\newpage

\section{Training and testing details}
\label{app:training}

% Once heterodyne measurement records are obtained, an output layer acts on these outputs to ideally return the corresponding state label $\sigma$; the result of the computation can therefore formally be written as
% \begin{align}
%     \sigmapred = \frc{\mathbf{y}} = \frc{\WO\bm{\vec{x}}^{(\sigma)}_{(n)} + \mathbf{b}}
%     \label{eq:trainW}
% \end{align}
% where $\WO$ is the matrix of trainable, time-independent output weights, $\mathbf{b}$ is a vector of trainable biases.

In this appendix, we analyze how optimal weights $\WOopt$ are learned from a training dataset in the \RC{} approach. 

\subsection{Cost function and learned weights}

We begin with the \RC{} map defined in the main text, Eq.~(\ref{eq:rcmap}):
\begin{align}
    \sigmapred = \frc{\mathbf{y}_{(n)}} = \frc{\WO \bm{\vec{x}}_{(n)} + \bv}
    \label{appeq:rcout}
\end{align}
now written to describe the mapping of a single instance $n$ of measured data, compiled in the vector $\vec{x}_{(n)}$, to a vector $\mathbf{y}_{(n)} \in \mathbb{R}^{C}$. The mapping is via a set of weights $\WO$ applied linearly to the data $\vec{x}$, and a set of weights that are additive, compiled in a column vector of biases $\bv \in \mathbb{R}^{C}$.

The vector $\bm{\vec{x}}$ lives in the \textit{joint} space of measurement records: $\vec{x}_{(n)} \in \mathbb{R}^{\NO\cdot\NT}$ is also a column vector, and can be written in the form:
\begin{align}
    \bm{\vec{x}}_{(n)}=
    \begin{pmatrix}
        \vec{x}_{1(n)} \\
        \vec{x}_{2(n)} \\
        \vdots \\
        \vec{x}_{\NO(n)}
    \end{pmatrix}
\end{align}
where each vector $\vec{x}_{m(n)} \in \mathbb{R}^{\NT}$ is a column vector describing the discretized records of $m \in [\NO]$ measurement observables, each with $\NT$ samples. Recall that for standard heterodyne readout, $\NO = 2$, where $\vec{x}_1 = \vec{I}$, $\vec{x}_2 = \vec{Q}$. From here on, we can work with this concatenated vector $\bm{\vec{x}}$.

In Eq.~(\ref{appeq:rcout}), $F[\cdot]$ is a function that maps the vector of measured heterodyne records to a discrete, scalar state label $\sigma \in [1,\ldots,C]$. This mapping is carried out via two operations. First, the measurement records $\bm{\vec{x}}^{(\sigma)}_{(n)}$ are mapped to an intermediate target vector $\mathbf{y}^{(\sigma)}_{(n)}$ employing a `one-hot' encoding (conventional for classification tasks): The $k$th element of this target vector $\mathbf{y}^{(\sigma)}_{(n)}$ is given by:
\begin{align}
    [\mathbf{y}^{(\sigma)}_{(n)}]_k = 
    \begin{cases}
    1~\text{if}~k = \sigma, \\
    0~\text{otherwise.}
    \end{cases}
\end{align}
Finally, a discriminator is used to map the target vector to a scalar state label.

With the key notation in place, we can discuss how the \RC{} training dataset is constructed. A training dataset of size $\NTrain$ consists of $n \in [\NTrain]$ heterodyne records for each of the $C$ states required to be distinguished in the classification task. We define a matrix $\mathbf{X} \in \mathbb{R}^{\NO\cdot\NT\times C\NTrain}$:
\begin{align}
    \mathbf{X} = 
    \begin{pmatrix}
        \bm{\vec{x}}^{(1)}_{(1)} & \bm{\vec{x}}^{(1)}_{(2)} & \cdots & \bm{\vec{x}}^{(1)}_{(\NTrain)} &  \cdots & \bm{\vec{x}}^{(C)}_{(1)} & \bm{\vec{x}}^{(C)}_{(2)} & \cdots & \bm{\vec{x}}^{(C)}_{(\NTrain)}
    \end{pmatrix}
\end{align}
We also define a matrix $\mathbf{Y}\in \mathbb{R}^{C\times C\NTrain}$ compiling the corresponding targets 
\begin{align}
    \mathbf{Y} =     \begin{pmatrix}
        \mathbf{y}^{(1)}_{(1)} & \mathbf{y}^{(1)}_{(2)} & \cdots & \mathbf{y}^{(1)}_{(\NTrain)} &  \cdots & \mathbf{y}^{(C)}_{(1)} & \mathbf{y}^{(C)}_{(2)} & \cdots & \mathbf{y}^{(C)}_{(\NTrain)}
    \end{pmatrix}
\end{align}
% It will prove useful to decompose the matrix of weights $\WO$ and the vector of biases $\bv$ as terms $\mathbf{W}^m$ and $\mathbf{b}^m$ applied to individual measured observables $m$:
% \begin{align}
%     \WO = 
%     \begin{pmatrix}
%         \WO^1 & \WO^2 & \cdots & \WO^{\NO} 
%     \end{pmatrix}
%     ,~
%     \bv = \sum^{\NO}_m \mathbf{b}^m 
% \end{align}
By further introducing $\vec{1} \in \mathbb{R}^{1\times C\NTrain}$ as a row vector containing all ones, Eq.~(\ref{appeq:rcout}) for all $C\NTrain$ records per measured observable can be written in the compact matrix form:
\begin{align}
    \mathbf{Y} = 
    \WO\mathbf{X}
    +\bv \vec{1} 
    % = 
    % \begin{pmatrix}
    %     \WO & \bv 
    % \end{pmatrix}
    % \begin{pmatrix}
    %     \mathbf{X} \\
    %     \vec{1} 
    % \end{pmatrix}
    % \equiv \bm{\mathcal{W}}\bm{\mathcal{X}}
    % \label{appeq:regression}
\end{align}
Before proceeding, we note that we have the freedom to introduce any invertible matrix $\mathbf{L} \in \mathbb{R}^{\NO\cdot\NT \times \NO\cdot\NT}$ as follows, without modifying the TPP map:
\begin{align}
    \mathbf{Y} = 
    \WO(\Lm^{-1}\Lm)\mathbf{X}
    +\bv \vec{1} = 
    \begin{pmatrix}
        \WO\Lm^{-1} & \bv 
    \end{pmatrix}
    \begin{pmatrix}
        \Lm\mathbf{X} \\
        \vec{1} 
    \end{pmatrix}
    \equiv \bm{\mathcal{W}}\bm{\mathcal{X}}
    \label{appeq:regression}
\end{align}
The auxiliary matrix $\Lm$ will prove convenient for our analysis later. 

Eq.~(\ref{appeq:regression}) helps us define $\bm{\mathcal{X}} \in \mathbb{R}^{(\NO\cdot\NT+1) \times C\NTrain}$ as a matrix which contains all measured records as well as a row of ones to account for the contribution of biases. Then, $\bm{\mathcal{W}} \in \mathbb{R}^{C \times (\NO\cdot\NT+1)}$ is the composite matrix of all learned weights. Eq.~(\ref{appeq:regression}) defines a regression problem that can be solved to obtain the optimal weights~\cite{larger_high-speed_2017},
\begin{align}
    \bm{\mathcal{W}}^{\rm opt} = \mathbf{Y}\bm{\mathcal{X}}^T(\bm{\mathcal{X}}\bm{\mathcal{X}}^T)^{-1} 
    \label{appeq:trainedWO}
\end{align}
% The optimal matrix of weights $\WO^{\rm opt}$ and biases $\mathbf{b}^{\rm opt}$ are thus determined by least squares optimization as defined in Eq.~(\ref{eq:wopt}) of the main text.

For convenience of the analysis to follow we introduce two new matrices: the \textit{mean} matrix $\mathbf{M} \in \mathbb{R}^{C \times (\NO\cdot\NT+1)}$,
\begin{align}
    \NTrain\mathbf{M} \equiv
        \mathbf{Y}\bm{\mathcal{X}}^T, \label{appeq:Mdef}
\end{align}
and the \textit{second-order moments matrix} $\mathbf{C} \in \mathbb{R}^{(\NO\cdot\NT+1)\times (\NO\cdot\NT+1)}$
\begin{align}
    \NTrain\mathbf{C} \equiv \bm{\mathcal{X}}\bm{\mathcal{X}}^T,
\end{align}
so that Eq.~(\ref{appeq:trainedWO}) can equivalently be written as:
\begin{align}
    \bm{\mathcal{W}}^{\rm opt} =\mathbf{M}\mathbf{C}^{-1}
    \label{appeq:trainedWO2}
\end{align}
where the factors of $\NTrain$ cancel out.

Note that the matrix $\mathbf{C} = \bm{\mathcal{X}}\bm{\mathcal{X}}^T$ can at times be ill-conditioned, making its inverse difficult to compute numerically. In such cases, we instead compute the quantity $\mathbf{C}^+$, related to the pseudoinverse of $\bm{\mathcal{X}}$, and defined by the following limit relation defining the pseudoinverse:
\begin{align}
    \mathbf{C}^+ = \lim_{\lambda\to 0}(\mathbf{C}-\lambda\mathbf{I})^{-1}
\end{align}
where $\mathbf{I}$ is the identity matrix on $\mathbb{R}^{(\NO\cdot\NT+1)\times (\NO\cdot\NT+1)}$ and $\lambda$ is typically referred to as a regularization parameter. If $\mathbf{C}$ is invertible, we have $\mathbf{C}^+ \to \mathbf{C}^{-1}$. We do emphasize that for the datasets analyzed in this paper, the intrinsic dataset noise serves as an effective regularizer, such that we can typically set $\lambda = 0$. 

\subsection{Testing via cross-validation}

For all classification infidelities calculated in the main text, we perform cross-validation. For a full dataset of $\NTraj$ records per state, a training set is constructed with $\NTrain < \NTraj$ records as described above. The remaining $\NTest = \NTraj - \NTrain$ records are used to construct a testing set. We use 80\% of the dataset for training, and the remaining 20\% for testing. This is consistent with training and testing set sizes for standard machine learning applications; for example, the MNIST handwritten digits classification task~\cite{deng_mnist_2012} uses 85.7\% of the total dataset for training, and the remaining 14.3\% for testing. Predicted state labels are obtained using this testing set via both the FGDA scheme, Eq.~(\ref{eq:standardmf}) of the main text, and the \RC{}, Eq.~(\ref{eq:rcmap}). This process is repeated until a total of $L=10$ iterations are completed: each time, a new set of weights $\bm{\mathcal{W}}^{\rm opt}$ is obtained from a distinct randomly chosen training set of the total $\NTraj$ records, and classification infidelities computed using the new random testing datasets. All classification fidelities are averaged to obtain the final values plotted in the main text. This cross-validation approach is standard in machine learning, and ensures that the observed performance is not unduly effected by variations due to the specific training or testing dataset used.

\subsection{Dependence on size and fidelity of training sets}

As the \RC{} deploys a supervised learning approach to training (not unlike a standard matched filter, as shown by Eq.~(\ref{eq:mfI}) in the main text), an important question is how its performance depends on the size of the available training dataset, as well as any possible errors in data labelling such as may arise due to qubit initialization errors for the case of qubit state readout.

To answer these questions, we consider again the case of measurement data experiencing only Gaussian white noise, and compare the general \RC{} and FGDA performance for binary classification of $p \in \{e,g\}$. Note that only using the \textit{general} \RC{} makes sense here, as the white noise \RC{} is exactly equal to the standard matched filter learned from a given training dataset in the special case of white noise. In Fig.~\ref{fig:ierror}(a), we thus plot the performance of the general \RC{} against the FGDA as in the main text, with a training set size of $\NTrain = 4000$ measurement records per class. We also consider the impact of qubit ground state $\ket{g}$ initialization error from 5\% upto 35\%: more opaque markers correspond to a lower qubit initialization error, and hence better classification performance. Fig.~\ref{fig:ierror}(b) repeats the same plot but now for a larger training set with $\NTrain = 8000$ measurement records per class.

%%%%%%%%%%%%%%%%%%%%%%%%%%%%%%%%%%%%%%%%%%%%%%%%%%%%%%%%%%%%%%%%%%%%%%%%%%%%%%%%%%%%%%%%%%%%%%%%%%

\begin{figure}[t]
    \centering
    \includegraphics[scale=1.0]{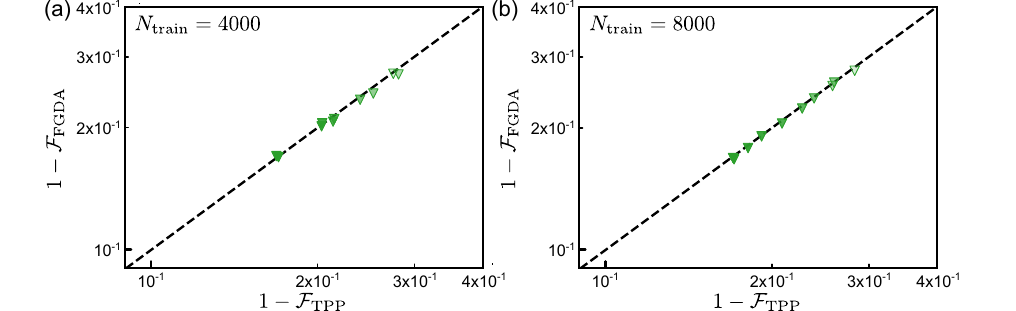}
    \caption{\textbf{Comparison of general \RC{} versus FGDA as a function of training set size and qubit initialization error.} Results are shown for training set sizes of (a) $\NTrain=4000$ measurement records and (b) $\NTrain=8000$ measurement records per class. More opaque markers indicate lower qubit initialization error.  }
    \label{fig:ierror}
\end{figure}

%%%%%%%%%%%%%%%%%%%%%%%%%%%%%%%%%%%%%%%%%%%%%%%%%%%%%%%%%%%%%%%%%%%%%%%%%%%%%%%%%%%%%%%%%%%%%%%%%%

We note that for the smaller training data set, the \RC{} very marginally underperforms in comparison to the FGDA for some of the data points. This is because the \RC{} has not yet converged to the optimal filter for this size of training data set. With increasing training set size, this difference becomes smaller and smaller. We also note that qubit initialization error appears to impact both schemes similarly, so that the \RC{} appears to not be unduly impact by data mislabeling.

Finally, we emphasize that the task considered here is one that most heavily favours the standard FGDA in contrast to the general \RC{}, as the measurement data actually satisfies the noise conditions assumed \textit{a priori} by the standard matched filter. Furthermore, the signal amplitudes used are weak (as indicated by the relatively low classification fidelity), so that measured data has a low signal-to-noise ratio and more data is needed to probe its statistics faithfully. If the true measured data exhibits noise statistics that deviate from this white noise case, even \RC{} filters learned using small training set sizes can already outperform the then sub-optimal standard MF trained on the same dataset.

\newpage

\section{\RC{} learned weights as optimal filters: analytic results}
\label{app:mf}

In this appendix, we will attempt to find an explicit form for the matrix $\bm{\mathcal{W}}^{\rm opt}$ from the previous seciton, under some assumptions on the form of the data contained in $\mathbf{X}$. 

\subsection{Measured data as stochastic random variables}

% For convenience of notation, in what follows we will no longer refer to the superscript $(n)$ indexing individual measurement records in the training set. The superscript index without parenthesis will refer to the position in the vector, which indexes temporal components. 

To make further progress, we must make some assumptions regarding the general form of measured data $\bm{\vec{x}}^{(c)}$. In particular, we assume:
\begin{align}
    \bm{\vec{x}}^{(c)} = \bm{\vec{s}}^{(c)} + \bm{\vec{\zeta}}^{(c)}
    \label{appeq:additiveNoise}
\end{align}
where $\bm{\vec{\zeta}}^{(c)}$ is a random noise process that contains the stochasticity of the data $\bm{\vec{x}}$. In particular, this includes contributions from heterodyne measurement noise $\vec{\xi}$, added classical noise $\vec{\xi}_{\rm cl}$, as well as quantum noise in conditional quantum trajectories. Without loss of generality, $\bm{\vec{\zeta}}$ can always be taken to have zero mean,
\begin{align}
    \mathbb{E}[ \bm{\vec{\zeta}}_j ] = 0~\forall~j
\end{align}
The random noise process can be defined by its covariance matrix,
\begin{align}
    \bm{\Sigma}^{(c)}_{jk} =  \mathbb{E}[ \bm{\vec{\zeta}}^{(c)}_j\bm{\vec{\zeta}}^{(c)}_k
    ].
\end{align}
The noise process will in general also possess non-zero higher-order cumulants, but these quantities will not make an appearance in our analysis here. 

Then, $\bm{\vec{s}}^{(c)}$ is simply equal to the expectation value of the random variable $\bm{\vec{x}}^{(c)}$ over an in principle infinite number of shots,
\begin{align}
    \bm{\vec{s}^{(c)}} = \mathbb{E}[\bm{\vec{x}}^{(c)}] 
    \label{appeq:meanS}
\end{align}

In practice, we will only have access to a finite number of shots $\NTrain$. Then, the above mean can be approximated using the estimator:
\begin{align}
    \bm{\vec{s}}^{(c)} \approx \frac{1}{\NTrain}\sum_{n=1}^{\NTrain} \bm{\vec{x}}^{(c)}_{(n)}.
    \label{appeq:meanSApprox}
\end{align}
Similarly, the covariance matrix of the noise process can be estimated via:
\begin{align}
    \bm{\Sigma}^{(c)}  \approx \frac{1}{\NTrain}\sum_{n=1}^{\NTrain} \bm{\vec{\zeta}}_{(n)}^{(c)}\bm{\vec{\zeta}}_{(n)}^{(c)T}.
    \label{appeq:meanVApprox}
\end{align}

% Here $\xi_{{\rm W},n}(t_i)$ is a discrete Gaussian random variable with statistical properties determined entirely by its mean and autocorrelation:
% \begin{align}
%     \mathbb{E}[\xi_{{\rm W},n}(t_i)] = 0,~\mathbb{E}[\xi_{{\rm W},n}(t_i)\xi_{{\rm W},n}(t_j)] = \sigma_{\rm W}^{(c)2} \delta_{ij}
% \end{align}

 Assuming the very general form of Eq.~(\ref{appeq:additiveNoise}), we can proceed to greatly simplify the matrices $\mathbf{M}$ and $\mathbf{C}$. 

\subsubsection{Simplification of mean matrix $\mathbf{M}$}
 
 The mean matrix $\mathbf{M}$, Eq.~(\ref{appeq:Mdef}), can be written explicitly as
\begin{align}
    \NTrain\mathbf{M}  &= 
    \mathbf{Y}
    \begin{pmatrix}
        \mathbf{X}^{T}\Lm^T &
        \vec{1}^T 
    \end{pmatrix} =
    \begin{pmatrix}
        \mathbf{Y}\mathbf{X}^{T}\Lm^T & \mathbf{Y}\vec{1}^T
    \end{pmatrix}
    \label{appeq:DDef}
\end{align}
We now proceed to simplify the general matrices $\mathbf{Y}\vec{1}^T$ and $\mathbf{Y}\mathbf{X}^T\Lm^T$. Starting with the former, which simply yields a column vector that is an element of $\mathbb{R}^{C \times 1}$, we find explicitly:
\begin{align}
    (\mathbf{Y}\vec{1}^T)_{l} = 
    \sum_{k=1}^{C\cdot\NTrain} \mathbf{Y}_{lk}\vec{1}^T_k = \sum_{n}\sum_{c}\mathbf{y}_l^{(c)} = \sum_{n}\sum_{c} \delta_{cl} = \NTrain
\end{align}
Here, we have used the fact that the sum over the columns of $\mathbf{Y}$ (and of $\mathbf{X}$), indexed by $k$, can be decomposed into two sums: over $\NTrain$ training records indexed by $n$, and over $C$ states indexed by $c$. From here on, we suppress the limits of these summations, for clarity.

Next we consider $\mathbf{Y}\mathbf{X}^T$, which can be expanded out explicitly,
\begin{align}
    \mathbf{Y}\mathbf{X}^T = \sum_{k} \mathbf{Y}_{lk}\mathbf{X}^T_{km} = \sum_{k} \mathbf{Y}_{lk}\mathbf{X}_{mk} = \sum_{n=1}^{\NTrain}\sum_{c=1}^C \delta_{lc}(\bm{\vec{x}}^{(c)}_{(n)})_m \simeq \NTrain \sum_{c=1}^C \delta_{lc}(\bm{\vec{s}}^{(c)})_m = \NTrain (\bm{\vec{s}}^{(l)})_m
\end{align}
where we have used Eq.~(\ref{appeq:meanSApprox}) in obtaining the final expression. Hence using Eq.~(\ref{appeq:DDef}), the matrix $\mathbf{M}$ takes the simple form (after the factors of $\NTrain$ cancel out):
\begin{align}
    \mathbf{M} = 
    \begin{pmatrix}
        (\Lm\bm{\vec{s}}^{(1)})^T & 1  \\
        \vdots & \vdots \\
        (\Lm\bm{\vec{s}}^{(C)})^T & 1
    \end{pmatrix} \equiv
    \begin{pmatrix}
        (\vec{S}^{(1)})^T \\
        \vdots \\
        (\vec{S}^{(C)})^T
    \end{pmatrix}
    ,
    \label{appeq:expmean}
\end{align}
which contains the mean traces for all measured observables over all states, explaining the nomenclature of the mean matrix. We have further introduced the vectors $\vec{S}^{(c)}$ which also include the contribution from the bias.

\subsubsection{Simplification of second-order moments matrix $\mathbf{C}$}

Simplifying the second-order correlation matrix $\mathbf{C}$ is more involved. We begin by expanding it to the form:
\begin{align}
    \NTrain\mathbf{C} \equiv \bm{\mathcal{X}}\bm{\mathcal{X}}^T = 
    \begin{pmatrix}
        \Lm\mathbf{X} \\
        \vec{1} 
    \end{pmatrix}
    \begin{pmatrix}
        \mathbf{X}^{T}\Lm^T &
        \vec{1}^T 
    \end{pmatrix}
    = 
    \begin{pmatrix}
        \Lm\mathbf{X}\mathbf{X}^T\Lm^T & \Lm\mathbf{X}\vec{1}^T \\
        \vec{1}\mathbf{X}^T\Lm^T & \vec{1}\vec{1}^T
    \end{pmatrix}
    \label{appeq:blockC}
\end{align}
Note that $\mathbf{X}\mathbf{X}^T$ is simply the two-time correlation matrix of the measured data.

We can further simplify $\mathbf{C}$, which has four components. Starting with the simplest, we note that:
\begin{align}
    \vec{1}\vec{1}^T = \sum_k \vec{1}_k \vec{1}^T_k = \sum_{n}\sum_{c} 1 = C\NTrain
\end{align}
Next, we consider the off-diagonal block term,
\begin{align}
    (\mathbf{X} \vec{1}^T)_{i} = \sum_{k} (\mathbf{X})_{ik} (\vec{1}^T)_k = \sum_c\sum_n (\bm{\vec{x}}^{(c)}_{(n)})_i \simeq \NTrain\sum_c [\bm{\vec{s}}^{m(c)}]_i
\end{align}
The other off-diagonal term is simply the transpose of the above.

Finally, we consider the block matrix,
\begin{align}
[\mathbf{X}\mathbf{X}^{T}]_{ij} = \sum_{k} [\mathbf{X}]_{ik} [\mathbf{X}^T]_{kj} = \sum_{k} [\mathbf{X}]_{ik} [\mathbf{X}]_{jk} = \sum_c\sum_n [\bm{\vec{x}}_{(n)}^{(c)}]_i[\bm{\vec{x}}_{(n)}^{(c)}]_j
\end{align}
To proceed further, we substitute Eq.~(\ref{appeq:additiveNoise}) into the final expression and expand:
\begin{align}[\mathbf{X}\mathbf{X}^{T}]_{ij} = \sum_{c} \sum_{n}[\bm{\vec{x}}_{(n)}^{(c)}]_i[\bm{\vec{x}}_{(n)}^{(c)}]_j = \sum_{c}\left\{ [\bm{\vec{s}}^{(c)}]_i[\bm{\vec{s}}^{(c)}]_j + \sum_{n}[\bm{\vec{\zeta}}_{(n)}^{(c)}]_i[\bm{\vec{s}}^{(c)}]_{j} + [\bm{\vec{s}}^{(c)}]_i\!\!\sum_{n}[\bm{\vec{\zeta}}_{(n)}^{l(c)}]_j +\sum_{n}[\bm{\vec{\zeta}}_{(n)}^{(c)}]_i[\bm{\vec{\zeta}}_{(n)}^{(c)}]_j \right\} 
\end{align}
Note that the sums indexed by $n$ over the training data are estimators of the statistics of the noise process. We can therefore write:
\begin{align}
    [\mathbf{X}\mathbf{X}^{T}]_{ij} = \NTrain\sum_c \left\{[\bm{\vec{s}}^{(c)}]_i[\bm{\vec{s}}^{(c)}]_j + \bm{\Sigma}_{ij}^{(c)} \right\}
\end{align}
It now proves useful to introduce two further matrices, the \textit{Gram} matrix $\mathbf{G}$:
\begin{align}
    \mathbf{G} = \sum_c \bm{\vec{s}}^{(c)}(\bm{\vec{s}}^{(c)})^T
\end{align}
and the empirical \textit{correlation} matrix $\mathbf{V}$:
\begin{align}
    \mathbf{V} = \sum_c \bm{\Sigma}^{(c)}
    \label{appeq:Vdef}
\end{align}

We can therefore write $\mathbf{C}$ in the simplified form,
\begin{align}
    \mathbf{C} = 
    \begin{pmatrix}
        \Lm(\mathbf{G}\!+\! \mathbf{V})\Lm^T & \sum_c \Lm\bm{\vec{s}}^{(c)} \\
        \sum_c (\bm{\vec{s}}^{(c)})^T\Lm^T & C
    \end{pmatrix}
    \label{appeq:blockCres}
\end{align}
and hence construct the full $\mathbf{C}$ via Eq.~(\ref{appeq:blockC}).

Having constructed explicit forms of $\mathbf{M}$ and $\mathbf{C}$, we are in principle positioned to evaluate the optimal weights and biases $\bm{\mathcal{W}}^{\rm opt}$ explicitly as well. To do so, it first again proves useful to interpret the learned weights in terms of optimal filters. 

\subsection{Constraints on \RC{} filters}

The learned matrix of weights can be written in vector form as:
\begin{align}
    \bm{\mathcal{W}}^{\rm opt} \equiv 
    \begin{pmatrix}
        (\bm{\vec{f}}_{1})^T\Lm^{-1} & b_{1} &  \\
        \vdots & \vdots  \\
        (\bm{\vec{f}}_{C})^T\Lm^{-1} & b_C\\
    \end{pmatrix} \equiv
    \begin{pmatrix}
        (\vec{F}_1)^T \\
        \vdots \\
        (\vec{F}_C)^T
    \end{pmatrix}  
    \label{appeq:formFC}
\end{align}
Next, using Eq.~(\ref{appeq:trainedWO}) together with the explicit form of the mean matrix $\mathbf{M}$ in Eq.~(\ref{appeq:expmean}), we arrive at the important relation:
\begin{align}
    \begin{pmatrix}
        (\vec{F}_1)^T \\
        \vdots \\
        (\vec{F}_C)^T
    \end{pmatrix} =
    \begin{pmatrix}
        (\vec{S}^{(1)})^T \\
        \vdots \\
        (\vec{S}^{(C)})^T
    \end{pmatrix}\mathbf{C}^{-1} \implies  \mathbf{C}^{-1}
    \begin{pmatrix}
        \vec{S}^{(1)} \cdots \vec{S}^{(C)}
    \end{pmatrix} = 
    \begin{pmatrix}
        \vec{F}_1 \cdots \vec{F}_C
    \end{pmatrix}
\end{align}
where we have used the fact that $\mathbf{C}$, and hence its inverse, is a symmetric matrix, and thereby computed the transpose of both sides. The above equation then implies:
\begin{align}
    \mathbf{C}^{-1}\vec{S}^{(c)} = \vec{F}_c\label{appeq:impFilter}
\end{align}

We note that the matrix $\mathbf{C}$ is very general as it is constructed for completely arbitrary measured signals; it is therefore generally dense and its inverse $\mathbf{C}^{-1}$ cannot be analytically determined. However, Eq.~(\ref{appeq:impFilter}) suggests that if we can find a way to work with quantities $\mathbf{C}^{-1}\vec{S}^{(c)}$ directly, we can avoid having to evaluate this regularized inverse of $\mathbf{C}$. This is our strategy to evaluate optimal filters analytically. 

We demonstrate this approach by considering the action of $\mathbf{C}$ on the constant inhomogeneous vector,
\begin{align}
    \vec{n} = 
    \begin{pmatrix}
        \bm{\vec{0}} \\
        1 
    \end{pmatrix}
    \label{appeq:n}
\end{align}
where $\bm{\vec{0}} \in \mathbb{R}^{\NO\cdot\NT}$ is a vector of zeros. In particular, we wish to evaluate $\mathbf{C}\vec{n}$. Using the block representation of $\mathbf{C}$, we have:
\begin{align}
    \mathbf{C}\vec{n} = 
    \begin{pmatrix}
        \Lm(\mathbf{G}\!+\! \mathbf{V})\Lm^T & \sum_c \Lm\bm{\vec{s}}^{(c)} \\
        \sum_c (\bm{\vec{s}}^{(c)})^T\Lm^T & C
    \end{pmatrix}
    \begin{pmatrix}
        \bm{\vec{0}} \\
        1
    \end{pmatrix}
    =  
    \begin{pmatrix}
        \sum_c \Lm\bm{\vec{s}}^{(c)} \\
        C
    \end{pmatrix}
    =
    \sum_c
    \begin{pmatrix}
        \Lm\bm{\vec{s}}^{(c)} \\
        1
    \end{pmatrix}
    =
    \sum_c \vec{S}^{(c)}
\end{align}
Most importantly, note that the right hand side is entirely independent of the covariance matrix $\mathbf{V}$, instead depending only on mean traces.

Now, using Eq.~(\ref{appeq:impFilter}), multiplying through by $\mathbf{C}^{-1}$ will allow us to work directly with the (unknown) optimal filters $\vec{F}^{(c)}$. We immediately find:
\begin{align}
    \sum_c \vec{F}_c = \vec{n}
    \label{appeq:constraint}
\end{align}

For completeness, we also consider the case where we instead require the calculation of $\mathbf{C}^+$. To this end, we add and subtract the regularization parameter $\lambda$,
\begin{align}
    (\mathbf{C}-\lambda\mathbf{I}) \vec{n} + \lambda \vec{n} =  \sum_c \vec{S}^{(c)} \implies \sum_c (\mathbf{C}-\lambda)^{-1}\vec{S}^{(c)} =  \vec{n} + \lambda (\mathbf{C}-\lambda\mathbf{I})^{-1}\vec{n}
\end{align}
or, finally,
\begin{align}
    \sum_c \vec{F}_c = \vec{n} + \lambda (\mathbf{C}-\lambda\mathbf{I})^{-1}\vec{n}
    \label{appeq:constraintPseudo}
\end{align}

The above defines a constraint on learned optimal filters, implying that they are not all linearly independent. Crucially, this constraint holds regardless of the correlation properties of the noise characterized by $\mathbf{V}$, and is hence very general.

% We use this relation to express any one of the desired vectors $\vec{F}^{(c)}$ in terms of the other $C-1$ unknown vectors, to obtain a fully-determined system that can be solved for the optimal filters.

\subsection{Analytically-calculable \RC{} filters: ``matched filters'' for arbitrary $C$}

Having obtained a useful constraint on \RC{}-learned filters, we will now take a step further and calculate semi-analytic expressions for these learned filters (eventually arriving at Eq.~(\ref{eq:analyticFilters}) of the main text).

The first step is to simplify the form of the matrix $\mathbf{C}$ in Eq.~(\ref{appeq:blockCres}), which we reproduce and expand below:
\begin{align}
    \mathbf{C} = 
    \begin{pmatrix}
        \Lm\mathbf{G}\Lm^T \!+\! \Lm\mathbf{V}\Lm^T & \sum_c \Lm\bm{\vec{s}}^{(c)} \\
        \sum_c (\bm{\vec{s}}^{(c)})^T\Lm^T & C
    \end{pmatrix}.
\end{align}
We have thus far allowed the auxiliary matrix $\Lm$ be completely general; we can now use it to simplify the form of $\mathbf{C}$. Note that $\mathbf{V}$ as defined in Eq.~(\ref{appeq:Vdef}) is the positive sum of individual positive-definite correlation matrices; as a result, it must also be positive-definite and real. Among the useful properties of such positive-definite matrices is that they admit a Cholesky decomposition. We choose the auxiliary matrix $\mathbf{L}$ such that it precisely determines the Cholesky decomposition of $\mathbf{V}$:
\begin{align}
    \mathbf{V} = \Lm^{-1}(\Lm^T)^{-1} \implies \mathbf{V}^{-1} = \Lm^T\Lm, 
    \label{appeq:Vcholesky}
\end{align}
where we have also used the fact that a positive-definite matrix is always invertible.

With this choice, we immediately find that $\mathbf{C}$ reduces to:
\begin{align}
    \mathbf{C} = 
    \begin{pmatrix}
        \Lm\mathbf{G}\Lm^T \!+\! \bar{\mathbf{I}} & \sum_c \Lm\bm{\vec{s}}^{(c)} \\
        \sum_c (\bm{\vec{s}}^{(c)})^T\Lm^T & C
    \end{pmatrix}.
    \label{appeq:blockCreswhite}
\end{align}
where $\bar{\mathbf{I}}$ is the identity matrix on $\mathbb{R}^{\NO\cdot\NT \times \NO\cdot\NT}$.

\subsubsection{Obtaining the linear system for filters}

To obtain a system of equations for the learned filters, we now consider the action of $\mathbf{C}$ on the vector $\vec{S}^{(c)}$. To do so, we will once again make use of the simplified block representation of $\mathbf{C}$, which allows us to write:
\begin{align}
    \mathbf{C}\vec{S}^{(c)} &= 
    \begin{pmatrix}
        \sum_{c'} \Lm\bm{\vec{s}}^{(c')}(\bm{\vec{s}}^{(c')})^T\Lm^T\!+\! \bar{\mathbf{I}} & \sum_{c'} \Lm\bm{\vec{s}}^{(c')} \\
        \sum_{c'} (\bm{\vec{s}}^{(c')})^T\Lm^T & C
    \end{pmatrix}
    \begin{pmatrix}
        \Lm\bm{\vec{s}}^{(c)} \\
        1 
    \end{pmatrix} \nonumber \\
    &= 
    \begin{pmatrix}
        \sum_{c'} \Lm\bm{\vec{s}}^{(c')}[(\bm{\vec{s}}^{(c')})^T\Lm^T\Lm\bm{\vec{s}}^{(c)}] +  \Lm\bm{\vec{s}}^{(c)} + \sum_{c'} \Lm\bm{\vec{s}}^{(c')}   \\
        \sum_{c'} [(\bm{\vec{s}}^{(c')})^T\Lm^T\Lm\bm{\vec{s}}^{(c)}] + C
    \end{pmatrix}
    \label{appeq:CS}
\end{align}
It proves useful to define the overlap of mean traces
\begin{align}
    O_{cc'} = (\bm{\vec{s}}^{(c')})^T\Lm^T\Lm\bm{\vec{s}}^{(c)} = (\bm{\vec{s}}^{(c')})^T\mathbf{V}^{-1}\bm{\vec{s}}^{(c)},
\end{align}
where we have used Eq.~(\ref{appeq:Vcholesky}). We can thus write:
\begin{align}
    \mathbf{C}\vec{S}^{(c)} &=     \begin{pmatrix}
        \sum_{c'} O_{cc'}\Lm\bm{\vec{s}}^{(c')} + \sum_{c'} \Lm\bm{\vec{s}}^{(c')} +   \Lm\bm{\vec{s}}^{(c)}  \\
        \sum_{c'} O_{cc'} + C
    \end{pmatrix} \nonumber \\
    &= 
    \begin{pmatrix}
        \sum_{c'} \left[ O_{cc'} + 1 + \delta_{cc'}\right]\Lm\bm{\vec{s}}^{(c')}  \\
        \sum_{c'} \left[O_{cc'} +1\right]
    \end{pmatrix} \nonumber \\
    &= 
    \begin{pmatrix}
        \sum_{c'} \left[ O_{cc'} + 1 + \delta_{cc'}\right]\Lm\bm{\vec{s}}^{(c')}  \\
        \sum_{c'} \left[O_{cc'} +1 + \delta_{cc'}\right]
    \end{pmatrix}
    -
    \begin{pmatrix}
        \bm{\vec{0}}  \\
        \sum_{c'}\delta_{cc'}
    \end{pmatrix} \nonumber \\
    &= \sum_{c'}\left[ O_{cc'} + 1 + \delta_{cc'}\right] 
    \begin{pmatrix}
        \Lm\bm{\vec{s}}^{(c')} \\
        1
    \end{pmatrix}
    - 
    \begin{pmatrix}
        \bm{\vec{0}}  \\
        1
    \end{pmatrix}
\end{align}
Finally, defining
\begin{align}
    M_{cc'} = \left[ O_{cc'} + 1 + \delta_{cc'}\right] 
\end{align}
and once again introducing $\vec{n}$ from Eq.~(\ref{appeq:n}), we arrive at the form:
\begin{align}
    \mathbf{C}\vec{S}^{(c)} = \sum_{c'} M_{cc'}\vec{S}^{(c')} - \vec{n}
\end{align}
Therefore, we find that the action of $\mathbf{C}$ on $\vec{S}^{(c)}$ can be expressed as a linear combination of the set of vectors $\{\vec{S}^{(c)}\}$, and a vector $\vec{n}$ that is independent of $c$.

We now wish to introduce the unknown filters $\vec{F}_c$ to the above system, using Eq.~(\ref{appeq:impFilter}). To do so, we add and subtract the regularization parameter $\lambda$, and multiply through by the regularized inverse of $\mathbf{C}$. This yields
\begin{align}
    \vec{S}^{(c)} &=
    \sum_{c'} (\mathbf{C}-\lambda\mathbf{I})^{-1}(M_{cc'} -\lambda\mathbf{I}\delta_{cc'}) \vec{S}^{(c')} -(\mathbf{C}-\lambda\mathbf{I})^{-1}
    \vec{n} \nonumber \\
    &= \sum_{c'} (M_{cc'} -\lambda\mathbf{I}\delta_{cc'})\vec{F}_{c'} -(\mathbf{C}-\lambda\mathbf{I})^{-1}
    \vec{n}, 
    \label{appeq:filterSystemInc}
\end{align}

% Note that this approach foregoes the calculation of the regularized inverse of $\mathbf{C}$ in the computation of the learned filters $\vec{F}_c$. We emphasize here that if we instead consider observable-\textit{dependent} Gaussian white noise, the terms $M_{cc'}$ are replaced by a block diagonal matrix in $\mathbb{R}^{(\NO\cdot\NT+1) \times (\NO\cdot\NT+1)}$. These matrices will not generally commute with $(\mathbf{C}-\lambda\mathbf{I})^{-1}$, preventing the transition from the first to the second line above, which is crucial to introducing $\vec{F}_c$ to the system. Filters for general situations such as that can always be obtained by evaluating Eq.~(\ref{appeq:trainedWO}) numerically.

However, Eq.~(\ref{appeq:filterSystemInc}) is not entirely free of the $(\mathbf{C}-\lambda\mathbf{I})^{-1}$ matrix, due to the inhomogeneous term. Fortunately, as the inhomogeneous term is constant, it can be removed by considering the difference of Eq.~(\ref{appeq:filterSystemInc}) for any two distinct $c$ values. For example, considering $c\neq c'' \in [1,\ldots,C]$:
\begin{align}
    \vec{S}^{(c)}-\vec{S}^{(c'')} = \sum_{c'} M_{cc'} \vec{F}_{c'}-\sum_{c'} M_{c''c'} \vec{F}_{c'} = \sum_{c'}\left[M_{cc'}-M_{c''c'} \right]\vec{F}_{c'}
    \label{appeq:spair}
\end{align}
This naturally introduces the difference of mean traces to the calculation of learned filters.

Finally, we recall that the unknown filters $\vec{F}_c$ are not all linearly independent. We therefore use the constraint Eq.~(\ref{appeq:constraint}) in the formal limit $\lambda \to 0$ to eliminate one of the unknown vectors, here taken to be $\vec{F}_C$:
\begin{align}
    \vec{F}_C = \vec{n} - \sum_{c'=1}^{C-1} \vec{F}_{c'}
\end{align}
Then Eq.~(\ref{appeq:spair}) can be rewritten as:
\begin{align}
    \vec{S}^{(c)}-\vec{S}^{(c'')} &= \sum_{c'=1}^{C-1} \left[M_{cc'}-M_{c''c'} \right]\vec{F}_{c'} + \left[M_{cC}-M_{c''C} \right]\vec{F}_C \nonumber \\
    &= \sum_{c'=1}^{C-1} \left[M_{cc'}-M_{c''c'} \right]\vec{F}_{c'} -\sum_{c'=1}^{C-1} \left[M_{cC}-M_{c''C} \right]\vec{F}_{c'} + \left[M_{cC}-M_{c''C} \right]\vec{n} \nonumber \\
    &= \sum_{c'=1}^{C-1} \left[ (M_{cc'}-M_{c''c'})-(M_{cC}-M_{c''C}) \right]\vec{F}_{c'} + \left[M_{cC}-M_{c''C} \right]\vec{n}
    \label{appeq:spair2}
\end{align}

Note that there are $C-1$ unknowns $\vec{F}_c$, and hence we require $C-1$ equations. These equations are simply provided by Eq.~(\ref{appeq:spair2}) by considering $C-1$ distinct pairs $[c,c'']$. For concreteness, we consider pairs $P_p = [c,c'']$ where $[c,c''] \in \{[1,2],[2,3],\ldots,[C-1,C]\}$ indexed by $p \in [1,\ldots,C-1]$. We also introduce notation to individually identify the states constituting the $p$th pair, for convenience: if $P_p = [c,c'']$, $P_p(1) = c, P_p(2) = c''$. We then define the difference of mean traces constituting a pair,
\begin{align}
    \vec{S}^{P_p} \equiv \vec{S}^{(P_p(1))} - \vec{S}^{(P_p(2))}
    \label{appeq:sdiffdef}
\end{align}
Each pair yields an equation of the form of Eq.~(\ref{appeq:spair2}); it is easily seen that the full set of $C-1$ equations can be compiled into the matrix system:
\begin{align}
\begin{pmatrix}
    \vec{S}^{P_1} \\
    \vdots \\
    \vec{S}^{P_{C-1}}
\end{pmatrix}
=
\left(\mathbf{Q} \otimes \mathbf{I} \right)
\begin{pmatrix} 
    \vec{F}_1 \\
    \vdots \\
    \vec{F}_{C-1}
\end{pmatrix}
+ \left(\mathbf{T} \otimes \mathbf{I} \right)
\begin{pmatrix}
    \vec{n} \\
    \vdots \\
    \vec{n}
\end{pmatrix}
\label{appeq:filterSystem}
\end{align}
using the properties of the Kronecker product. Here $\mathbf{I}$ is the identity matrix on $\mathbb{R}^{\NO(\NT+1)\times\NO(\NT+1)}$ as before, while both $\mathbf{Q}$ and $\mathbf{T}$ are elements of the much smaller space $\mathbb{R}^{(C-1)\times(C-1)}$. In particular, their matrix elements are given by:
\begin{align}
    \mathbf{Q}_{pc} = \left[ (M_{P_p(1)c}-M_{P_p(2)c})-(M_{P_p(1)C}-M_{P_p(2)C}) \right],~\mathbf{T}_{pc} = \delta_{pc} \left[M_{P_p(1)C}-M_{P_p(2)C} \right]
    \label{appeq:qcpdef}
\end{align}
Note further that $\mathbf{T}$ is a diagonal matrix.

\subsubsection{Solving the linear system for filters}

Being a simple linear system, Eq.~(\ref{appeq:filterSystem}) has the formal solution,
\begin{align}
    \begin{pmatrix} 
    \vec{F}_1 \\
    \vdots \\
    \vec{F}_{C-1}
\end{pmatrix}=
\left(\mathbf{Q}^{-1} \otimes \mathbf{I} \right)
\begin{pmatrix}
    \vec{S}^{P_1} \\
    \vdots \\
    \vec{S}^{P_{C-1}}
\end{pmatrix}
-\left(\mathbf{Q}^{-1}\otimes \mathbf{I}\right)\left(\mathbf{T} \otimes \mathbf{I}\right)
\begin{pmatrix}
    \vec{n} \\
    \vdots \\
    \vec{n}
\end{pmatrix}.
\label{appeq:filterSol}
\end{align}
We can now simply read off the solution for the unknown vector $\vec{F}_c$:
\begin{align}
    \vec{F}_c = \sum_{p=1}^{C-1} \mathbf{Q}^{-1}_{cp}\vec{S}^{P_p} -\sum_{p=1}^{C-1} \mathbf{Q}^{-1}_{cp}\mathbf{T}_{pp} \vec{n}
\end{align}
The first term on the right hand side completely defines the filter components in $\vec{F}_c$, as they have a zero at the position corresponding to the bias component. The second term then entirely defines the bias. Using the form of $\vec{F}_c$ from Eq.~(\ref{appeq:formFC}), we can immediately read off the individual filters for each measured quadrature:
\begin{align}
    (\Lm^{-1})^T\bm{\vec{f}}_c = \sum_p \mathbf{Q}^{-1}_{cp}\Lm\bm{\vec{s}}^{(P_p)} 
\end{align}
which simplifies to:
\begin{align}
    \bm{\vec{f}}_c = \sum_p \mathbf{Q}^{-1}_{cp}\Lm^T\Lm\bm{\vec{s}}^{(P_p)} \implies \bm{\vec{f}}_c = \sum_p \mathbf{Q}^{-1}_{cp}\mathbf{V}^{-1}\bm{\vec{s}}^{(P_p)}
\end{align}
where we have again used Eq.~(\ref{appeq:Vcholesky}). The bias terms are finally given by:
\begin{align}
    \bv_c = -\sum_p \mathbf{Q}^{-1}_{cp}\mathbf{T}_{pp}
\end{align}
The remaining learned filter and bias is then given by the constraint, Eq.~(\ref{appeq:constraint}).

An alternative, more practical form of the learned filters can be extracted by transitioning from the representation in terms of difference vectors $\vec{S}^{P_p}$, to the individual traces $\vec{S}^{(c)}$, using Eq.~(\ref{appeq:sdiffdef}). We find:
\begin{align}
    \bm{\vec{f}}_c = \mathbf{Q}^{-1}_{c1}\mathbf{V}^{-1}\bm{\vec{s}}^{(1)} + \sum_{p=2}^{C-1} \left[\mathbf{Q}^{-1}_{cp}-\mathbf{Q}^{-1}_{c(p-1)}\right]\mathbf{V}^{-1}\bm{\vec{s}}^{(p)} - \mathbf{Q}^{-1}_{c(C-1)}\mathbf{V}^{-1}\bm{\vec{s}}^{(C)}
\end{align}
which provides the learned filters as a linear combination of mean signals corresponding to each state to be classified. Comparing with Eq.~(\ref{eq:analyticFilters}) from the main text, we have:
\begin{align}
    \bm{\vec{f}}_c = \sum_{p=1}^{C} C_{cp} \mathbf{V}^{-1}\bm{\vec{s}}^{(p)},~C_{cp} = 
    \begin{cases}
        +\mathbf{Q}^{-1}_{c1}~~~~&{\rm if}~p = 1, \\
        - \mathbf{Q}^{-1}_{c(C-1)}~~~~&{\rm if}~p = C, \\
        \mathbf{Q}^{-1}_{cp}-\mathbf{Q}^{-1}_{c(p-1)}~~~~&{\rm otherwise}.
    \end{cases}
    \label{appeq:ckpdef}
\end{align}

\subsection{Reduction to standard matched filter for binary classification ($C=2$)}

For $C=2$, the matrix system in Eq.~(\ref{appeq:filterSystem}) reduces to a single equation:
\begin{align}
    \vec{S}^{(1)}-\vec{S}^{(2)} = \left[M_{11}- M_{21}-(M_{12}-M_{22}) \right]\vec{F}_1 + \left[M_{21}-M_{22} \right]\vec{n}
\end{align}
From here we can directly read off the filter and bias term:
\begin{align}
    \begin{pmatrix}
    \bm{\vec{f}}_1 \\
    \bv_1 
    \end{pmatrix} 
    = \frac{ \mathbf{V}^{-1}} {M_{11}- M_{21}-(M_{12}-M_{22}) }
    \begin{pmatrix}
    \bm{\vec{s}}^{(1)}-\bm{\vec{s}}^{(2)} \\
    0
    \end{pmatrix}
    - \frac{M_{21}-M_{22} }{M_{11}- M_{21}-(M_{12}-M_{22}) }
    \begin{pmatrix}
        \bm{\vec{0}} \\
        1 
    \end{pmatrix}
\end{align}

\subsection{Example: analytic construction of \RC{}-learned filters for three-state classification ($C=3$)}

We now provide an example of the construction of \RC{}-learned optimal filters for $C=3$ state classification. To compute these filters using Eq.~(\ref{appeq:ckpdef}), we simply require knowledge of the matrix $\mathbf{Q}$, whose matrix elements are given by Eq.~(\ref{appeq:qcpdef}). For $C=3$, $\mathbf{Q} \in \mathbb{R}^{2\times 2}$, and the distinct state pairs $P_p$ for $p=1,2$ are given by $P_1 = [1,2], P_2 = [2,3]$. Then, $\mathbf{Q}$ takes the form:
\begin{align}
    \mathbf{Q} = 
    \begin{pmatrix}
        M_{11}-M_{21}-(M_{13}-M_{23}) & & M_{12}-M_{22}-(M_{13}-M_{23}) \\
        M_{21}-M_{31}-(M_{23}-M_{33}) & & M_{22}-M_{32}-(M_{23}-M_{33})
    \end{pmatrix}
\end{align}
and its inverse can hence be easily computed:
\begin{align}
    \mathbf{Q}^{-1} = 
    \frac{1}{\det \mathbf{Q}}
    \begin{pmatrix}
        M_{22}-M_{32}-(M_{23}-M_{33}) & & (M_{13}-M_{23})-(M_{12}-M_{22}) \\
        (M_{23}-M_{33})-(M_{21}-M_{31}) & & M_{11}-M_{21}-(M_{13}-M_{23}) 
    \end{pmatrix}
\end{align}
Using Eq.~(\ref{appeq:ckpdef}), we can therefore write for the non-trivial \RC{}-learned filters:
\begin{subequations}
\begin{align}
    \bm{\vec{f}}_1 &= \frac{\mathbf{V}^{-1}}{\det \mathbf{Q}} \left\{ \left[M_{22}-M_{32}-(M_{23}-M_{33})\right]\bm{\vec{s}}^{(1)} + \left[M_{13}-M_{12}-(M_{33}-M_{32}) \right]\bm{\vec{s}}^{(2)} + \left[M_{12}-M_{22}-(M_{13}-M_{23})\right]\bm{\vec{s}}^{(3)}  \right\} \\
    \bm{\vec{f}}_2 &= \frac{\mathbf{V}^{-1}}{\det \mathbf{Q}} \left\{ \left[ M_{23}-M_{33}-(M_{21}-M_{31}) \right]\bm{\vec{s}}^{(1)} + \left[M_{11}-M_{13}-(M_{31}-M_{33}) \right]\bm{\vec{s}}^{(2)} + \left[M_{13}-M_{23}-(M_{11}-M_{21}) \right]\bm{\vec{s}}^{(3)} \right\}
\end{align}
\end{subequations}
Note that the final filter $\bm{\vec{f}}_3$ must be defined by the constraint Eq.~(\ref{eq:constraint}) of the main text (or equivalently Eq.~(\ref{appeq:constraint}) of this appendix section).

\subsection{\RC{}-learned optimal filters for multi-state classification under Gaussian white noise}
\label{subsec:white}

We now present an example of \RC{}-learned optimal filters for dispersive qubit readout where the dominant noise source is additive Gaussian white noise. This is ensured via a theoretical simulation of Eq.~(\ref{eq:sme}) as discussed in Sec.~\ref{subsec:whitenoisesim} of the main text. These simulations yield single-shot measurement records for any number of transmon states. Examples of these records are then shown in Fig.~\ref{fig:analyticFilters} for four distinct transmon states $p \in \{e,g,f,h\}$; for ease of visualization we only consider the $I$ quadrature. We use this simulated dataset as a training set to determine the \RC{}-learned filters under the white noise assumption, as defined by Eq.~(\ref{eq:analyticFilters}) in the main text with $\mathbf{V} \propto \bar{\mathbf{I}}$. While the individual measurement records are obscured by white noise, the empirically-calculated mean traces in the top right of Fig.~\ref{fig:analyticFilters} illustrate the physics at play. The mean traces grow once the measurement tone is turned on past $\mathcal{T}_{\rm on}$, and settle to a steady state depending on the induced dispersive shift $\chi_p$ and the measurement amplitude. The traces begin to fall beyond $\mathcal{T}_{\rm off}$ and eventually settle to background levels. These means, together with an estimate of the variances, determine the coefficients $C_{kp}$ that define the contribution of the mean trace $\vec{s}^{(p)}$ to the $k$th filter, and are hence sufficient to calculate optimal filters for the classification of any subset of states. 

%%%%%%%%%%%%%%%%%%%%%%%%%%%%%%%%%%%%%%%%%%%%%%%%%%%%%%%%%%%%%%%%%%%%%%%%%%%%%%%%%%%%%%%%%%%%%%%%%%%

\begin{figure*}[t]
    \centering
    \includegraphics[scale=1.0]{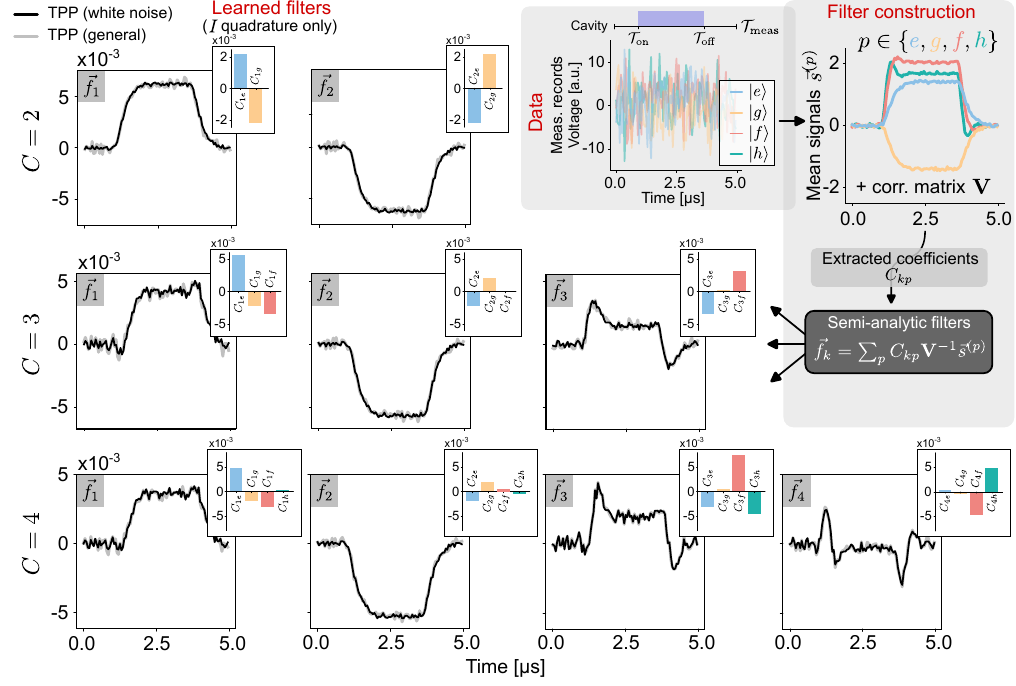}
    \caption{\textbf{\RC{}-learned optimal filters for simulated multi-state classification under Gaussian white noise conditions.} Top right: single-shot measurement records obtained under the indicated measurement tone, and empirical mean traces of several heterodyne records of the cavity $I$ quadrature corresponding to multi-level atom states $\ket{p}$ where $p \in \{e,g,f,h\}$. For a transmon $\chi_p/\kappa \in \{-\chi, \chi, -3\chi,  -5\chi\}$, $\chi/\kappa = 0.195$, and $\kappa/2\pi = 1.54~{\rm MHz}$. Rows: \RC{}-learned optimal filters for classifying states $p \in \{e,g\}$ ($C=2$), $\{e,g,f\}$ ($C=3$), and $\{e,g,f,h\}$ ($C=4$). Black curves are filters learned under the white noise assumption, calculated analytically using Eq.~(\ref{eq:analyticFilters}). Bar plots show the coefficients $C_{kp}$ applied to respective mean traces in calculating these filters. Gray curves are general filters calculated by numerically solving Eq.~(\ref{eq:wopt}). Both analytically-computed white noise filters and general filters can be extended to arbitrary $C$. }
    \label{fig:analyticFilters}
\end{figure*}

%%%%%%%%%%%%%%%%%%%%%%%%%%%%%%%%%%%%%%%%%%%%%%%%%%%%%%%%%%%%%%%%%%%%%%%%%%%%%%%%%%%%%%%%%%%%%%%%%%%

For the standard binary classification task ($C=2$) of distinguishing $\{e,g\}$ states, the learned filters are shown in black in the top row of Fig.~\ref{fig:analyticFilters}, together with bar plots showing the coefficients $C_{kp}$. Again for visualization, we only show filters $\vec{f}_k \in \mathbb{R}^{\NT}$ for $I$ quadrature data; the complete vector $\bm{\vec{f}}_k$ includes filters for all $\NO$ observables. For the binary case, the $k=1$ \RC{}-learned filter \textit{always} satisfies $C_{1e} = -C_{1g}$. Hence it is simply proportional to the difference of mean traces for the two states, $\vec{f}_1 \propto \vec{s}^{(e)}-\vec{s}^{(g)}$, making it exactly equivalent to the standard matched filter for binary classification, as discussed in the previous subsection. We note that the second filter ($k=2$) is simply the negative of the first, as demanded by Eq.~(\ref{eq:constraint}).

Crucially, the \RC{} approach now provides the generalization of such matched filters to the classification of an arbitrary number of states. For three-state ($C=3$) classification of $\{e,g,f\}$ states, the three \RC{}-learned filters are plotted in the middle row, while the last row shows the four filters for the classification of $C=4$ states $\{e,g,f,h\}$. Filters for the classification of an arbitrary number of states $C$ can be constructed similarly. The bar plots of $C_{kp}$ show how these filters typically have non-zero contributions from the mean traces for \textit{all} states. This emphasizes that the \RC{}-learned filters are not simply a collection of binary matched filters, but a more non-trivial construction. Most importantly, our analytic approach enables this construction by inverting a matrix in $\mathbb{R}^{(C-1) \times (C-1)}$ to determine $C_{kp}$. This is a substantially lower complexity relative to the pseudoinverse calculation demanded by Eq.~(\ref{eq:wopt}), which requires inverting a much larger matrix $\mathbf{C} \in \mathbb{R}^{\NO\cdot\NT \times \NO\cdot\NT}$.

% We note that while the \RC{} framework can be cast as the learning and application of optimal filters, the second stage of standard post-processing - the fitting of Gaussian profiles and calculation of separatrices - is also completely forgone. Under the hood, such fitting also relies on the calculation of mean traces and variances, precisely the same quantities that determine the bias terms in the \RC{} framework. In addition to the efficiency of requiring only a number of filters that scales linearly with the number of states $C$ to be classified, the \RC{} framework provides the conceptual simplicity of predict class labels via a simple linear transformation that can natively be generalized to arbitrary $C$. 

Of course, the latter approach of obtaining $\WOopt$ and hence \RC{} filters using Eq.~(\ref{eq:wopt}) can also be employed for learning using the same training data. Here, it yields the underlying filters in gray. The resulting filters appear to simply be noisier versions of the analytically calculated filters. The reason for this straightforward: the fact that the noise in the measurement data is additive Gaussian white noise is a key piece of information used in calculating the \RC{} filters via Eq.~(\ref{eq:analyticFilters}), but is not \textit{a priori} known to the general \RC{}. The latter makes no assumptions regarding the underlying noise statistics of the dataset. Instead, the training procedure itself enables the \RC{} to learn the statistics of the noise and adjust $\WOopt$ accordingly. The fact that the general \RC{} filters approach the white noise filters shows this learning in practice. This ability to extract noise statistics from data is a key feature that makes \RC{} learning useful under more general noise conditions, as was demonstrated in Secs.~\ref{sec:rcexpt},~\ref{sec:corr} of the main text.

\subsection{Comparison of \RC{} against $C-1$ instances of FGDA}
\label{app:multifgda}

In Sec.~\ref{subsec:whitenoisesim} of the main text, the performance of \RC{}-learned optimal filters was compared against standard FGDA implementations where a single matched filter is used. However, as the \RC{} uses $C-1$ independent filters, it is natural to ask for $C>2$ state classification tasks whether employing multiple instances of FGDA with distinct filters could provide an improvement in performance. In other words, is the improvement in \RC{} performance observed in Fig.~\ref{fig:analyticFiltersPerf} of the main text arising simply because the \RC{} is using more filters, or is due to the learned optimal filters being able to extract more useful information from the noisy temporal measurement data?

To investigate this, we consider the $C=3$ state classification task from Sec.~\ref{subsec:whitenoisesim}, but now compare the \RC{} against $C-1$ instances of the FGDA. A standard approach to do so is to consider one-versus-all classification. Here, for a single instance, an FGDA is trained to process temporal data and to output a state label as being $p$, or \textit{not} p, (or $!p$ for short), instead of predicting a precise state label in the $!p$ case. The `filter' portion of this FGDA can be labelled a one-versus-all matched filter, and can be constructed for example as:
\begin{align}
    \vec{h}_{I,p} = \frac{1}{\NTrain}\sum_{n=1}^{\NTrain} \left(\vec{I}_{(n)}^{(p)} - \frac{1}{C-1}\sum_{p'\neq p} \vec{I}_{(n)}^{(p')} \right)  
    \label{appeq:ovafilter}
\end{align}
Next, a second instance of the FGDA processes the same temporal data, but using a one-versus-all matched filter constructed for a \textit{different} state label $q$, and hence now predicts the state label as being $q$ or $!q$`. FGDA instances are used to process temporal data until $C-1$ instances have been used, and hence one of $2^{(C-1)}$ possible outcomes has been obtained. A concrete example of the possible outcomes for $C=3$ state classification is shown in Fig.~\ref{fig:multiFGDA}(a) for one-versus-all filters constructed for $p=g, q=e$. Depending on the possible joint outcome, a state label can finally be assigned: for example, the result $g$ and $!e$ is consistent with the state label $g$, $!g$ and $e$ implies $e$, and $!g$ and $!e$ implies $f$. Note that the final outcome $g$ and $e$ is ambiguous; here we use a random choice to assign a state label.

Note that using $C-1$ instances of the FGDA introduces more ambiguities than using just a single filter: different choices of $p$ and $q$ can be made, as indicated by the other tables in Fig.~\ref{fig:multiFGDA}(a) where we either choose $p=e,q=f$ or $p=g,q=f$. There is even greater ambiguity about the choice of the $C-1$ one-versus-all matched filters, Eq.~(\ref{appeq:ovafilter}), where the prefactors of each mean trace can be chosen arbitrarily. Even before exploring the performance of $C-1$ FGDA instances, we note that the \RC{} already provides a unique set of filters, determined by coefficients $C_{kp}$ as given by Eq.~(\ref{appeq:ckpdef}).

%%%%%%%%%%%%%%%%%%%%%%%%%%%%%%%%%%%%%%%%%%%%%%%%%%%%%%%%%%%%%%%%%%%%%%%%%%%%%%%%%%%%%%%%%%%%%%%%%%

\begin{figure}[t]
    \centering
    \includegraphics[scale=1.0]{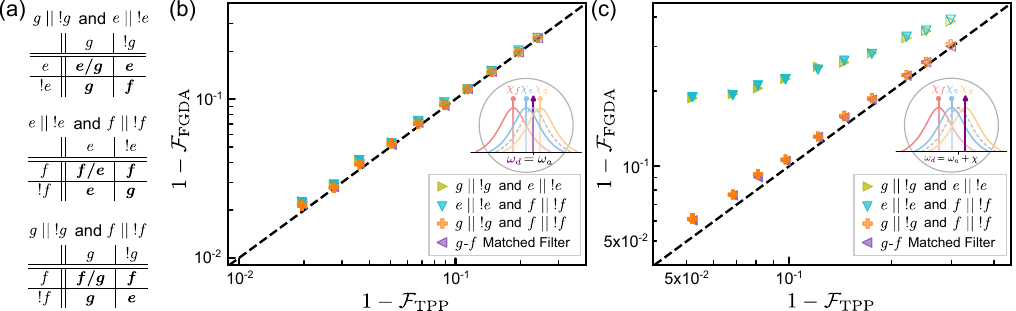}
    \caption{\textbf{Multi-state ($C=3$) classification performance of \RC{} versus $C-1$ instances of FGDA under Gaussian white noise conditions.} (a) Three distinct schemes (each corresponding to a different table) to implement $C=3$ state classification using $C-1$ instances of FGDA. Each instance predicts outcomes given by headers of rows and columns respectively, while bold labels indicate final predicted labels based on joint outcomes; see text for details. Performance comparison for (b) the same readout conditions as Fig.~\ref{fig:analyticFiltersPerf} of the main text, and (c) for readout conditions where the measurement drive is resonant with the dispersively-shifted cavity when the transmon qubit is in state $\ket{g}$. The \RC{} still outperforms $C-1$ FGDA instances, with the latter's performance also varying depending on readout conditions. }
    \label{fig:multiFGDA}
\end{figure}

%%%%%%%%%%%%%%%%%%%%%%%%%%%%%%%%%%%%%%%%%%%%%%%%%%%%%%%%%%%%%%%%%%%%%%%%%%%%%%%%%%%%%%%%%%%%%%%%%%

We now compare the performance of the \RC{} against the three distinct $C-1$ FGDA instance implementations shown in Fig.~\ref{fig:multiFGDA}(a), first for the readout conditions from Fig.~\ref{fig:analyticFiltersPerf} of the main text; the results are shown in Fig.~\ref{fig:multiFGDA}(b). Also shown is the use of a single $g$-$f$ matched filter, which was found to match the performance of the \RC{} in this case. We clearly see that the performance of the three distinct $C-1$ FGDA instances matches the \RC{} performance much more closely than the single matched filters used in Fig.~\ref{fig:analyticFiltersPerf} of the main text. However, if the readout conditions are modified, for example if the cavity readout drive is now resonant with the cavity frequency when the qubit is in the ground state $\ket{g}$, the performance can vary significantly, as shown in Fig.~\ref{fig:multiFGDA}(c). Now, all $C-1$ FGDA instances have a higher classification infidelity than the \RC{}, with certain instances faring much worse than others. 

It is therefore clear that the improvement in classification fidelity provided by the \RC{} is \textit{not} due only to its use of more than a single filter: $C-1$ FGDA instances using the same number of independent filters as the \RC{} do not always match its performance. This emphasizes the need to optimize the individual filters used; the \RC{} provides an autonomous, model-free approach to achieve precisely this objective for the classification of an arbitrary number of states.

\subsection{Semi-analytic \RC{}-learned optimal filters beyond Gaussian white noise conditions}

As shown in the main text, a key feature of the \RC{} is that applies to post-processing of temporal data experiencing more general noise conditions than simply uniform, observable-independent Gaussian white noise. In the main text, we compared numerically-calculated general \RC{} filters to semi-analytic filters computed under the white noise approximation. In this subsection we also show the semi-analytic but general \RC{} filters, as defined by Eq.~(\ref{eq:analyticFilters}) of the main text for a general correlation matrix $\mathbf{V}$. 

We start with a simple case where the required $\mathbf{V}^{-1}$ can be computed analytically. Consider the case of heterodyne measurement $\NO = 2$ but where the two measured observables (quadrature time series $\vec{I}$ and $\vec{Q}$) have stationary but \textit{distinct} variances $\sigma_I^2, \sigma_Q^2$ respectively; for concreteness we assume $\sigma_Q^2 > \sigma_I^2$. In this case, $\mathbf{V}$ takes the simple form:
\begin{align}
    \mathbf{V} = 
    \begin{pmatrix}
        \sigma_I^2 \tilde{\mathbf{I}} & \mathbf{0} \\
        \mathbf{0} & \sigma_Q^2 \tilde{\mathbf{I}}
    \end{pmatrix}
\end{align}
where $\tilde{\mathbf{I}}$ is the identity matrix in $\mathbb{R}^{\NT\ times \NT}$. Of course, this form of $\mathbf{V}$ can be straightforwardly inverted:
\begin{align}
    \mathbf{V}^{-1} = 
    \begin{pmatrix}
        \frac{1}{\sigma_I^2} \tilde{\mathbf{I}} & \mathbf{0} \\
        \mathbf{0} & \frac{1}{\sigma_Q^2} \tilde{\mathbf{I}}
    \end{pmatrix}
\end{align}
For convenience, we define filters and mean traces for each quadrature as $\bm{\vec{f}}_k = 
\begin{psmallmatrix} \vec{f}_k^I \\ \vec{f}_k^Q \end{psmallmatrix}$,~$\bm{\vec{s}}^{(p)}_k =\begin{psmallmatrix} \vec{s}_k^{I(p)} \\ \vec{s}_k^{Q(p)} \end{psmallmatrix}$ respectively. To calculate the semi-analytic general filters, we then simply use Eq.~(\ref{eq:analyticFilters}) from the main text to immediately find:
\begin{align}
    \vec{f}_k^I &= \sum_k C_{kp}(\mathbf{V})\frac{1}{\sigma_I^2}\vec{s}_k^{I(p)} \nonumber \\
    \vec{f}_k^Q &= \sum_k C_{kp}(\mathbf{V})\frac{1}{\sigma_Q^2}\vec{s}_k^{Q(p)}
    \label{appeq:genFilterIQ}
\end{align}
We see that there is now a relative weighting of the filters in accordance with their variance: noisier observables are suppressed relative to less noisy observables. Additionally, the coefficients $C_{kp}$ also depend on $\mathbf{V}^{-1}$. In Fig.~\ref{fig:genAnalyticFilters}(a) we plot the resulting filters for the readout conditions considered in Fig.~\ref{fig:multiFGDA}(b) (this ensures both $I$ and $Q$ quadratures have non-zero mean signal values), using both the semi-analytic general \RC{} filters given by Eq.~(\ref{appeq:genFilterIQ}), as well as the exact general \RC{} filters; the latter are shown with thicker lines deliberately to highlight differences between the two (to be expanded upon in due course). Finally, also shown are the filters assuming uniform white noise across the measured quadratures, which are clearly distinct from the general filters and do not penalize the noisier $Q$ quadrature.

Secondly, in Fig.~\ref{fig:genAnalyticFilters}(b), we consider the case of correlated quantum noise added by a finite-bandwidth phase-preserving quantum amplifier from Sec.~\ref{subsec:amp} of the main text, now also showing calculated semi-analytic general filters. We see that for all cases the semi-analytic general \RC{} filters show only very small differences when compared to the exact general \RC{} filter. As both schemes use empirically-calculated mean traces to construct the Gram matrix $\mathbf{G}$ and empirically-estimate the correlation matrix $\mathbf{V}$, the residual differences can be attributed to the fact that the semi-analytic \RC{} filter assumes the noise terms have zero mean, while the exact general filter does not make such an assumption. 

We will also emphasize that computing the exact general \RC{} filter requires the inversion of the matrix $\mathbf{C} \in \mathbb{R}^{(\NO\cdot\NT+1)\times(\NO\cdot\NT+1)}$, while the semi-analytic general \RC{} requires the inversion of $\mathbf{V} \in \mathbb{R}^{(\NO\cdot\NT)\times(\NO\cdot\NT)}$. In the general case, the numerical advantage in inverting the slightly smaller matrix $\mathbf{V}$ is not as significant as it is in the special case where $\mathbf{V}$ is proportional to the identity matrix. However, in cases where an analytic form for $\mathbf{V}$ and more importantly its inverse is similarly known, the semi-analytic general \RC{} filter can be more numerically-efficient than calculating these filters exactly.

%%%%%%%%%%%%%%%%%%%%%%%%%%%%%%%%%%%%%%%%%%%%%%%%%%%%%%%%%%%%%%%%%%%%%%%%%%%%%%%%%%%%%%%%%%%%%%%%%%

\begin{figure}[t]
    \centering
    \includegraphics[scale=1.0]{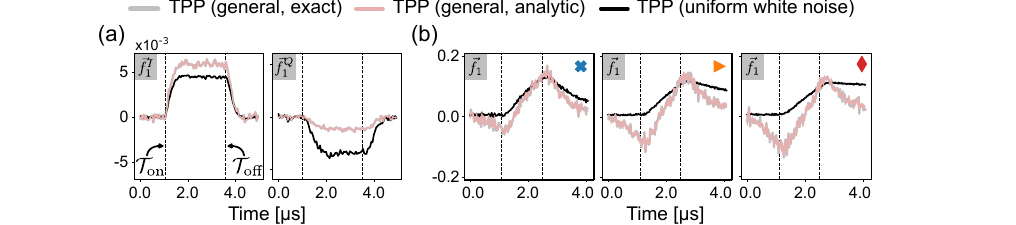}
    \caption{\textbf{Comparison of semi-analytic and general \RC{} filters under general noise correlation conditions.} Filters are shown for (a) simulated data where different quadrature time series $\vec{I}, \vec{Q}$ have different variances $\sigma^2_I,\sigma^2_Q$ respectively, and (b) simulated data from a phase-preserving quantum-limited amplifier, as in Sec.~\ref{subsec:amp} of the main text. Excellent agreement is observed between the exact general \RC{} filter and the semi-analytic general \RC{} filter, while both are markedly different from the black curves, which show \RC{} filters under the assumption of uniform (namely, observable-independent) Gaussian white noise. }
    \label{fig:genAnalyticFilters}
\end{figure}

%%%%%%%%%%%%%%%%%%%%%%%%%%%%%%%%%%%%%%%%%%%%%%%%%%%%%%%%%%%%%%%%%%%%%%%%%%%%%%%%%%%%%%%%%%%%%%%%%%

\newpage

\section{Supplementary classification results}

\subsection{Classification performance versus increasing signal amplitude for real qubit readout}
\label{app:fidvamp}

In theory, namely ignoring measurement chain non-idealities and qubit transitions discussed in Sec.~\ref{sec:rcfilter} of the main text, increase in measurement tone amplitude leads to an improvement in qubit classification fidelity. However, for real qubits, additional effects can create complex readout conditions such that increasing the measurement tone amplitude may not uniformly improve readout fidelity. To analyze whether increased readout power facilitates improved readout fidelity for the real qubit readout data collected in this work, we analyze the data in Fig.~\ref{fig:exptResults} in a slightly different form. We first introduce the quantity:
\begin{align}
    \mathcal{N}_j(s) = \frac{1-\mathcal{F}_j(s)}{1-\mathcal{F}_j(s_0)} 
    \label{appeq:ndef}
\end{align}
where $j \in \{{\rm FGDA}, {\rm \RC{}} \}$ depending on the used classification scheme, while $s$ denotes signal amplitude and $s_0$ is the smallest signal amplitude for a given dataset. As a result, $\mathcal{N}_j$ is simply the infidelity as a function of measurement tone amplitude, normalized by the infidelity at the smallest amplitude; we therefore require $\mathcal{N}_j < 1$ for a reduction in readout \textit{in}fidelity (and hence an improvement in readout \textit{fidelity}) for increasing readout power.

In Fig.~\ref{fig:FidVSigAmp}, we now plot $\mathcal{N}_{\rm FGDA}$ against $\mathcal{N}_{\rm TPP}$ for the two dispersive qubit-cavity systems that were analyzed as a function of measurement tone amplitude in the main text. The datapoint at $\mathcal{N}_{\rm FGDA} = \mathcal{N}_{\rm TPP} = 1$ for each dataset corresponds to the lowest amplitude, by construction of Eq.~(\ref{appeq:ndef}). We next note that a few datapoints fall into the category where both $\mathcal{N}_{\rm TPP} > 1$ and $\mathcal{N}_{\rm FGDA} > 1$, indicating increased \textit{in}fidelity with increasing signal amplitude using either classification scheme. Here, non-idealities such as enhanced qubit transitions degrade the measured data, which neither classification scheme is able to overcome. We note that here we still have $\mathcal{N}_{\rm FGDA} > \mathcal{N}_{\rm TPP}$, so that the FGDA performance is worse than the \RC{}.

Of all the other datapoints corresponding to higher readout amplitudes, a majority lie in the blue shaded regions of the plot, where $\mathcal{N}_{\rm TPP} < 1$. These are qubit readout conditions for which increasing the measurement tone amplitude leads to improved classification performance when using the \RC{}. Of these points, just over half also have $\mathcal{N}_{\rm FGDA} < 1$, implying that using either the FGDA or the \RC{} provides an improvement. Again, $\mathcal{N}_{\rm FGDA} > \mathcal{N}_{\rm TPP}$, so the improvement is larger with the \RC{}.

Crucially, the other half of the datapoints are such that $\mathcal{N}_{\rm TPP} < 1$ while $\mathcal{N}_{\rm FGDA} > 1$. For these regimes, the use of the \RC{} is necessary to extract an advantage in readout fidelity with increasing signal amplitude. Equally as importantly, \textit{none} of the datapoints fall in the category where $\mathcal{N}_{\rm TPP} > 1$ while $\mathcal{N}_{\rm FGDA} < 1$; this indicates that there is in general no disadvantage to deploying the \RC{} instead of the FGDA at higher signal amplitudes, as the \RC{} will not be outperformed by the FGDA. Together, these results supplement findings in the main text that the \RC{} can provide a robust classification scheme to extract maximum performance in complex readout regimes at high powers.

%%%%%%%%%%%%%%%%%%%%%%%%%%%%%%%%%%%%%%%%%%%%%%%%%%%%%%%%%%%%%%%%%%%%%%%%%%%%%%%%%%%%%%%%%%%%%%%%%%

\begin{figure}[t]
    \centering
    \includegraphics[scale=1.0]{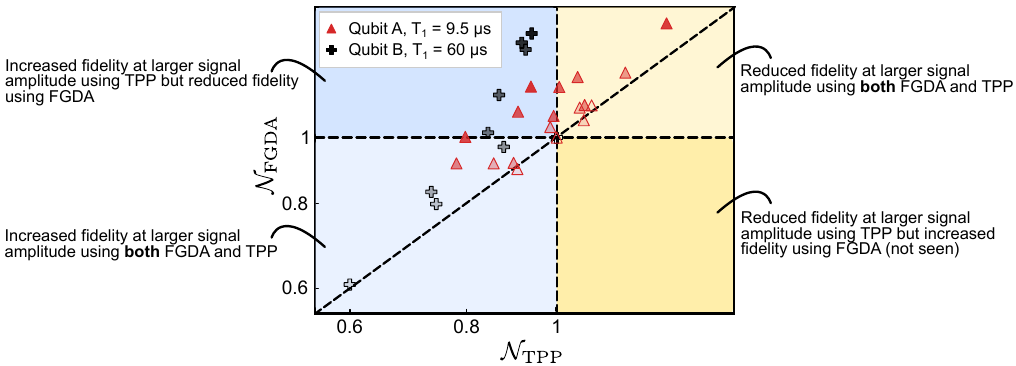}
    \caption{\textbf{Classification performance of \RC{} versus FGDA as a function of increasing measurement tone (signal) amplitude for readout of real qubits.} Same data as Fig.~\ref{fig:exptResults} of the main text, but now plotting $\mathcal{N}$, the classification infidelity normalized by infidelity at lowest signal amplitude for each dispersive qubit-cavity system shown; see Eq.~(\ref{appeq:ndef}). As before, more opaque markers again indicate stronger measurement tone amplitudes.  }
    \label{fig:FidVSigAmp}
\end{figure}

%%%%%%%%%%%%%%%%%%%%%%%%%%%%%%%%%%%%%%%%%%%%%%%%%%%%%%%%%%%%%%%%%%%%%%%%%%%%%%%%%%%%%%%%%%%%%%%%%%

\subsection{3-state classification results for real qubit readout}
\label{app:3state}

In this appendix section we include some supplementary results to Fig.~\ref{fig:exptResults} of the main text, now comparing classification performance for multi-state ($C=3$) classification for real qubit readout of $p \in \{e,g,f\}$. The results are shown in Fig.~\ref{fig:3State} for the readout of Qubit B.

The standard FGDA is deployed here using the $g$-$f$ matched filter, introduced in Sec.~\ref{sec:rcfilter}; as discussed in Appendix.~\ref{app:multifgda}, this provides the best performance amongst other single matched filters, while $C-1$ matched filters do not provide a marked improvement in this readout configuration.  We note again that the \RC{} outperforms the FGDA for almost all data points, and the performance difference increases at stronger measurement tone amplitudes. The underperformance at the lowest measurement tone amplitude can again be attributed to the fact that under these simpler readout conditions, the optimal filter is close to the white noise filter (see Fig.~\ref{fig:filterComparisons} of the main text); the general \RC{} does not know this \textit{a priori}, and must learn this information from a finite training dataset, whose size limits constrains the fidelity of the learned filter. At higher signal amplitudes, the \RC{} outperforms the FGDA inspite of this training cost. Overall, we see that the \RC{} provides a better classification scheme for multi-state readout of real qubits, supplementing the improvement in performance demonstrated for binary classification of real qubits in the main text.

%%%%%%%%%%%%%%%%%%%%%%%%%%%%%%%%%%%%%%%%%%%%%%%%%%%%%%%%%%%%%%%%%%%%%%%%%%%%%%%%%%%%%%%%%%%%%%%%%%

\begin{figure}[t]
    \centering
    \includegraphics[scale=1.0]{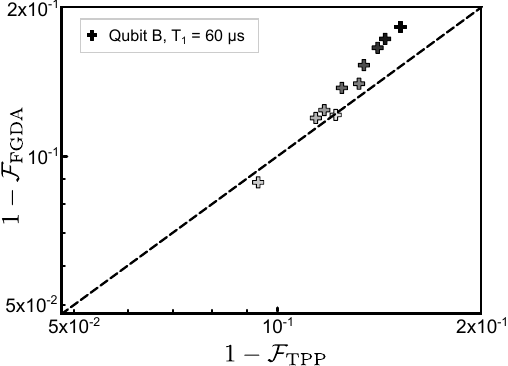}
    \caption{\textbf{Multi-state ($C=3$) classification performance of \RC{} versus FGDA for readout of real qubits.} Classification infidelities using both schemes are plotted against each other for one of the three dispersive qubit-cavity systems analyzed in the main text.  The dashed line marks $1-\fidmf = 1-\mathcal{F}_{\rm TPP}$. As before, more opaque markers indicate stronger measurement tone amplitudes.  }
    \label{fig:3State}
\end{figure}

%%%%%%%%%%%%%%%%%%%%%%%%%%%%%%%%%%%%%%%%%%%%%%%%%%%%%%%%%%%%%%%%%%%%%%%%%%%%%%%%%%%%%%%%%%%%%%%%%%

\subsection{\RC{} learning of correlated \textit{classical} noise}
\label{app:coloredclassical}

In this appendix section, we use a further example to demonstrate the ability of \RC{}-based learning to extract correlations from measured data, to supplement simulations in Sec.~\ref{sec:corr}. Like Sec.~\ref{subsec:amp}, we again consider simulated datasets of measured heterodyne records from a measurement chain of a qubit-cavity-amplifier setup, as in Appendix~\ref{subsec:white}. Now, however, we consider the excess classical noise added by the measurement process to also possess a component with a colored spectrum (suppressing quadrature labels for clarity):
\begin{align}
   \xi^{\rm cl}(t_i) = \sigma^{\rm W}\xi^{\rm W}(t_i) + \sigma^{\rm P}\xi^{\rm P}(t_i) 
    \label{eq:clNoise}
\end{align}
where $\xi^{\rm W}(t_i)$ describes white noise as before, while $\xi^{\rm P}(t_i)$ describes $1/f$ (or pink) noise. The power spectral density of the noise processes is given by the Fourier transform of their steady-state autocorrelation function (by the Wiener-Khinchin theorem),  $S_{\rm N}[f] = \int d\tau~e^{-i2\pi f \tau}\E{\xi^{\rm N}(0)\xi^{\rm N}(\tau)}$ for ${\rm N}\in\{\rm W,P\}$. The noise processes are normalized so that the total noise power, $\int df~|S_{\rm N}[f]|$ is the same for any of the considered noise processes; hence the relative magnitude $(\sigma^{\rm P}\!/\sigma^{\rm W})^2$ determines the relative strength of the noise processes with different correlation statistics. 

We restrict ourselves again to binary classification of states $\ket{e}$ and $\ket{g}$. In Fig.~\ref{fig:simResults}, we plot the calculated infidelities using the MF and \RC{} approaches against each other in logscale for different noise conditions parameterized by $(\sigma^{\rm P}\!/\sigma^{\rm W})^2$, and as a function of the coherent input tone power: darker markers correspond to readout with stronger input tones.

We immediately see that if the excess classical noise is purely white, the FGDA and \RC{} exhibit very similar performance: both lie along the dashed line of equal infidelities. However, the situation is very different if the added noise is colored, namely $(\sigma^{\rm P}\!/\sigma^{\rm W})^2 \neq 0$, and hence has a non-zero correlation timescale. We immediately note that even when the colored noise power is only a fraction of the white noise power, the \RC{}-learned filters provide a non-negligible improvement over the standard FGDA scheme using matched filters.

%%%%%%%%%%%%%%%%%%%%%%%%%%%%%%%%%%%%%%%%%%%%%%%%%%%%%%%%%%%%%%%%%%%%%%%%%%%%%%%%%%%%%%%%%%%%%%%%%%

\begin{figure}[t]
    \centering
    \includegraphics[scale=1.0]{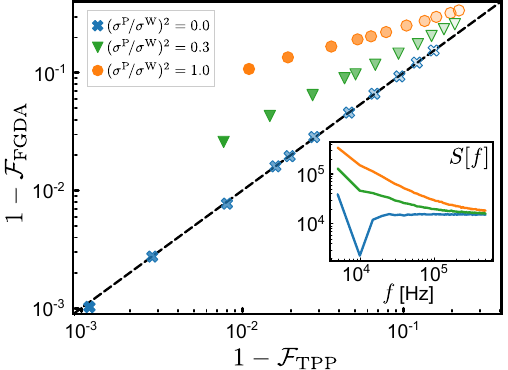}
    \caption{\textbf{Comparative classification performance of FGDA versus \RC{} in the presence of classical correlated noise.} We consider a $C=2$ (binary) dispersive qubit readout task using simulated data and for different colored noise conditions. Darker markers indicate stronger measurement tone amplitudes. The dashed line indicates $1-\fidmf = 1-\mathcal{F}_{\rm TPP}$. The inset plots the corresponding noise spectral density $S[f]$, which remains unchanged with coherent input power.}
    \label{fig:simResults}
\end{figure}

%%%%%%%%%%%%%%%%%%%%%%%%%%%%%%%%%%%%%%%%%%%%%%%%%%%%%%%%%%%%%%%%%%%%%%%%%%%%%%%%%%%%%%%%%%%%%%%%%%

\newpage

\section{Time-shuffled data}
\label{app:shuffle}

As discussed in Appendix~\ref{app:training}, the trained weights $\WO$ take the form of Eq.~(\ref{appeq:trainedWO}),
\begin{align}
    \bm{\mathcal{W}}^{\rm opt} = \mathbf{Y}\bm{\mathcal{X}}^T(\bm{\mathcal{X}}\bm{\mathcal{X}}^T-\lambda\mathbf{I})^{-1}
\end{align}
We now consider the operation of a matrix $\mathbf{J}$ on $\mathbf{X}$ that serves to re-order the time indices of measurement records; this amounts to an exchange of specific rows of $\mathbf{X}$ and is therefore referred to as an exchange matrix, a special case of the more general permutation matrix in standard linear algebra. As $\bm{\mathcal{X}} \in \mathbb{R}^{(\NO\cdot\NT+1)\times C\NTrain}$ and the exchange matrix is intended to switch \textit{rows} of the data, $\mathbf{J} \in \mathbb{R}^{(\NO\cdot\NT+1)\times(\NO\cdot\NT+1)}$. Furthermore, the exchange matrix satisfies the properties: $\mathbf{J}^{-1} = \mathbf{J} = \mathbf{J}^T$, so that $\mathbf{J}\mathbf{J} = \mathbf{I}$. 

We therefore define a new data matrix $\bm{\mathcal{X}}_J$ with exchanged rows under the action of the exchange matrix:
\begin{align}
    \bm{\mathcal{X}}_J = \mathbf{J}\bm{\mathcal{X}} \implies \bm{\mathcal{X}} = \mathbf{J}\bm{\mathcal{X}}_J
\end{align}
where we have used the property that $\mathbf{J}^{-1} = \mathbf{J}$. Note that the target matrix $\mathbf{Y}$ is unchanged, since the particular class a measurement record belongs to should not be related to time ordering of the measurement records.

Then, the trained weights can equivalently be written as:
\begin{align}
    \bm{\mathcal{W}}^{\rm opt} = \mathbf{Y}(\mathbf{J}\bm{\mathcal{X}}_J)^T(\mathbf{J}\bm{\mathcal{X}}_J\bm{\mathcal{X}}_J^T\mathbf{J}^T-\lambda\mathbf{I})^{-1}
\end{align}
which, after some simplification and using $\mathbf{J}^T = \mathbf{J}$ reduces to:
\begin{align}
    \bm{\mathcal{W}}^{\rm opt} = \mathbf{Y}\bm{\mathcal{X}}_J^T\mathbf{J}\mathbf{J}(\bm{\mathcal{X}}_J\bm{\mathcal{X}}_J^T)^{-1}\mathbf{J} = \left[\mathbf{Y}\bm{\mathcal{X}}_J^T(\bm{\mathcal{X}}_J\bm{\mathcal{X}}_J^T-\lambda\mathbf{I})^{-1}\right]\mathbf{J} 
\end{align}
The term in square brackets is simply the new trained weights when using the exchanged data matrix $\bm{\mathcal{X}}_J$; we label this $(\bm{\mathcal{W}}^{\rm opt})_J$. We therefore find:
\begin{align}
    (\bm{\mathcal{W}}^{\rm opt})_J = \bm{\mathcal{W}}^{\rm opt}\mathbf{J}
\end{align}
which simply indicates that the new trained weights are simply exchanged versions of the previous trained weights.

\begin{center}
    \rule{10cm}{1pt}
\end{center}

\end{widetext}

\end{document}